\documentclass{ws-ijmpa}
\usepackage[super,compress]{cite}
\usepackage{graphicx}
\usepackage{enumerate}
\usepackage{amsmath}
\usepackage{slashed}
\usepackage{tensor,amssymb}
\usepackage{epsfig,bm}
\usepackage{hyperref}
\usepackage{xcolor}

\newcommand{\G}{\textbf}
\newcommand{\be}{\begin{equation}}\newcommand{\ee}{\end{equation}}
\newcommand{\bea}{\begin{eqnarray}}\newcommand{\eea}{\end{eqnarray}}
\newcommand{\brr}{\begin{array}}\newcommand{\err}{\end{array}}
\newcommand{\bit}{\begin{itemize}}\newcommand{\eit}{\end{itemize}}
\newcommand{\ben}{\begin{enumerate}}\newcommand{\een}{\end{enumerate}}

\newcommand{\bbm}{\begin{bmatrix}}\newcommand{\ebm}{\end{bmatrix}}
\newcommand{\ba}{\begin{array}}
\newcommand{\ea}{\end{array}}

\newcommand{\bthe}{\begin{theorem}} \newcommand{\ethe}{\end{theorem}}
\newcommand{\ble}{\begin{Lemma}} \newcommand{\ele}{\end{Lemma}}

\newcommand{\dr}{\mathrm{d}}

\def\ha{\frac{1}{2}}

\def\intx{\int \! \! \mathrm{d}^3 \textbf{x}}

\def\ph{\varphi}
\def\lab{\label}\def\lan{\langle}
\def\lf{\left}

\def\non{\nonumber}\def\pa{\partial}\def\ran{\rangle}
\def\rar{\rightarrow}
\def\ri{\right}
\def\al{\alpha}\def\bt{\beta}\def\ga{\gamma}\def\Ga{\Gamma}
\def\de{\delta}\def\De{\Delta}

\def\la{\lambda}\def\La{\Lambda}\def\si{\sigma}
\def\om{\omega}\def\Om{\Omega}

\def\CP{{_{C\!P}}}
\def\mass{{_{1,2}}}
\def\flav{{e,\mu}}\def\1{{_{1}}}\def\2{{_{2}}}
\def\bk{{\bf {k}}}\def\bx{{\bf {x}}}

\newcommand{\ide}{1\hspace{-1mm}{\rm I}}

\def\noHe0{:\;\!\!\;\!\!:H_e(0):\;\!\!\;\!\!:}
\def\noHm0{:\;\!\!\;\!\!:H_\mu(0):\;\!\!\;\!\!:}

\def\boldsymbol#1{{\bm #1}}

\def\lab{\label}
\def\lan{\langle}
\def\lf{\left}

\def\non{\nonumber}
\def\pa{\partial}\def\ran{\rangle}
\def\rar{\rightarrow}
\def\ri{\right}

\def\al{\alpha}\def\bt{\beta}\def\ga{\gamma}
\def\Ga{\Gamma}\def\de{\delta}\def\De{\Delta}

\def\la{\lambda}
\def\La{\Lambda}\def\si{\sigma}
\def\om{\omega}\def\Om{\Omega}

\def\CP{{_{C\!P}}}
\def\mass{{_{1,2}}}
\def\flav{{e,\mu}}\def\1{{_{1}}}\def\2{{_{2}}}
\begin{document}
\markboth{M.Blasone and L.Smaldone}{Perturbative and nonperturbative aspect of flavor oscillations in quantum field theory}

%
\catchline{}{}{}{}{}
%

\title{Perturbative and nonperturbative aspects of neutrino oscillations in quantum field theory}

\author{Massimo Blasone and Luca Smaldone}

\address{Dipartimento di Fisica, Universit\`a di Salerno, Via Giovanni Paolo II 132, 84084 Fisciano (SA), Italy\\ 
INFN Sezione di Napoli, Gruppo collegato di Salerno, Italy
\\
blasone@sa.infn.it; lsmaldone@unisa.it}



\maketitle


\begin{abstract}
In this work, we present a comprehensive and pedagogical review of two quantum field theoretical approaches to neutrino flavor mixing and oscillations: the non-perturbative flavor Fock space formalism and the perturbative interaction picture framework. Starting from a minimally extended Standard Model, where neutrino masses and mixing are treated analogously to the quark sector, we derive the explicit form of leptonic flavor charges from the charged-current  Lagrangian in the spontaneously broken phase.
We present the standard quantum mechanical treatment of neutrino oscillations and an alternative derivation based on the first-quantized Dirac equation. We then review the construction of the flavor Fock space, in which flavor states emerge as eigenstates of the flavor charges, and show how oscillation probabilities can be computed both from charge and currents expectation values and from Green's functions. The non-trivial vacuum structure associated with this approach leads to key results such as the conservation of lepton number at tree level and a rigorous derivation of the time–energy uncertainty relation in neutrino oscillations. We also discuss the extension to the three-flavor case and the entangled nature of flavor states.
In parallel, we explore a perturbative approach that treats flavor mixing as an interaction. Starting from simple quantum mechanical models and extending to bosonic field theories, we show how neutrino oscillation probabilities can be derived via Dyson expansion in the interaction picture. Remarkably, this method reproduces the same oscillation formulas as the non-perturbative approach, within the expected limits. We emphasize the necessity of working at finite time, rather than in the asymptotic 
$S$-matrix framework.
Our analysis highlights the conceptual and structural unity of these approaches and offers a solid framework for further developments in the field of neutrino physics.
\end{abstract}

\ccode{PACS numbers:}

\tableofcontents

\section{Introduction}

Neutrinos are the most elusive particles known to date. Their existence was first proposed by Pauli in 1930 to resolve the apparent violation of energy conservation in the beta decay. Fermi later incorporated neutrinos into his theory of beta decay, which also marked the introduction of the weak interaction \cite{Fermi1934}. However, it took more than two decades before their existence was experimentally confirmed \cite{Cowan1956}.
In the original formulation of the Standard Model, neutrinos were assumed to be massless \cite{PhysRevLett.19.1264, Salam:1968rm}, and were thus regarded as a paradigmatic example of Weyl fermions \cite{Weyl1929}.

A turning point came with the Homestake experiment, which measured the flux of neutrinos produced in nuclear reactions in the Sun. The observed flux was only about one-third of the predicted value, revealing a  discrepancy that became known as the \emph{solar neutrino puzzle} \cite{Bahcall1976}.
Pontecorvo, who had earlier proposed the possibility of neutrino-antineutrino oscillations \cite{Pontecorvo:1957cp,Pontecorvo:1957qd}, together with his collaborators, offered an explanation: the neutrino flavor states, such as electron, muon, and tau, are not identical to the neutrino mass (energy) eigenstates \cite{Pontecorvo:1967fh,Gribov:1968kq,Bilenky:1975tb,Bilenky:1976yj,Bilenky:1977ne}. Instead, each flavor state is a linear superposition of mass eigenstates. Because the mass eigenstates evolve differently over time due to the mass differences, the original superposition turns into a different one, resulting in a non-zero probability of detecting a different flavor than the one originally produced.
In the case of solar neutrinos, this means that the electron neutrinos produced in the Sun can oscillate into other flavors (primarily muon neutrinos) which the Homestake detector was not designed to detect. The hypothesis proposed by Pontecorvo and his collaborators has since been confirmed by numerous experiments and is now widely accepted \cite{Vogel:2015wua,IceCube:2017lak,PhysRevD.98.030001,Nakano:2020lol,OPERA:2021xtu} (for an historical perspective also look at Ref. \cite{Bilenky:2016pep}).

For neutrino oscillations to occur, neutrinos must have different masses, implying that at least some of them cannot be massless. As a result, the Standard Model must be extended to account for non-zero neutrino masses. However, the precise mechanism by which neutrino masses arise, as well as their nature, Dirac or Majorana, remains unclear, and several theoretical frameworks have been proposed to explain their origin \cite{PhysRevD.22.2860,PhysRevD.98.055007,King:2017guk,Chaber:2018cbi,RevModPhys.82.2701,King:2008vg}.
One of the most widely studied proposals is the \emph{see-saw mechanism} \cite{MINKOWSKI1977421,Yanagida:1980xy}, originally introduced in the context of grand unified theories. In this review, however, we focus on a minimal extension of the Standard Model, in which (Dirac) neutrino masses are generated through Yukawa couplings to the Higgs field. As with quarks, the resulting mass matrices in the flavor basis are generally non-diagonal, meaning that weak interactions mix different mass eigenstates \cite{Miransky:1994vk}. This neutrino mixing is the underlying cause  for the phenomenon of neutrino oscillations.

Although Pontecorvo and his collaborators began with the hypothesis of field mixing \cite{Bilenky:1978nj}, their original treatment was ultimately based on a single-particle quantum mechanical (QM) framework, rather than being derived directly from the formalism of quantum field theory (QFT). Over the years, many efforts have been made to develop a consistent QFT-based description of neutrino mixing and oscillations \cite{PhysRevD.37.1935,PhysRevD.45.2414,PhysRevD.48.4310,Blasone:1995zc,Grimus:1996av,PhysRevD.59.113003,Beuthe:2001rc,giunti2007fundamentals,Ho:2012yja,Lobanov:2015esa,Grimus:2019hlq,Tureanu:2019pui,Blasone:2020wer,Blasone:2023brf}.
All these approaches converge to the conclusion that Pontecorvo’s theory provides an essentially correct description in the case of relativistic neutrinos (see, for example, Ref.\cite{Fantini:2018itu} for a comprehensive review of the standard theory). At the same time, they all predict the presence of non-relativistic corrections to the neutrino oscillation probability that arise naturally within a proper QFT treatment. However, different approaches can lead to very different corrections. For example, in Ref. \cite{PhysRevD.45.2414} it was argued that flavor states should, in general, be defined in relation to the specific production and detection processes. In this framework, the oscillation probability inherently depends on the details of how the neutrinos are produced and detected. A similar conclusion was reached in Ref.\cite{PhysRevD.48.4310}, where it was proposed that neutrinos should be treated as internal lines in Feynman diagrams, with the external lines corresponding to the particles involved in the production and in the detection. These approaches emphasize that neutrinos are not directly observed, but inferred through the interaction with associated particles.

Another approach was proposed in Ref. \cite{Blasone:1995zc} and later thoroughly developed (see e.g. Refs. \cite{PhysRevD.60.111302,Blasone:1998bx,BHV99,Hannabuss:2000hy,PhysRevD.64.013011,PhysRevD.65.096015,Lee:2017cqf}): one explicitly builds the Fock space for flavor fields and then flavor states are viewed as excitations of a \emph{flavor vacuum}, which is annhilated by flavor annihilation operators. The advantage of such formalism is that the flavor states so-defined turn out to be exact eigenstates of the neutrino flavor (lepton) charges. Because the lepton charges are non-conserved due to the flavor mixing, the approach is intrinsically a finite-time QFT approach \cite{Blasone2019}.  

The flavor Fock space approach, which is the main topic of the present review, builds upon a distinctive feature of QFT: the existence of infinitely many unitarily inequivalent representations of the canonical anticommutation relations (CAR). In systems with a finite number of degrees of freedom, such as in QM, the Stone–von Neumann theorem guarantees that all representations of the canonical commutation relations (CCR, i.e., the Weyl algebra) are unitarily equivalent, and therefore physically indistinguishable \cite{Stone1930,vonNeumann1931}. In contrast, QFT admits infinitely many unitarily inequivalent representations of both CCR and CAR on orthogonal Hilbert spaces. This property underlies a variety of nontrivial physical phenomena, such as spontaneous symmetry breaking \cite{FabPic,Miransky:1994vk}, superconductivity, and ferromagnetism \cite{umezawa1982thermo,blasone2011quantum}, and is formalized in relativistic QFT by Haag’s theorem \cite{Haagqft,haag1996local,Bogolyubov:1990kw}.
In the context of neutrino mixing, it turns out that the flavor and mass representations of the CAR are unitarily inequivalent. These two representations are related by an improper canonical transformation (a Bogoliubov transformation) which induces a non-trivial condensate structure in the flavor vacuum. This   closely resembles the vacuum condensates encountered in superconductivity \cite{PhysRev.122.345,PhysRev.124.246}. Interestingly, it has been proposed that such a condensate structure could be dynamically generated in certain $D$-foam models inspired by string theory \cite{Mavromatos:2009rf,Mavromatos:2012us,bigs1}, leading to a non-trivial Poincar\'e structure of vacuum \cite{Blasone:2010zn,bigs2}.

Within this QFT framework, the resulting oscillation formula includes both the standard Pontecorvo contribution and an additional high-frequency oscillating term. Each component is weighted by the squared coefficients of the Bogoliubov transformation discussed above. The high-frequency term becomes significant only in the non-relativistic regime, while in the relativistic limit it effectively vanishes, recovering the standard oscillation formula.

A recent approach to neutrino mixing and oscillations treats mixing itself as an interaction between different flavor fields \cite{Blasone:2023brf}. By employing the Dyson expansion of the time-evolution operator in the mixing coupling, one can compute flavor transition and survival probabilities within the interaction (Dirac) picture. Remarkably, the resulting oscillation formula coincides with that obtained via the flavor Fock space approach, at least within the perturbative regime.
As already emphasized, this framework is based on evaluating the matrix elements of the time-evolution operator rather than those of the $S$-matrix. This is not merely a matter of choice, but a conceptual necessity. In fact, it becomes clear that no flavor transitions can occur in the asymptotic regime when the energy is exactly resolved \cite{Blasone:2024nuf}. This is deeply connected to the time–energy uncertainty relation (TEUR), which imposes a fundamental condition for neutrino oscillations to take place \cite{Bilenky:2005hv,Bilenky2008,Akhmedov:2008zz,Bilenky:2008dk,Bilenky:2009zz,Bilenky:2011pk,Blasone2019, Blasone2020} (see also the comprehensive review \cite{Luciano:2023}).
As in the case of the non-perturbative flavor Fock space formalism, a consistent treatment demands the development of a finite-time QFT. The situation closely parallels that encountered in the study of unstable particles \cite{Blasone:2024zsn}, where finite-time methods have been employed to evaluate decay probabilities beyond the standard exponential (long-time) approximation, and to investigate phenomena such as the quantum Zeno effect. Notably, the structure of finite-time Feynman diagrams is strikingly similar in both the neutrino and unstable particle cases.

The aim of this work is to provide a comprehensive and pedagogical review of both the non-perturbative flavor Fock space approach and the perturbative interaction picture approach to neutrino mixing and oscillations. We begin by introducing a minimal extension of the Standard Model, in which neutrino masses and mixing are treated on the same footing as those of quarks. We then show that, by studying the charged-current lepton Lagrangian in the Goldstone (i.e., spontaneously broken) phase and working in the flavor basis, one can explicitly derive the form of the flavor charges. These foundational aspects are presented in Section~\ref{sectionw}.
In Section~\ref{sectionqm}, we review how neutrino flavor states and oscillations are described within quantum mechanics. We include both the standard derivation of the oscillation formula and an alternative treatment based on the first-quantized Dirac equation.
Section~\ref{1bfmixing} is dedicated to the non-perturbative flavor Fock space approach. We begin with the construction of the flavor Fock space and demonstrate that the resulting flavor states are eigenstates of the flavor charges introduced earlier. We then compute the oscillation probabilities using two different methods: (i) as the expectation values of flavor charges evaluated on flavor states at a fixed reference time, and (ii) via the Green's functions in the flavor basis. We thus show how a general treatment, based on flavor currents, can be employed to derive a more realistic oscillations formula. We also show that the fact that flavor states are eigenstates of the corresponding flavor charges ensures lepton number conservation at tree level for short times. Furthermore, we explore the entangled nature of flavor states and compute the TEUR for neutrino oscillations, a central theme of this review. Finally, we briefly discuss the extension to the three-flavor case in Section~\ref{3flavorAppA}.
In Section~\ref{perneu}, we turn to the perturbative interaction picture approach. We begin with a quantum mechanical toy model to illustrate the key features of the method. We then consider a boson mixing model that captures the essential QFT aspects, and finally derive the neutrino flavor transition and survival probabilities within this framework.
We conclude in Section~\ref{conclusion} with a summary of our results and a discussion of open problems and future directions. For the reader's convenience, we also include \ref{ineqcar}, where we review the concept of unitarily inequivalent representations for fields with different masses.

Along this paper, we work in natural units \(\hbar=c=1\).

\section{Weak interactions and flavor mixing} \label{sectionw}

In this Section, we briefly review a minimal extension of the  Standard Model containing neutrino masses and mixing. This is then used for the derivation of the neutrino flavor charges, on which our subsequent discussion of neutrino oscillations is based.

\subsection{Minimally extended Standard Model and field mixing}

The Lagrangian for a minimal extension of the Standard Model  can be easily written down \cite{Miransky:1994vk} (see also Refs. \cite{FRITSCH197672,Bilenky:1978nj,Bilenky:1987ty})
\be
\mathcal{L}_{SM} \ = \ \mathcal{L}_{f g} \, + \, \mathcal{L}_{hf}+\mathcal{L}_{h}+\mathcal{L}_{g} \, , 
\ee
where $\mathcal{L}_{f g}$ describes the interaction of fermions with the $SU(2)_L \times U(1)_Y$ gauge fields
\be
\mathcal{L}_{f g} \ = \ \sum_\si \lf( i \bar{q}_{\si,L} \ga^\mu \, D_\mu q_{\si,L} + i \bar{q}_{\si,R} \ga^\mu \, D_\mu q_{\si,R} + i \bar{l}_{\si,L} \ga^\mu \, D_\mu l_{\si,L} + i \bar{l}_{\si,R} \ga^\mu \, D_\mu l_{\si,R} \ri)\, , 
\ee
where we have introduced the quark and the lepton vectors
\be
q_{\si} \ = \ \begin{pmatrix} u'_\si \\ d'_\si \end{pmatrix} \, , \qquad  l_{\si} \ = \ \begin{pmatrix} \nu'_\si \\ e'_\si \end{pmatrix} \, , 
\ee
with the index $\si$ indicating the flavor. The reason of superscript comma will be clear below. Thus $u_\si$ runs over the \emph{up, charm and top} quarks, $d_\si$ runs over the \emph{down, strange and bottom} quarks, while $e_\si$ runs over the \emph{electron, muon and tauon} and similarly for neutrinos $\nu_\si$. The covariant derivative action on left and right (Weyl) fields is defined as
\bea
D_\mu \psi_L & \equiv & \pa_\mu \psi_L +i \, g_1 Y B_\mu \psi_L + i \, g_2 \, \sum^3_{a=1} W_\mu^a \frac{\si^a}{2} \psi_L \, , \\[2mm]
D_\mu \psi_R & \equiv & \pa_\mu \psi_R +i \, g_1 Y B_\mu \psi_R \, , 
\eea
where $W^a_u$ and $B_\mu$ are the gauge fields, $Y$ is the \emph{hypercharge}, $\si^a$ are the Pauli matrices and $\psi$ indicates both quarks and lepton fields. $\mathcal{L}_{hf}$ describes the Yukawa interaction
\bea \non
\mathcal{L}_{hf} & = & -\sqrt{2}\sum_{\si,\rho} \lf(\overline{u}'_{\si,L} R^{\si \rho}_u \Phi  u'_{\rho,R}+\overline{d}'_{\si,L} R^{\si \rho}_d \Phi  d'_{\rho,R}
\ri.\\
&&\qquad \qquad \lf.
+\overline{\nu}'_{\si,L} R^{\si \rho}_\nu \Phi \nu'_{\rho,R}+\overline{e}'_{\si,L} R^{\si \rho}_e \Phi e'_{\rho,R}+ h.c.\ri) \, , 
\eea
where
\be
\Phi \ = \ \frac{1}{\sqrt{2}} \lf(h + i \sum^3_{a=1}\si^a \pi^a\ri)
\ee
and $R_u,R_d, R_\nu,R_e$ are, generally non-diagonal, matrices in the flavor space. The fields $h$ and $\pi^1,\pi^2, \pi^3$ are real scalar fields. $h$ is known as the \emph{Higgs field}. 
 $\mathcal{L}_{h}$ is the Lagrangian for the scalar fields:
\be
 \mathcal{L}_{h} \ = \ \frac{1}{2} D_\mu h D^\mu h+ \frac{1}{2} D_\mu \boldsymbol{\pi} \cdot D^\mu \boldsymbol{\pi}  - \frac{\mu^2}{2} \lf(h^2 + |\boldsymbol{\pi} |^2 \ri)- \frac{\la}{4 !} \lf(h^2 + |\boldsymbol{\pi} |^2 \ri)^2
\ee
where we introduced the vector $\boldsymbol{\pi}  = (\pi^1,\pi^2,\pi^3)$, and the covariant derivative acts on both $h$ and $\boldsymbol{\pi}$ as
\be
D_\mu  \ = \ \pa_\mu - i g_1 B_\mu + i g_2  \sum^3_{a=1}\frac{\si_a}{2} W_\mu^a \, .
\ee
Finally, $\mathcal{L}_g$ is the kinetic Lagrangian of the gauge fields
\be
\mathcal{L}_g \ = \ -\frac{1}{4} \lf(B_{\mu\nu} B^{\mu\nu} + \sum^3_{a=1} W^a_{\mu\nu} W^{\mu\nu;a}\ri) \, ,
\ee
with\cite{Miransky:1994vk}
\bea
B_{\mu\nu}&=&\pa_\mu B_\nu - \pa_\nu B_\mu \\
W^a_{\mu\nu}&=&\pa_\mu W^a_\nu - \pa_\nu W^a_\mu - g_2 \, \epsilon^{abc} \, W^b_\mu W^c_\nu
\eea

The Nambu--Goldstone phase of the model is obtained when $\mu^2<0$ and
\be
\lan \Phi \ran = \frac{1}{\sqrt{2}}v \ide \, .
\ee
The effective Yukawa Lagrangian in such phase thus reads
\be \label{ndmass}
\mathcal{L}_{hf} \ = \ -\sum_{\si,\rho} \lf(\overline{u}'_{\si,L} M^{\si \rho}_u  u'_{\rho,R}+\overline{d}'_{\si,L} M^{\si \rho}_d   d'_{\rho,R}+\overline{\nu}'_{\si,L} M^{\si \rho}_\nu  \nu'_{\rho,R}+\overline{e}'_{\si,L} M^{\si \rho}_e  e'_{\rho,R}+ h.c.\ri) \, , 
\ee
where we introduced the notation $M^{\si \rho}=R^{\si \rho} v$. The above terms can be diagonalized by means of bi-unitary transformations
\be\label{massM}
M_{diag} \ = \ U^\dag_L \, M \, U_R \, , 
\ee
where, with $M$ we meant each one of the four matrices, while the $U_L$ and $U_R$ transformations are generally different in the four pieces. The Lagrangian \eqref{ndmass} thus describe the mass terms for quarks and leptons
\be \label{dmass}
\mathcal{L}_{hf} \ = \ -\sum_{j} \lf(\overline{u}_{j} m_{u,j}  u_{j}+\overline{d}_{j} m_{d,j}    d_{j}+\overline{\nu}_{j} m_{\nu,j}   \nu_{j}+\overline{e}_{j} m_{e,j}   e_{j}\ri) \, . 
\ee
Here we indicated the lepton fields with a definite mass as
\be
l_{j, L} \ = \ \sum_\si U_{L; j \si} l'_{\si,L} \, ,
\ee
and the same for the right fields and quarks.

Let us now see how the weak interaction term in Goldstone phase is affected by the bi-unitary transformation. We define, as usual
\bea
W_\mu^{\pm} & = & \frac{W_\mu^{1} \mp i \, W_\mu^{2}}{\sqrt{2}} \, \\[2mm]
Z_\mu & = & \cos \theta_W \, W_\mu^{3}- \sin \theta_W \, B_\mu \, , \\[2mm]
A_\mu & = & \sin \theta_W \, W_\mu^{3}+ \cos \theta_W \, B_\mu \, , 
\eea
with $\cos \theta_W = \frac{g_2}{g_1^2+g_2^2}$. The $W^\pm$ boson mediate charged-current weak interactions, while the $Z$ boson mediates neutral current weak interactions and the $A$ field is the usual photon field, mediating electromagnetic interactions. In particular, in this work, we are interested in charged-current weak interactions of leptons. The corresponding effective Lagrangian reads
\be
\mathcal{L}_{CC} \ = \  \frac{g_2}{2\sqrt{2}} \, \sum_{k,j,\rho} \,
\lf [ W_{\mu}^{+}\,
\overline{\nu}_j\, U_{L, j \rho}^{\nu \dag} U_{L,\rho k}^e \,  \gamma^{\mu}\,(1-\gamma^{5})\,e_k +
h.c. \ri] \, , 
\ee
where the  Latin indices indicate the mass, while the Greek ones indicate the flavor. This convention will be followed along the entire review. It is evident that we can apply both transformations to the neutrino fields only. Then, we can identify the charged leptons fields in the flavor basis with the ones in the mass basis. We thus write
\be \label{Lcc}
\mathcal{L}_{CC} \ = \  \frac{g_2}{2\sqrt{2}} \, \sum_{\si} \,
\lf [ W_{\mu}^{+}\,
\overline{\nu}_\si  \gamma^{\mu}\,(1-\gamma^{5})\,e_\si \, + \,
h.c. \ri] \, , 
\ee
where we have defined the neutrino fields in the flavor basis as
\bea \label{mixtra}
\nu_\si(x) \ = \  \sum_{k} U_{ \si  j } \nu_j(x) \, . 
\eea
The matrix 
\be \label{genmix}
U \equiv \  U_L^{\dag e} \, U_L^{\nu} \, , 
\ee
is known as the \emph{mixing matrix}. It is a unitary matrix. In the two-flavor case this can be simply written as a rotation
\be \label{2fmix}
U \ = \ \begin{pmatrix} \cos \theta & \sin \theta \\ -\sin \theta & \cos \theta \end{pmatrix} \, . 
\ee
In the case of three flavors, one has the \emph{Pontecorvo--Maki--Nakagawa--Sakata} (PMNS) matrix \cite{10.1143/PTP.28.870} 
\bea \label{pmns}
U \, =\lf(\begin{array}{ccc}
  c_{12}c_{13} & s_{12}c_{13} & s_{13}e^{-i\delta} \\
  -s_{12}c_{23}-c_{12}s_{23}s_{13}e^{i\delta} &
  c_{12}c_{23}-s_{12}s_{23}s_{13}e^{i\delta} & s_{23}c_{13} \\
  s_{12}s_{23}-c_{12}c_{23}s_{13}e^{i\delta} &
  -c_{12}s_{23}-s_{12}c_{23}s_{13}e^{i\delta} & c_{23}c_{13}
\end{array}\ri) \, , 
 \eea
with $c_{ij}=\cos\theta_{ij},  s_{ij}=\sin\theta_{ij}$, $i, j=1,2,3$ and $\de$ is the $CP$-violating phase. Similar considerations hold for the quark sector. In that case the $\theta$ in the two-flavor case (see Eq.\eqref{2fmix}) is known as \emph{Cabibbo angle}, while the three flavor mixing matrix \eqref{pmns} is known as \emph{Cabibbo--Kobayashi--Maskawa} (CKM) matrix.

Then, in the flavor basis the weak interaction is diagonal (this is also true for neutral weak interactions). However, the mass term \eqref{dmass} is now non-diagonal. In the next subsection, we study the two-flavor effective Lagrangian, and we show how Noether's theorem permits to identify the lepton charges in the presence of mixing.

\subsection{The two-flavor effective Lagrangian and the lepton charges}
Let us now consider the Goldstone phase. In the following we will mostly limit to the case of two-lepton flavors $\si=e, \mu$. Actually we will completely disregard the quark sector.

The two-flavor effective Lagrangian is
\be
\label{Lagrangian} \mathcal{L}=\mathcal{L}_0+ \mathcal{L}_{CC} \, .
\ee
where $ \mathcal{L}_0  =  {\cal L}_{\nu}+{\cal L}_{e} $, and
\bea \lab{neutr}
\label{Lnu}
{\cal L}_{\nu} & = & \overline{\nu}_e  \lf( i \ga_\mu \pa^\mu - m_e \ri)\nu_e +\overline{\nu}_e  \lf( i \ga_\mu \pa^\mu - m_\mu \ri)\nu_e + m_{e \mu} \lf( \overline{\nu}_e \nu_\mu+\overline{\nu}_\mu \nu_e\ri)\, ,  \\[2mm]
{\cal L}_{e} &  =  & \overline{e} \lf( i \ga_\mu \pa^\mu - \tilde{m}_{e} \ri) e + \overline{\mu} \lf( i \ga_\mu \pa^\mu - \tilde{m}_{\mu} \ri) \mu  \lab{lept} \, , \\[2mm]
 {\cal L}_{CC} &  = &  \frac{g}{2\sqrt{2}}
\lf [ W_{\mu}^{+}\,
\overline{\nu}_e\,\gamma^{\mu}\,(1-\gamma^{5})\,e + W_{\mu}^{+}\,
\overline{\nu}_\mu \,\gamma^{\mu}\,(1-\gamma^{5})\,\mu +
h.c. \ri] \, .
\label{Linteract}
\eea
Here we adopted the common notation $g_2=g$.
In the present case the mixing angle $\theta$ and the neutrino masses $m_j \equiv m_{\nu,j}$, $j=1,2$ are related to the parameters of $M_\nu$ by
\bea \label{met}
m_e & = & m_1 \cos^2 \theta + m_2 \sin^2 \theta \, ,\\[2mm] \label{mmut}
m_\mu & = & m_1 \sin^2 \theta + m_2 \cos^2 \theta \, , \\[2mm]
\tan 2 \theta & = & 2 m_{e \mu}/(m_\mu-m_e) \, .
\eea
The term $m_{e \mu}$ in ${\cal L}_{\nu}$ will be referred to as the  \emph{mixing term}.

The Lagrangian $\mathcal{L}$ is invariant under the global $U(1)$ transformations
$\nu_\si \rightarrow e^{i \alpha} \nu_\si$ and $e_\si \rightarrow e^{i \alpha} e_\si$, $\si=e,\mu$,
leading to the conservation of the total  flavor charge $Q_{l}^{tot}$,  corresponding to the lepton-number conservation~\cite{Bilenky:1987ty}. This can be written in terms of the flavor charges for neutrinos and charged leptons~\cite{Blasone:2001qa}, which are obtained by $SU(2)$  transformations
\be
Q_{l}^{tot} =  \sum_{\si=e,\mu} Q_\si^{tot}(t) \,,\quad   Q_{\si}^{tot} (t) = Q_{\nu_{\si}}(t) + Q_{\si}(t)\,,
\ee
with
\bea
Q_{e} & = &  \intx \,
e^{\dag}(x)e(x) \,, \qquad Q_{\nu_{e}} (t) =  \intx \,
\nu_{e}^{\dag}(x)\nu_{e}(x)\,,
\nonumber \\ [2mm]
Q_{\mu} & = &   \intx \,
 \mu^{\dag}(x) \mu(x)\,, \qquad Q_{\nu_{\mu}} (t)= \intx \, \nu_{\mu}^{\dag}(x) \nu_{\mu}(x)\,  .
 \label{QflavLept}
\eea
Notice that, because of the mixing term, the flavor neutrino charges are time-dependent.

Since \([ \mathcal{L}_{CC}({\bf x}, t), Q_\si^{\text{tot}}(t) ] = 0\), neutrinos are produced and detected in states with definite flavor~\cite{PhysRevD.45.2414, giunti2007fundamentals, BLASONE200937}. However, the commutator \([ (\mathcal{L}_{\nu} + \mathcal{L}_{e})({\bf x}, t), Q_\si^{\text{tot}}(t) ] \neq 0\), indicates that flavor is not conserved during neutrino propagation. As a result, \emph{flavor oscillations} occur: a neutrino created with a definite flavor \(\si\) at time \(t = 0\) may later be detected with a different flavor \(\rho\) at time \(t\).
Quoting Ref. \cite{close2010neutrino}, when neutrinos are produced and detected they ``\emph{carry identity cards}'', i.e. a definite flavor and ``\emph{can surreptitiously change them if given the right opportunity}''. 

It is evident that, because of the presence of the mixing term, flavor neutrino states cannot be  built with the usual techniques of QFT for free-fields. Then, the problem arises of properly defining flavor states and to compute the oscillation probability
\be \label{oscprob}
P_{\si \to \rho}(t) \ = \ |\lan \nu_\rho(t)| \nu_\si\ran|^2 \, .
\ee
Before the presentation of the QFT treatment of such problems, in the next section we will show how these issues were studied in QM. 
\section{Neutrino oscillations in quantum mechanics} \label{sectionqm}

In this Section, we review the standard derivation of neutrino oscillations both in QM and by means of Dirac equation.

\subsection{Standard derivation} \label{stder}
Neutrinos are produced with a definite flavor, i.e. as eigenstates of $\mathcal{L}_{CC}$. which we indicate as $|\nu_\si\ran$, with $\si=e,\mu,\ldots$. On the other hand, the propagation is described by the eigenstates of $\mathcal{L}_0$ - the \emph{mass eigenstates} - which are denoted as $|\nu_j\ran$, with $j=1,2,\ldots$.

The usual assumption is that flavor and mass eigenstates are related by
\be \label{stflavstates}
|\nu_\si \ran \ = \ \sum_j \, U^*_{\si j} |\nu_j\ran \, , 
\ee
where $U$ is the mixing matrix we introduced in Eq.\eqref{genmix}. In order to derive the flavor-transition probability, which is given by Eq.\eqref{oscprob}, we assume that mass eigenstates are just standard plane waves, so their time-evolution is simply
\be
|\nu_j(t) \ran \ = \ e^{-i E_j t} |\nu_j\ran \, , 
\ee
so that
\be
|\nu_\si(t) \ran \ = \  \sum_j \, U^*_{\si j} e^{-i E_j t} |\nu_j\ran \, .
\ee
Then
\be
\lan \nu_\rho(t)| \nu_\si\ran \ = \ \sum_{j} \, U_{\rho j} U^*_{\si j} e^{i E_j t}  \, , 
\ee
where we used the orthogonality of mass eigenstates. Therefore, the probability is found to be
\be \label{qmfor}
P_{\si \to \rho}(t)  \ = \ \sum_{j,k} \, U_{ \rho j} U^*_{\si j} U^*_{ \rho k} U_{\si k} e^{-i (E_k-E_j) t}  \, , 
\ee
In the two-flavor case, which will be the typical case for a large part of this review, the formula simplifies to
\bea  \label{sform}
P_{\si \to \rho}(t)  & = & \sin^2(2 \theta) \sin^2 \lf(\frac{E_2-E_1}{2}t\ri)  \, , \si \neq \rho \\[2mm]
P_{\si \to \si}(t)  & = & 1-\sin^2(2 \theta) \sin^2 \lf(\frac{E_2-E_1}{2} t\ri)  \, . 
\eea
In the relativistic limit, we can use the approximation
\be
E_j \ \approx \ E + \frac{m_j^2}{2 E} \, , 
\ee
where $E$ is the energy of a massless neutrino. Therefore the general formula \eqref{qmfor} becomes
\be 
P_{\si \to \rho}(t)  \ = \ \sum_{j,k} \, U_{ \rho j} U^*_{\si j} U^*_{ \rho k} U_{\si k} e^{-i \frac{\de m^2_{j k}}{2 E} t}  \, , \label{psirho}
\ee
where we introduced, as usual, $\de m_{j k}^2 \ \equiv \ m_j^2-m_k^2$. Actually, in the experiments time is not directly measured. However, the distance between the source and the detector is known. Then, a time-to-space conversion is usually performed \cite{Lipkin:2005kg}. The simplest recipe, although not really rigorous, is just to put $t \approx L$. Then, we get
\be 
P_{\si \to \rho}(L)  \ = \ \sum_{j,k} \, U_{ \rho j} U^*_{\si j} U^*_{ \rho k} U_{\si k} e^{-i \frac{ 2 \pi L}{L_{j k}}}  \, , 
\ee
where $L_{j k}\equiv 4\pi E / \delta m_{ij}^2$ is the \emph{oscillation length} \cite{giunti2007fundamentals}.
Finally, in the two-flavor case such expression reads
\bea  \label{stoscfor}
P_{e\rightarrow \mu}(L) \ = \  \sin^2 (2 \theta)\sin^2\lf(\frac{\pi L}{L_{osc}}\ri) \,,  \quad \si \neq \rho \, ,
\eea
with $L_{osc} = L_{1 2}$.

Although the above derivation is very standard, it has many drawbacks. For example, it deals with neutrinos as plane-waves, while physically they should be treated as wave-packets. This issue was largely discussed in literature. The treatment in terms of wave-packets introduces a damping term in the oscillation formula, over a characteristic length scale which is called \emph{coherence length} \cite{GIUNTI199287,PhysRevD.58.017301}. This point will be discussed, in the QFT treatment, in Section \ref{seccur}. 

Another problem  is that the spinor nature of neutrinos is not used in the above derivation. A more rigorous  derivation of the flavor oscillation probability should be based on relativistic QM and on the Dirac equation. This approach was developed in Refs.\cite{PhysRevD.73.053013,Bernardini:2004wr, PhysRevD.71.076008,bernardini2011quantum} and leads to correction to the above oscillation formula. Later on, we will see that such corrections are the same which are also computed in a rigorous QFT approach.
\subsection{First quantized oscillation formula} \label{DiracEq}

In the following we specialize to the two-flavor case. The starting point is the Dirac equation\cite{bernardini2011quantum}
\be
\lf(i \ga^\mu \, \pa_\mu \otimes \ide_2- \ide_4 \otimes M_\nu \ri) \, \Psi(x) \ = \ 0 \, ,
\ee
where $M_\nu$ is the neutrino mass matrix in the flavor basis
\be
M_\nu \ = \ \begin{pmatrix} m_e & m_{e \mu} \\ m_{e \mu} & m_\mu \end{pmatrix} \, .
\ee

For simplicity we consider a neutrino moving along the $z$-axis. Moreover, we introduce the wavefunctions of neutrino with definite masses
\be
\lf(i \ga^0 \pa_0+i \ga^3 \pa_3-m_j\ri) \, \psi_j(z,t) \ = \ 0 \,, \qquad j=1,2 \, ,
\ee
and the eigenstates of the mass matrix
\be
M_\nu \, \nu_j \ = \ m_j \, \nu_j \, , \qquad j=1,2 \, .
\ee
The general neutrino wavepacket $\Psi$ is thus
\bea\hspace{-4mm}
\Psi(z,t) & = & \cos \theta \, \psi_1(z,t)  \otimes \nu_1 \, + \, \sin \theta \, \psi_2(z,t)  \otimes \nu_2 \non \\[2mm]
& = & \lf[\psi_1(z,t) \, \cos^2 \theta \, + \, \psi_2(z,t) \, \sin^2 \theta \ri]  \otimes  \nu_e + \sin \theta \, \cos \theta \, \lf[\psi_1(z,t)-\psi_2(z,t)\ri] \otimes  \nu_\mu \non \\[2mm] \label{flavfunc}
& \equiv & \psi_e(z,t) 	 \otimes \nu_e \, + \, \psi_\mu(z,t)  \otimes  \nu_\mu \, .
\eea
Here we have defined the flavor eigenstates $\nu_e, \nu_\mu$ as
\bea
\nu_e & = & \cos \theta \, \nu_1+ \sin \theta \, \nu_2 \\[2mm]
\nu_\mu & = & \cos \theta \, \nu_2 -\sin \theta \, \nu_1  \, .
\eea
and the corresponding wave functions
\bea
\psi_e(z,t) & = & \psi_1(z,t) \, \cos^2 \theta \, + \, \psi_2(z,t) \, \sin^2 \theta \, , \\[2mm]
\psi_\mu(z,t) & = & \sin \theta \, \cos \theta \, \Big[\psi_1(z,t)-\psi_2(z,t)\Big]  \, . \label{muwav}
\eea
Let us stress that here $\nu_j$, and then $\nu_\si$, are just two-component vectors, which encode the flavor degrees of freedom of the Dirac equation.

Consider the case where a neutrino is produced with flavor $e$. In such case $\psi_1(z,0)= \psi_2(z,0)=\psi_e(z,0)$. The oscillation formula is then obtained by taking the probability of measuring the neutrino in the $\mu$ state:
\be
P_{e \rightarrow \mu}(t) \ = \ \int^{+\infty}_{-\infty} \!\! \dr z \, \psi^\dag_\mu(z,t) \, \psi_\mu(z,t) \, .
\ee
By using Eq.\eqref{muwav} we find
\be
P_{e \rightarrow \mu}(t) \ = \ \frac{1}{2} \sin^2 2 \theta\, \Big[1-I_{{}_{12}}(t)\Big] \, ,
\ee
where the interference term is given by
\be 	\label{interf}
I_{{}_{12}}(t) \ = \ \Re e \lf[\int^{+\infty}_{-\infty} \!\! \dr z \, \psi^\dag_1(z,t) \, \psi_2(z,t)\ri] \, .
\ee
Note that, unlike the standard QM treatment, which includes only positive frequency modes, this analysis explicitly shows that negative frequency contributions must also be taken into account when computing the interference term~\eqref{interf}. To illustrate this, let us consider the Fourier expansion of $\psi_j(z,t)$, with $j = 1, 2$:
\begin{eqnarray}
\psi _{j}(x) =  \sum_r \, \int^{+\infty}_{-\infty} \!\! \frac{\dr p_z}{2 \pi}  \Big( u_{p_z,j}^{r} \, \alpha _{p_z,j}^{r} \, e^{-i \, \om_{p_z,j} \, t} +  \ v_{-p_z,j}^{r}  \beta _{-p_z,j}^{r*
}\, e^{i \, \om_{p_z,j} \, t}\Big)  e^{i \, p_z \, z}   .
\label{psiex}
\end{eqnarray}
with $\om_{p_z,j}=\sqrt{p_z^2+m_j^2}$. 
The requirement that neutrino is produced with definite flavor (see above), assumes the form:
\be
\sum_r\, \lf[u_{p_z,j}^{r} \, \alpha_{p_z,j}^{r}  +  \ v_{-p_z,j}^{r}  \beta_{-p_z,j}^{r*
} \ri] \ = \ \ph_e(p_z-p_0) \, w \, ,
\ee
where $\ph_\si(p_z-p_0)$ is the flavor neutrino distribution in the momentum space, at $t=0$, $p_0$ is the mean momentum of mass wavepackets and $w$ is a constant spinor, satisfying $w^\dag w =1$. By means of the orthogonality conditions for the Dirac spinors we derive the relations
\bea
\alpha_{p_z,j}^{r} & = & \ph_e(p_z-p_0) \, u_{p_z,j}^{r\dag} \, w \, , \\[2mm]
\bt_{-p_z,j}^{r*} & = & \ph_e(p_z-p_0) \, v_{-p_z,j}^{r\dag} \, w \, .
\eea
Substituting in Eq.\eqref{psiex} and then in Eq.\eqref{interf} we finally arrive at~\cite{PhysRevD.73.053013,Bernardini:2004wr, PhysRevD.71.076008,bernardini2011quantum}
\be
I_{{}_{12}}(t) \ = \ \int^{+\infty}_{-\infty} \!\! \frac{\dr p_z}{2 \pi} \, \ph^2_\si(p_z-p_0) \, \Big[|U_{p_z}|^2 \, \cos (2 \, \Om^-_{p_z} t) \, + \, |V_{p_z}|^2 \, \cos (2 \, \Om^+_{p_z} t)\Big] \, ,
\ee
where
\bea
\Om^{_\pm}_{p_z} & = & \frac{\om_{p_z,1} \pm \om_{p_z,2}}{2} \, , \\[2mm]
|V_{p_z}|^2 & = & 1-|U_{p_z}|^2 \ = \ \frac{\om_{p_z,1} \, \om_{p_z,2}-p_z^2-m_1 m_2}{2 \om_{p_z,1} \, \om_{p_z,2}} \, .
\eea
For plane waves, $\ph_e(p_z-p_0) \ = \ \de(p_z-p_0)$. The oscillation probability thus reads
\be \label{fqosc}
P_{e \rightarrow \nu_\mu} \ = \ \sin^2 2 \theta \, \Big[|U_{p_0}|^2 \, \sin^2 \lf(\Om^{_-}_{p_z} t\ri) \, + \, |V_{p_0}|^2 \, \sin^2 \lf(\Om^{_+}_{p_0} t\ri)\Big] \, .
\ee
The main difference compared to the oscillation formula in Eq.\eqref{sform} is the appearance of a rapidly oscillating term with frequency $\Om^{+}{p_0}/2$. This is reminiscent of the Zitterbewegung phenomenon in atomic physics, which gives rise to the Darwin term in the fine structure of the hydrogen atom\cite{itzykson2012quantum}. In our context, this effect becomes negligible when $p_0 \gg \sqrt{m_1 m_2}$: in this limit, $|U_{p_0}|^2 \rightarrow 1$ and $|V_{p_0}|^2 \rightarrow 0$, and the oscillation probability reduces to the standard form in Eq.~\eqref{sform}.

It is worth noting that the standard QM treatment discussed in the previous subsection, which excludes negative frequency components, resembles the rotating wave approximation~\cite{Kurcz:2009gh} commonly used in quantum optics and atomic physics. In the rotating wave approximation, rapidly oscillating terms in the Hamiltonian are neglected to allow for exact solutions of the eigenvalue problem. However, in the context of neutrino oscillations, there is generally no justification for neglecting the contributions involving $\Om^+_{p_0}$ -- except in the limit $p_0 \gg \sqrt{m_1 m_2}$, where such terms are indeed suppressed.


\section{Neutrino mixing in quantum field theory: a non-perturbative approach} \label{1bfmixing}

In this section, we review the non perturbative approach to flavor mixing and oscillations, usually known as \emph{flavor Fock space} approach \cite{Blasone:1995zc, PhysRevD.59.113003,Blasone:1998bx, Hannabuss:2000hy, PhysRevD.64.013011, PhysRevD.65.096015, Hannabuss:2002cv,Lee:2017cqf}. We derive the oscillation formula (see Eqs.\eqref{oscfor} and \eqref{oscfor2}) \cite{BHV99}. Remarkably, the final result matches the one obtained in the previous section using the first-quantized Dirac equation. We then explore various conceptual and phenomenological implications of this construction.

In the following, for simplicity, we will consider the two-flavor case and  Dirac neutrinos. The above treatment can be extended to the case of three flavors,  as sketched in Section \ref{3flavorAppA}. Also, we remark that similar conclusions hold for the case of Majorana neutrinos, see Ref.\cite{Blasone:2003hh}. In fact, these features are quite generic, also occurring in the case of boson field mixing\cite{Blasone:2001du} and for an arbitrary number of flavors\cite{Hannabuss:2000hy,Hannabuss:2002cv}. 


\subsection{Field mixing transformation and flavor vacuum} \label{QFT}
Let us focus on \( \mathcal{L}_\nu \) (see Eq.~\eqref{Lnu}). In what follows, we neglect weak interactions. This approximation is justified if we regard the neutrino fields as describing asymptotic states (e.g., \( \text{out} \) fields). However, as we will discuss later, due to mixing, the fields are never truly asymptotic—that is, we never work in the very-far future (or past), i.e. $t \to \infty$ ($t \to -\infty$). What matters, instead, is that the fields are defined sufficiently far from the interaction region so that weak interactions can be neglected. In this respect, see also the discussion in Section \ref{lepnum}.

In the mass basis, the neutrino fields satisfy free Dirac equations
\be \label{direq}
\lf(i \slashed{\pa}-m_j\ri) \nu_j(x) \ = \ 0 \, , \qquad j=1,2.
\ee
They can be thus expanded as
\begin{eqnarray}
\nu _{j}(x) =  \sum_r \,  \int \!\! \frac{\dr^3 k}{(2 \pi)^{\frac{3}{2}}} \, \left[ u_{{\bf k},j}^{r}(t) \, \alpha _{{\bf k},j}^{r} +  \ v_{-{\bf k},j}^{r}(t)  \,\beta _{-{\bf k},j}^{r\dagger
}\right]  e^{i{\bf k}\cdot {\bf x}}  \, .
\label{fieldex}
\end{eqnarray}
with $u^r_{{\bf k},j}(t) \,= \, e^{- i \om_{\G k,j} t}\, u^r_{{\bf k},j}\;$,
$\;v^r_{{\bf k},j}(t) \,= \, e^{ i \om_{\G k,j} t}\, v^r_{{\bf k},j}$,
 $\om_{\G k,j}=\sqrt{|\G k|^2 + m_j^2}$. In the Dirac representation
\bea
u_{{\bf k},i}^{1}&=&A_{k,i}\left[
\begin{array}{l}
1 \\
0 \\
\frac{k_{3}}{\omega _{k,i}+m_{i}} \\
\frac{k_{1}+ik_{2}}{\omega _{k,i}+m_{i}}
\end{array}
\right] ;\qquad u_{{\bf k},i}^{2}=A_{k,i}\left[
\begin{array}{l}
0 \\
1 \\
\frac{k_{1}-ik_{2}}{\omega _{k,i}+m_{i}} \\
\frac{-k_{3}}{\omega _{k,i}+m_{i}}
\end{array}
\right] \\[2mm]
v_{-{\bf k},i}^{1}&=&A_{k,i}\left[
\begin{array}{l}
\frac{-k_{3}}{\omega _{k,i}+m_{i}} \\
\frac{-k_{1}-ik_{2}}{\omega _{k,i}+m_{i}} \\
1 \\
0
\end{array}
\right] ;\qquad v_{-{\bf k} ,i}^{2}=A_{k,i}\left[
\begin{array}{l}
\frac{-k_{1}+ik_{2}}{\omega _{k,i}+m_{i}} \\
\frac{k_{3}}{\omega _{k,i}+m_{i}} \\
0 \\
1
\end{array}
\right]
\eea
and
\be
A_{k,i}=\left( \frac{\omega_{k,i}+m_{i}}{2\omega _{k,i}}\right) ^{\frac{1}{2}} \,  .
\ee
As usual $\al^r_{\G k, j}$ and $\beta _{{\bf k},j}^{r}$ are the annihilation operators for neutrinos and antineutrinos, respectively, which define the \emph{mass vacuum} $|0\rangle_{1,2}$:
\be \label{vacm}
\al^r_{\G k, j}|0 \rangle_{1,2} = 0 = \beta _{{\bf k},j}^{r} |0 \rangle_{1,2} \  .
\ee
The equal-time anticommutation relations now read
\be \label{CAR} \{\nu^{\al}_{i}(x), \nu^{\bt\dag }_{j}(y)\}_{t_x=t_y} =
\de^{3}({\bf x}-{\bf y})
\de _{\al\bt} \de_{ij} \ee
\be  \label{CAR2} \{\al ^r_{{\bf k},i}, \al ^{s\dag }_{{\bf q},j}\} = \de
_{\bf k q}\de _{rs}\de _{ij}  \quad ; \quad \{\bt^r_{{\bf k},i},
\bt^{s\dag }_{{\bf q},j}\} =
\de _{\bf k q} \de _{rs}\de _{ij}, \ee
and the orthonormality and completeness relations are:
\bea \label{orth}
u^{r\dag}_{{\bf k},i} u^{s}_{{\bf k},i} =
v^{r\dag}_{{\bf k},i} v^{s}_{{\bf k},i} = \de_{rs}
\, , \quad u^{r\dag}_{{\bf k},i} v^{s}_{-{\bf k},i} = 0
\,, \quad \sum_{r}(u^{r\l*}_{{\bf k},i} u^{r\bt}_{{\bf k},i} +
v^{r\al*}_{-{\bf k},i} v^{r\bt}_{-{\bf k},i}) = \de_{\al\bt}\;.
\eea
It is crucial to stress that
\be
\sum_{r,s}  v^{r\dag}_{{\bf k},i} u^{s}_{-{\bf k},j} \neq 0 \, , \qquad i \neq j \, , \quad m_i \neq m_j \, ,
\ee
 and similarly for other spinor products.  

We then denote the Fock space where the fields $\nu_1$,
$\nu_2$ are defined as
${\cal H}_\mass = \lf\{ \al_\mass^{\dag}\;,\;
\bt_\mass^{\dag}\;,\; |0\ran_\mass \ri\}$, whose basis can be obtained by a cyclic application of creation operators on the vacuum
\be
|n_{1}(\G k, j, r) \, n_{2}(\G p, i, s) \,  \ldots \ran \ = \ (\al^{r\dag}_{\G k,j})^{n_1} (\al^{s\dag}_{\G p,i})^{n_2} \, \ldots |0\ran_{1,2} \, , 
\ee
with the occupation numbers being Boolean variables.

The key observation of the present approach is that mixing transformation \eqref{mixtra} can be formally rewritten as \cite{Blasone:1995zc}
\bea  \non
\nu_{e}(x)  &=& G^{-1}_{\theta}(t)
\nu_{1}(x)
G_{\theta}(t) \\[2mm]
\nu_{\mu}(x) &=& G^{-1}_{\theta}(t)
\nu_{2}(x) \;
 G_{\theta}(t) \label{Gmix2}
\eea
with the generator given by:
\bea G_\theta(t) \ = \ \exp \lf[\theta \int d^{3}{\bf x}\, \lf(\nu_{1}^{\dag}(x) \, \nu_{2}(x)-\nu_{2}^{\dag}(x) \, 
\nu_{1}(x)\ri)\ri]\, .
\eea
In fact, employing the anticommutation relations \eqref{CAR} we obtain, e.g., for $\nu_e$
\be
\frac{d^2}{d\theta^2}\,\nu_{e}(x)\,=
\,-\nu_{e} (x)
\ee
with the initial conditions
\be
\lf.\nu_{e}(x)\ri|_{\theta=0}=\nu_{1}(x) \quad, \quad
\lf.\frac{d}{d\theta}\nu_{e}(x)\ri|_{\theta=0}=\nu_{2}(x)
\ee
and similarly for $\nu_{\mu}$. The solution of the above equations give
\be
\begin{pmatrix} \nu_e(x) \\ \nu_\mu(x) \end{pmatrix} \ = \ \begin{pmatrix} \cos \theta & \sin \theta \\ -\sin \theta & \cos \theta \end{pmatrix}  \, \begin{pmatrix} \nu_1(x) \\ \nu_2(x) \end{pmatrix} \, ,
\ee
which is exactly the mixing transformation \eqref{mixtra} in the two-flavor case.

Another crucial remark is that the mass vacuum $|0 \ran_\mass$
 is not invariant
under the action of the generator:
\be \label{timedep}
|0 (t)\ran_\flav \equiv G^{-1}_\theta(t)\; |0 \ran_\mass \, .
\ee
The state \eqref{timedep} is known as \emph{flavor vacuum} and it is annihilated by the operators $\al_{\sigma}(t)$ and $\bt_{\sigma}(t)$, defined by:
\be\al_e(t)|0(t)\ran_\flav \ \equiv \ G^{-1}_\theta(t)\al_1
{ G_\theta(t) \; G^{-1}_\theta(t)}|0\ran_\mass \ = \ 0, \ee
and similarly for $\bt_{\sigma}(t)$.
Their explicit form is\footnote{Here and in the following we will always work in a frame where $\G k=(0,0,|\G k|).$}
\bea \lab{operat1}
\al^r_{{\bf k},e}(t)=\cos\theta\,\al^r_{{\bf k},1} +
\sin\theta \lf(
 U_{{\bf k}}^{*}(t)\, \al^r_{{\bf k},2}
 + \epsilon^r
V_{{\bf k}}(t)\, \bt^{r\dag}_{-{\bf k},2}\ri)\quad &&\\
\lab{operat2}
\al^r_{{\bf k},\mu}(t)=\cos\theta\,\al^r_{{\bf
k},2}-\sin\theta \lf(
 U_{{\bf k}}(t)\, \al^r_{{\bf k},1}
 - \epsilon^r
V_{{\bf k}}(t)\, \bt^{r\dag}_{-{\bf k},1}\ri)\quad  &&\\
\lab{operat3}
\!\!\bt^r_{-{\bf k},e}(t)=\cos\theta\,\bt^r_{-{\bf
k},1}+\sin\theta\lf(
U_{{\bf k}}^{*}(t)\, \bt^r_{-{\bf k},2}
 -\epsilon^r
V_{{\bf k}}(t)\, \al^{r\dag}_{{\bf k},2}\ri) \;\; &&\\ \lab{operat4}
\!\!\bt^r_{-{\bf k},\mu}(t)=\cos\theta \, \bt^r_{-{\bf k},2} -
\sin\theta \lf(
 U_{\bf k}(t)\, \bt^r_{-{\bf k},1}  + \epsilon^r
V_{{\bf k}}(t) \, \al^{r\dag}_{{\bf k},1} \ri) \;\; &&
\eea
In Eqs.(\ref{operat1})-(\ref{operat4}), $\epsilon^r \equiv (-1)^r$, and $U_{\bf k}\,$ and $\,V_{\bf k}$ are the \emph{Bogoliubov
coefficients}:
\begin{eqnarray}
U_{{\bf k}}(t)& \equiv & u^{r\dag}_{{\bf k},2}u^r_{{\bf k},1}\;
e^{i(\om_{\G k,2}-\om_{\G k,1})t} \ = \ |U_\G k| \, e^{i(\om_{\G k,2}-\om_{\G k,1})t} \,  \, ,  \\[2mm]
V_{{\bf k}}(t) & \equiv & \epsilon^r\; u^{r\dag}_{{\bf k},1}v^r_{-{\bf k},2}\;
e^{i(\om_{\G k,2}+\om_{\G k,1})t} \ = \ |V_\G k| \, e^{i(\om_{\G k,2}+\om_{\G k,1})t} \, .
\end{eqnarray}
Explicitly
\bea \non
|U_\G k| & \equiv & u^{r\dag}_{{\bf k},2} \, u^{r}_{{\bf k},1} \ = \  v^{r\dag}_{-{\bf k},1} \, v^{r}_{-{\bf k},2} \non \\[2mm] \label{ucoff}
& = & \left(\frac{\omega_{\G k,1}+m_{1}}{2\omega_{\G k,1}}\right)^{\frac{1}{2}}
\left(\frac{\omega_{\G k,2}+m_{2}}{2\omega_{\G k,2}}\right)^{\frac{1}{2}}
\left(1+\frac{{\bf k}^{2}}{(\omega_{\G k,1}+m_{1})(\omega_{\G k,2}+m_{2})}\right) \, . \label{uk}\\[2mm]
|V_\G k| & = & \epsilon^r\; u^{r\dag}_{{\bf
k},1} \, v^{r}_{-{\bf k},2} \ = \  -\epsilon^r\, u^{r\dag}_{{\bf
k},2} \, v^{r}_{-{\bf k},1} \non \\[2mm] \label{vcoff}
& = &  \frac{|\G k|}{\sqrt{4 \om_{\G k,1}\om_{\G k,2}}}
\lf(\sqrt{\frac{\om_{\G k,2}+m_2}{\om_{\G k,1}+m_1}}-\sqrt{\frac{\om_{\G k,1}+m_1}{\om_{\G k,2}+m_2}}\ri) \, .
\eea
Notice that
\be
|U_{\bf k}|^2 + |V_{\bf k}|^2 \ = \ 1 \, .
\ee
In the relativistic limit $\omega_{\G k,j} \approx |\G k| $, $|U_\G k| \rightarrow 1$ and $|V_{{\bf k}}| \rightarrow 0$. One can also verify that $|V_{{\bf k}}|=0$  when $m_1=m_2$  and/or $\theta=0$, i.e. when no mixing occurs. 

\begin{figure}[t]
\begin{center}
\includegraphics*[width=9cm]{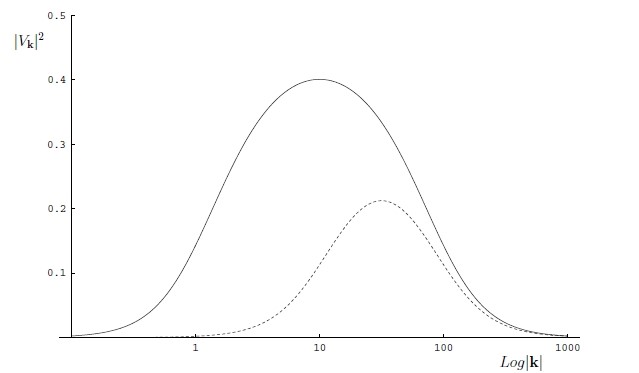}
\end{center}
\caption{$|V_\G k|^2$ for sample values of masses. The solid line corresponds to $m_1=1$ and $m_2=100$, while the dashed line corresponds to $m_1=10$ and $m_2=100$ \cite{blasone2011quantum}.}
	\label{fig:fermioncond}
\end{figure}

The function $|V_{\bf k}|^2$ has its maximum  at $|\G k|=\sqrt{m_1 m_2}$  with $|V_\G k|^2_{max}\ \rar\  1/2$ for $\frac{( m_{2}-m_{1})^2}{m_{1} m_{2}} \rar \infty$, and $|V_{{\bf k}}|^2\simeq \frac{(m_2 -m_1)^2}{4 |\G k|^2}$
 for $ |\G k|\gg\sqrt{m_1 m_2}$ (see Figure \ref{fig:fermioncond}). The function $|V_{\bf k}|^2$ is associated to the energy of the flavor vacuum: implications for cosmology of  this vacuum energy were discussed in Ref.\cite{Blasone:2004yh}.

 The flavor fields can be thus expanded as:
\begin{eqnarray} \lab{nue}
\nu_{e}(x)
&=& \sum_{{\bf k},r}  \frac{e^{i {\bf k}\cdot{\bf x}}}{\sqrt{V}}
  \lf[ u^r_{{\bf k},1}(t) \,\al^r_{{\bf k},e}(t)\, +\,
v^r_{-{\bf k},1}(t)\,\bt^{r\dag}_{-{\bf k},e}(t)
\ri] \, ,
\\ [2mm]
 \lab{numu}  \nu_{\mu}(x)
&=& \sum_{{\bf k},r}  \frac{e^{i {\bf k}\cdot{\bf x}}}{\sqrt{V}}
\lf[ u^r_{{\bf k},2}(t)\, \al^r_{{\bf k},\mu}(t)\, + \,
v^r_{-{\bf k},2}(t)\,\bt^{r\dag}_{-{\bf k},\mu}(t)
\ri] \, ,
\end{eqnarray}
and the \emph{flavor Hilbert space} is defined as ${\cal H}_{e,\mu} = \lf\{ \al_{e,\mu}^{\dag}\;,\;
\bt_{e,\mu}^{\dag}\;,\; |0\ran_{e,\mu} \ri\}$. As in the case of mass-representation, a basis can be constructed acting with creation operators on vacuum\footnote{For multi-particle states, we cannot work the simplified frame where $\G k = (0,0,\G k)$. However, in this work we will not consider multiparticle states in the non-perturbative treatment.}
\be \label{flavon}
|n_{1}(\G k, \si, r) \, n_{2}(\G p, \rho, s) \,  \ldots ; t\ran \ = \ (\al^{r\dag}_{\G k,\si}(t))^{n_1} (\al^{s\dag}_{\G p,\rho} (t))^{n_2} \, \ldots |0(t)\ran_{e,\mu} \, , 
\ee
with the occupation numbers being Boolean variables. These occupation numbers describe how many neutrinos with a definite spin, momentum and \emph{flavor}, are excited from the flavor vacuum.

The mixing generator can be decomposed as \cite{Blasone:1995zc,BLASONE2016104}:
\bea \label{mavidec}
G_{\theta}  \; = \;  B(m_{1},m_{2})  \;  \; R(\theta)  \; \; B^{-1}(m_{1},m_{2}) \, ,
\eea
where $ B(m_1,m_2)\equiv B_1(m_1) \, B_2(m_2) $, with
\bea
R(\theta) & \equiv & \exp \lf\{\theta  \sum_{{\bf k},r} \Big[\lf(\alpha^{r\dagger}_{{\bf k},1} \alpha^{r}_{{\bf k},2}+ \beta^{r\dagger}_{- {\bf k},1} \beta^{r}_{-{\bf k},2} \ri) e^{i\psi_{\G k}}
- h.c. \Big]\ri\} \, , \\[2mm]
B_i(m_i) & \equiv & \exp{\Big\{  }   \sum_{{\bf k},r} \Theta_{{\bf k},i} \;\epsilon^r  \Big[  \alpha^{r}_{{\bf k},i} \beta^{r}_{-{\bf k},i} e^{-i\phi_{{\G k}\!,\!i}} - \beta^{r\dagger}_{-{\bf k},i} \alpha^{r\dagger}_{{\bf k},i}e^{i\phi_{{\G k},i} }\Big]\Big\} \, , \quad i=1,2 \, .
\eea
Here $\Theta_{\G k,i}= 1/2 \, \cot^{-1}(|\G k|/m_i)$,  $ \psi_{\G k} = (\om_{{\G k},1}-\om_{{\G k},2})t$ and $\phi_{{\G k},i}=2 \om_{{\G k},i}t$. $B_i(\Theta_{{\bf k},i})$, $ i=1,2$ generate the Bogoliubov transformations which are related to mass shifts ($m_1 \neq m_2$) (see Eqs.\eqref{bogfer1},\eqref{bogfer2}), and $ R(\theta)$ generates a rotation. Note that
\be
 R^{-1}(\theta)  |0\ran_{1,2} \ = \ |0\ran_{1,2} \, ,
\ee
and the Bogoliubov transformations induce a condensate structure in the vacuum:
\bea \lab{Bogol}
| \widetilde{0}\ran_{1,2} \ \equiv  \ B^{-1}(m_{1},,m_{2}) |0\ran_{1,2} \ = \ \prod_{{\bf k},r,i} \Big[ \cos{\Theta_{{\bf k},i}} +\epsilon^r \sin{\Theta_{{\bf k},i}} \alpha^{r \dagger}_{{\bf k},i} \beta^{r \dagger}_{-{\bf k},i} \Big] |0\ran_{1,2} \, .
\eea
It is interesting to note that the vacuum with the condensate structure Eq. (\ref{Bogol}) looks like the well known superconductivity vacuum and the (mean-field) vacuum of Nambu--Jona Lasinio model \cite{PhysRev.122.345,PhysRev.124.246}. It is an entangled state for the $\alpha^{r}_{{\bf k},i}$   and $\beta^{r}_{-{\bf k},i}$ modes, for any ${\bf k}, r$ and $i =1,2$.

The decomposition Eq. (\ref{mavidec}) clearly exhibits that a \emph{a rotation of fields generally does not correspond to a simple rotation of creation and annihilation operators}. This   is well exploited by Eqs.(\ref{operat1})-(\ref{operat4}) and is rooted in the fact that the above rotation and Bogoliubov transformations do not  commute with each other: $[R,B_i]\neq 0$.

Another interesting observation is that flavor  vacuum $|0(t) \ran_\flav$  is an $SU(2)$
generalized coherent state \cite{perelomov2012generalized}. Its explicit condensate structure is given by \cite{Blasone:1995zc}
\begin{eqnarray}
&& \label{vacuumflav}
|0\ran_\flav= \prod_{{\bf k},r} \lf[
(1-\sin^2\theta\,|V_{{\bf
k}}|^2)
 \ri.-\,\epsilon^r\sin\theta\,\cos\theta\, |V_{{\bf k}}|
\, (\al^{r\dag}_{{\bf k},1}\bt^{r\dag}_{-{\bf k},2}+
\al^{r\dag}_{{\bf k},2} \bt^{r\dag}_{-{\bf k},1})\\[0mm] \non
&&\lf.
+\,\epsilon^r\sin^2\theta \,|V_{{\bf k}}| |U_{{\bf k}}| \,(
\al^{r\dag}_{{\bf k},1}\bt^{r\dag}_{-{\bf k},1} -
\al^{r\dag}_{{\bf k},2}\bt^{r\dag}_{-{\bf k},2} )
+\,\sin^2\theta \, |V_{{\bf k}}|^2
\, \al^{r\dag}_{{\bf k},1}\bt^{r\dag}_{-{\bf k},2}
\al^{r\dag}_{{\bf k},2}\bt^{r\dag}_{-{\bf
k},1}\ri]|0\ran_\mass \, .
\end{eqnarray}
Here we have chosen $t=0$ as reference time. Moreover we abbreviated $|0\ran_{e ,\mu} \equiv |0(t=0)\ran_{e ,\mu}$. We will often use such abbreviation and $\al^{r\dag}_{\G k,\si} \equiv \al^{r\dag}_{\G k,\si}(0)$.

We see from Eq.\eqref{vacuumflav} that flavor vacuum presents four kinds of condensate
particle-antiparticle pairs bringing zero momentum and
spin. The condensation density for the $\al^r_{{\bf k},j}$ particle is
\be \lab{condens}
\;_\flav\lan0(t)|\al^{r\dag}_{{\bf k},j}\al^r_{{\bf k},j}
|0(t)\ran_\flav = \sin^2\theta \; |V_{{\bf k}}|^2
\ee
vanishing for $m_\1=m_\2$  and/or $\theta=0$
(in both cases no mixing occurs).
The same result holds for $\bt_j$.

When ve remove the finite volume (infrared) regularization, we discover that flavor vacuum $|0(t)\ran_\flav$ is orthogonal to the mass vacuum $|0 \ran_\mass$ \cite{Blasone:1995zc,PhysRevD.59.113003, Hannabuss:2000hy, PhysRevD.64.013011, PhysRevD.65.096015, Hannabuss:2002cv,Lee:2017cqf}:
\be
\lim_{V \rar \infty}\, _\mass\lan0|0(t)\ran_\flav =
\lim_{V \rar \infty}\, e^{\!V \int
\frac{d^{3}{\bf k}}{(2\pi)^{3}}
\,\ln\,\lf(1- \sin^2\theta\,|V_{\bf k}|^2\ri)^2 }= 0 \label{ineqrep}
\ee
i.e. \emph{flavor and mass representations are unitarily inequivalent representations of the CAR}. Similarly, for $t \neq t'$,
\be \label{orthtt}  \lim_{V \rar \infty}\, _\flav \lan0(t')|0(t)\ran_\flav = 0 \, , 
\ee
i.e. flavor representations at different times are unitarily inequivalent.  This is feature which was previously encountered in other context as in the study of unstable particles \cite{DeFilippo:1977bk,PhysRevD.28.2621}, in QFT in curved space-times  \cite{Martellini:1978sm}, and of quantum dissipative systems \cite{Celeghini:1991yv}. We will come back to the analogy between oscillating neutrinos and unstable particles in Section \ref{TEUR} and in Section \ref{perneu}.


 \subsection{Flavor eigenstates and neutrino oscillations} \label{bfmixing}

In the above formalism, the normal ordered neutrino flavor charges (see Eq.\eqref{QflavLept}) can be expanded as
\be \label{normcharge}
:Q_{\nu_\si}(t): \ = \ \sum_{\G k,r} \lf(\al^{r\dag}_{\G k,\si}(t) \al^{r}_{\G k,\si}(t)-\bt^{r\dag}_{\G k,\si}(t) \bt^{r}_{\G k,\si}(t)\ri) \, .
\ee
In the following we will omit the normal ordering symbol.

Now we are able to build flavor states $|\nu^{r}_{\G k,\si}\ran \ $ as excitations of the flavor vacuum\cite{PhysRevD.60.111302} (see Eq.\eqref{flavon})
\be \label{bvflavstate}
|\nu^{r}_{\G k,\si}\ran \equiv \ |1(\G k, \si, r) ; t=0\ran  \ = \ \al^{r\dag}_{\G k,\si} |0\ran_{e ,\mu}  \, .
\ee
and similarly for the antineutrinos\footnote{Again we choose $t=0$ as reference time. Moreover we abbreviated $|0\ran_{e ,\mu} \equiv |0(t=0)\ran_{e ,\mu}$, $\al^{r\dag}_{\G k,\si} \equiv \al^{r\dag}_{\G k,\si}(0)$. We will use often use such abbreviations along the manuscript.} ($|\bt^{r}_{\G k,\si} \ran \equiv \bt^{r\dag}_{\G k,\si} |0\ran_{e ,\mu}$). It is easy to see that such flavor states are eigenstates of the charge operators:
\be
Q_{\nu_\si}(0) |\nu^r_{\G k,\si}\ran \ = \ \sum_{\G p,s} \al^{s\dag}_{\G p,\si} \al^{s}_{\G p,\si} \al^{r\dag}_{\G k,\si} |0\ran_\flav \ = \  |\nu^r_{\G k,\si}\ran \, ,
\ee
where we employed CAR, the expansion \eqref{normcharge} and the definition \eqref{bvflavstate}.

In the relativistic limit $m_i/|\G k| \rightarrow 0$ 
\bea \lab{operat1r}
\al^r_{{\bf k},e} & \approx & \cos\theta\,\al^r_{{\bf k},1} +
\sin\theta \, \al^r_{{\bf k},2}\\[2mm]
\lab{operat2r}
\al^r_{{\bf k},\mu} & \approx & \cos\theta\,\al^r_{{\bf
k},2}-\sin\theta \, \al^r_{{\bf k},1} \\[2mm]
\lab{operat3r}
\!\!\bt^r_{-{\bf k},e} & \approx &\cos\theta\,\bt^r_{-{\bf
k},1}+\sin\theta \, \bt^r_{-{\bf k},2} \\[2mm] \lab{operat4r}
\!\!\bt^r_{-{\bf k},\mu} & \approx & \cos\theta \, \bt^r_{-{\bf k},2} -
\sin\theta  \, \bt^r_{-{\bf k},1}  \, ,
\eea
(see Eqs.(\ref{operat1})-(\ref{operat4})), and 
\be
\al^{r\dag}_{\G k,\si}|0\ran_\flav \approx \ \al^{r\dag}_{\G k,\si}|0\ran_\mass \, . 
\ee
We thus find
\bea
|\nu^r_{\G k,e} \ran & \approx & |\nu^r_{\G k,e} \ran_{_P} \ \equiv \ \cos \theta |\nu^r_{\G k,1}\ran+ \sin \theta |\nu^r_{\G k,2} \ran \, , \\[2mm] \label{postate}
 |\nu^r_{\G k,\mu} \ran & \approx & |\nu^r_{\G k,\mu} \ran_{_P} \ \equiv \ -\sin \theta |\nu^r_{\G k,1}\ran+ \cos \theta |\nu^r_{\G k,2} \ran \, ,
\eea
which are the flavor states \eqref{stflavstates} we have encountered in the QM treatment, with the equal-momenta hypothesis. These will be here called the \emph{Pontecorvo states}~\cite{Gribov:1968kq,Bilenky:1975tb,Bilenky:1976yj,Bilenky:1977ne,Bilenky:1987ty}. Now we can show they are not appropriate states in a QFT description. In fact, at fixed time, they are eigenstates of flavor charges only for high momenta:
\be \label{relei}
\lim_{m_i/|\G k|\rightarrow 0} \, Q_{\nu_\si}(0)|\nu^r_{\G k,\si} \ran_{_P} \ = \ |\nu^r_{\G k,\si} \ran_{_P} \, .
\ee
Actually, this is not valid at all energy scales; therefore, the states are not true flavor eigenstates unless one assumes relativistic neutrinos. However, since this is typically the case in current oscillation experiments, the Pontecorvo description, and hence the oscillation formula~\eqref{stoscfor}, proves to be remarkably successful.

It is interesting to note that in Ref.~\cite{Bilenky:1978nj}, it is explicitly stated that the form~\eqref{stflavstates} of the flavor states follows directly from the field mixing transformation~\eqref{mixtra}. This is an unjustified (and generally incorrect) assertion, which only holds in the relativistic limit, where the flavor states can indeed be approximated by Eq.~\eqref{postate}. Unfortunately, this statement has led to considerable confusion among particle physicists not directly involved in the study of the QFT aspects of neutrino oscillations.
Nonetheless, in the well-known review~\cite{Bilenky:1987ty}, it is clearly acknowledged that Pontecorvo states are derived from the mixing transformation in the relativistic limit, although the derivation and general structure of flavor states are not discussed in that work.

We now can derive QFT flavor oscillation formula by means of flavor states \eqref{bvflavstate}. One would be tempted to compute the amplitude $ \lan \nu^r_{\G k,\si}| U(t) |\nu^r_{\G k,\si}\ran$, where $U$ is the time-evolution operator. However, as we have said above, such an operator is not a proper unitary operator on the flavor Hilbert space at $t=0$ (see Eq.\eqref{orthtt}) and such amplitude is null! Actually, it has been shown that flavor oscillation probability can be more properly computed by taking the expectation value of the flavor charges with respect to a definite flavor state~\cite{BHV99}
\be
\mathcal{Q}_{\si \rightarrow \rho}(t) \ = \ \lan Q_{\nu_\rho}(t) \ran_\si \, ,
\ee
where $\langle \cdots\rangle_\si \equiv \lan \nu^r_{\G k,\si}| \cdots |\nu^r_{\G k,\si}\ran$. The result can be expressed in terms of anticommutators of flavor ladder operators at different times. For example
\bea \label{qee}
\mathcal{Q}_{e\rightarrow e}(t) & = & 
|\{\al^r_{\G k, e}(t),\al^{r\dag}_{\G k, e}(0)\}|^2+|\{\bt^r_{-\G k, e}(t),\al^{r\dag}_{\G k, e}(0)\}|^2 \, , \\[2mm] \label{qemu}
\mathcal{Q}_{e\rightarrow \mu}(t) & = & 
|\{\al^r_{\G k, \mu}(t),\al^{r\dag}_{\G k, e}(0)\}|^2+|\{\bt^{r\dag}_{-\G k, \mu}(t),\al^{r\dag}_{\G k, e}(0)\}|^2 \, , 
\eea

Explicitly
\bea \label{oscfor}
&& \mbox{\hspace{-4mm}}\mathcal{Q}_{\si\rightarrow \rho}(t)  =   \sin^2 (2 \theta)\Big[|U_\G k|^2\sin^2\lf(\Om_{\G k}^{_-}t\ri)+  |V_\G k|^2\sin^2\lf(\Om_{\G k}^{_+}t\ri)\Big]  , \quad \si \neq \rho \, ,  \\[1mm]  \label{oscfor2}
&& \mbox{\hspace{-4mm}}\mathcal{Q}_{\si\rightarrow \si}(t)  =  1 \ - \ \mathcal{Q}_{\si\rightarrow \rho}(t) \, , \quad \si \neq \rho \, ,
\eea
where $\Om_{\G k}^{_{\pm}}\equiv (\om_{\G k,2}\pm\om_{\G k,1})/2$. Notice that this is the same as Eq.\eqref{fqosc}, that we derived in the first quantized approach based on the Dirac equation. Therefore, the same considerations we made in that case are still valid. In the relativistic limit, the QFT flavor states reduce to the Pontecorvo states \eqref{postate}, and we recover the standard formula \eqref{stoscfor}.
In the next section, we will compute some Green's functions of flavor fields on the flavor vacuum, and we will see that the above formula can be also derived starting from such objects.

At this stage, we delve deeper into the differences between the Pontecorvo framework and our full-fledged QFT construction. Eq.\ref{ineqrep}) implies that
\be \label{neutort}
\lim_{V \rightarrow \infty}\lan \nu^r_{\G k,i}| \nu^r_{\G k,\si}\ran \ = \ 0 \, , \qquad  i=1,2 \, ,
\ee
that is, flavor neutrino eigenstates, produced in charged current weak decays, cannot generally be expressed as linear combinations of single-particle mass eigenstates.
In contrast, this orthogonality condition does not hold for the Pontecorvo states defined in Eq.(\ref{postate}):
\be \label{neutortpon}
\lim_{V \rightarrow \infty}\lan \nu^r_{\G k,1}| \nu^r_{\G k,e}\ran_P \ = \ \cos \theta \, .
\ee
This contradiction is overcome by the observation that
\be
\lim_{m_i/|\G k|\rightarrow 0}\,\,\lim_{V \rightarrow \infty} \ \neq \ \lim_{V \rightarrow \infty}\,\, \lim_{m_i/|\G k|\rightarrow 0} \, ,
\ee
which means that the relativistic ${m_i/|\G k|\rightarrow 0}$ cannot be taken after the thermodynamic QFT limit; it can only be considered within the single-particle approximation, which neglects the inherently multi-particle nature of QFT. Eq.~(\ref{neutort}) should be thus understood as
\begin{eqnarray}
&&\mbox{\hspace{-9mm}}\lan \nu^r_{\G k,i}| \nu^r_{\G k,\si}\ran \! =\! \ {}_{1,2}\lan 0_{\G k}|\al^r_{\G k,1} \al^{r \dag}_{\G k,e}|0_{\G k}\ran_{e,\mu}   \prod_{\G p \neq \G k}\!\! {}_{1,2}\lan 0_{\G p}|0_{\G p}\ran_{e,\mu} \, .
\end{eqnarray}
Here we have used the fact that the QFT Hilbert space factorizes into tensor products of states with different momenta \cite{berezin1966method}. The first term on the right-hand side corresponds to Eq.\eqref{neutortpon}, and remains finite. However, as discussed above, this merely selects a single-particle subspace from the full Hilbert space. Put differently, the Pontecorvo definition of a neutrino state fails beyond the QM, single-particle picture.
\subsection{Two-point Green's functions for flavor fields} \label{greensec}
Following Ref.\cite{BHV99}, we show that Eq.\eqref{oscfor} can also be derived by means of the flavor fields Wightman functions and Green's functions on the flavor vacuum. 

We start from the propagator constructed on the mass vacuum\footnote{Here, to maintain the notation of Ref. \cite{BHV99}, we use a different definition of the propagator with respect to Eq.\eqref{perprop}. The two definitions just differ by a multiplicative factor $i$.}
\bea \label{massprop}
S_f(x,y)
& = &
\begin{pmatrix}
 S^{\al \bt}_{e e}(x,y) &  S^{\al \bt}_{e \mu}(x,y) \\ S^{\al \bt}_{\mu e}(x,y) &  S^{\al \bt}_{\mu \mu}(x,y)
\end{pmatrix}
\\ [2mm] \non
& = & \begin{pmatrix}
{}_{1,2}\lan 0 |T\lf[\nu^\al_e(x) \, \overline{\nu}^\bt_e(y)\ri]|0\ran_{1,2} & \, \,  {}_{1,2}\lan 0 |T\lf[\nu^\al_e(x) \, \overline{\nu}^\bt_\mu(y)\ri]|0\ran_{1,2} \\[2mm] {}_{1,2}\lan 0 |T\lf[\nu^\al_\mu(x) \, \overline{\nu}^\bt_e(y)\ri]|0\ran_{1,2} & \, \,  {}_{1,2}\lan 0 |T\lf[\nu^\al_\mu(x) \, \overline{\nu}^\bt_\mu(y)\ri]|0\ran_{1,2}
\end{pmatrix} \, .
\eea
This can be expanded in terms of the propagators of mass fields on the mass vacuum
\be
S_f(x,y)
=
\left[
\begin{array}{cc}
 S^{\al \bt}_1(x,y) \cos^2 \theta+S^{\al \bt}_2(x,y) \sin ^2 \theta  &  \, \,  (S^{\al \bt}_2(x,y)-S^{\al \bt}_1(x,y)) \cos \theta \sin \theta  \\
 (S^{\al \bt}_2(x,y)-S^{\al \bt}_1(x,y)) \cos \theta  \sin \theta  &  \, \,   S^{\al \bt}_2(x,y)\cos^2 \theta+S^{\al \bt}_1(x,y) \sin ^2\theta  \\
\end{array}
\right] \, ,
\ee
where
\be
S_j^{\al \bt} \ = \ i \int \! \frac{\dr^4 k}{(2 \pi)^4} \, e^{-i k \cdot (x-y)} \, \frac{\slashed{k}+m_j}{k^2-m^2_j+i \varepsilon} \, , \qquad j=1,2 \, .
\ee
In Ref.\cite{BHV99} it was defined the survival probability amplitude for an electron neutrino $\nu_e$ created at $t=0$ as
\be
\mathcal{P}^{>}_{ee}(\G k, t) \ = \ i \, u^{r\dag}_{\G k,1} \, e^{i \om_{\G k,1} t} \, S^{>}_{ee}(\G k,t) \, \ga^0 \,  u^{r}_{\G k,1} \, ,
\ee
where $S^{>}_{ee}(\G k,t)$ is the Fourier transform of the Wightman function
\be
S^{>}_{ee}(t,\G x; 0,\G y)= {}_{1,2}\lan 0 |\nu_e(t,\G x) \, \overline{\nu}_e(0,\G y)|0\ran_{1,2} \, .
\ee
In other words, the electron neutrino is created with a wave-function $u^{r}_{\G k,1}$ (see the expansion \eqref{nue}), it propagates and it is later detected in the same state.

Explicitly
\be
\mathcal{P}^{>}_{ee}(\G k, t) \ = \ \cos^2 \theta \ + \ \sin^2 \theta \, |U_\G k|^2 \, e^{-i (\om_{\G k,2}-\om_{\G k,1})t} \, .
\ee
However this result cannot be accepted because it has a wrong initial condition:
\be \label{wic}
\mathcal{P}^{>}_{ee}(\G k, 0^+) \ = \ \cos^2 \theta \ + \ \sin^2 \theta |U_\G k|^2 \ < \ 1\, .
\ee
This is due to the fact that we cannot generally build flavor eigenstates on the mass vacuum. As we have seen above, this is only possible in the relativistic limit. In that case $|U_\G k| \approx 1$ and the above inconsistency disappears.

In order to solve the problem we compute, instead, the propagator on the flavor vacuum:
\bea \label{flavprop}
\mathcal{G}_f(x,y)
& = &
\begin{pmatrix}
\mathcal{G}^{\al \bt}_{e e}(x,y) & \mathcal{G}^{\al \bt}_{e \mu}(x,y) \\ \mathcal{G}^{\al \bt}_{\mu e}(x,y) & \mathcal{G}^{\al \bt}_{\mu \mu}(x,y)
\end{pmatrix}
\\ \non
& = & \begin{pmatrix}
{}_{e,\mu}\lan 0 |T\lf[\nu^\al_e(x) \, \overline{\nu}^\bt_e(y)\ri]|0 \ran_{e,\mu} &  \, \,   {}_{e,\mu}\lan 0 |T\lf[\nu^\al_e(x) \, \overline{\nu}^\bt_\mu(y)\ri]|0 \ran_{e,\mu} \\[2mm] _{e,\mu}\lan 0 |T\lf[\nu^\al_\mu(x) \, \overline{\nu}^\bt_e(y)\ri]|0\ran_{e,\mu} &   \, \,  {}_{e,\mu}\lan 0|T\lf[\nu^\al_\mu(x) \, \overline{\nu}^\bt_\mu(y)\ri]|0\ran_{e,\mu}
\end{pmatrix} \, ,
\eea
with $y_0=0$.

One can check that $\mathcal{G}_f(x,y)$ and $S_f(x,y)$ are both Green's functions of the operator $i \slashed{\pa}-M_\nu$, and then they only differ by a boundary term. For example, the Fourier transform of $\mathcal{G}_{e e}(x,y)$ reads
\bea \non
\mathcal{G}_{e e}(\G k,t) & = & S_{e e}(\G k,t) \ + \ 2 \pi i \sin^2 \theta\Big[|V_\G k|^2(\slashed{k}+m_2)\de(k^2-m^2_2) \\  \lab{diff} & - & |U_\G k||V_\G k| \sum_r\lf(\epsilon^r \, u^r_{\G k, 2}\overline{v}_{-{\bf k},2}^{r}\de(k_0-\om_2)+\epsilon^r v^r_{-\G k, 2}\overline{u}_{{\bf k},2}^{r}\de(k_0+\om_2)\ri)\Big]  \, , 
\eea
where the imaginary piece is the above mentioned boundary term.

Employing the Wightman function $\mathcal{G}^{>}_{ee}(\G k,t)$, we can compute the survival probability amplitude as
\be
\mathcal{P}^{>}_{ee}(\G k, t) \ = \ i \, u^{r\dag}_{\G k,1} \, e^{i \om_{\G k,1} t} \, \mathcal{G}^{>}_{ee}(\G k,t) \, \ga^0 \,  u^{r}_{\G k,1} \ = \ \{\al^r_{\G k, e}(t),\al^{r\dag}_{\G k, e}(0)\}\, .
\ee
Explicitly
\be
\mathcal{P}^{>}_{ee}(\G k, t) \ = \ \cos^2 \theta \ + \ \sin^2 \theta\lf( |U_\G k|^2 \, e^{-i (\om_{\G k,2}-\om_{\G k,1})t}+|V_\G k|^2 \, e^{i (\om_{\G k,1}+\om_{\G k,2})t}\ri) \, .
\ee
$\mathcal{P}^{>}_{ee}(\G k, t)$ satisfies the correct initial condition:
\be
\mathcal{P}^{>}_{ee}(\G k, 0^+) \ = \ 1 \, .
\ee

In the same way one can compute the amplitudes
\bea \non 
\mathcal{P}^{>}_{\bar{e}e}(\G k, t) & = & i \, v^{r\dag}_{-\G k,1} \, e^{-i \om_{\G k,1} t} \, \mathcal{G}^{>}_{ee}(\G k,t) \, \ga^0 \,  u^{r}_{\G k,1} \ = \ \{\bt^{r\dag}_{-\G k, e}(t),\al^{r\dag}_{\G k, e}(0)\} \\[2mm]
& = & \epsilon^r \, |V_\G k| \, |U_\G k| \, \lf(e^{i \lf(\om_{\G k,2}-\om_{\G k,1}\ri)t}-e^{-i \lf(\om_{\G k,2}+\om_{\G k,1}\ri)t}\ri) \, , \\[2mm]
\non 
\mathcal{P}^{>}_{\mu e}(\G k, t) & = & i \, u^{r\dag}_{\G k,2} \, e^{i \om_{\G k,2} t} \, \mathcal{G}^{>}_{\mu e}(\G k,t) \, \ga^0 \,  u^{r}_{\G k,1} \ = \ \{\al^{r}_{\G k, \mu}(t),\al^{r\dag}_{\G k, e}(0)\} \\[2mm]
& = &   |U_\G k| \, \lf(1-e^{i \lf(\om_{\G k,2}-\om_{\G k,1}\ri)t}\ri) \, , \\[2mm] \non
\mathcal{P}^{>}_{\bar{\mu} e}(\G k, t) & = & i \, v^{r\dag}_{-\G k,2} \, e^{-i \om_{\G k,2} t} \, \mathcal{G}^{>}_{\mu e}(\G k,t) \, \ga^0 \,  u^{r}_{\G k,1} \ = \ \{\bt^{r\dag}_{-\G k, \mu}(t),\al^{r\dag}_{\G k, e}(0)\} \\[2mm]
& = & \epsilon^r \,  |V_\G k| \, \lf(1-e^{-i \lf(\om_{\G k,2}+\om_{\G k,1}\ri)t}\ri) \, .
\eea

Notice that the above amplitudes are the same appearing in Eqs.\eqref{qee},\eqref{qemu}. Then, we recognize
\bea \label{probcond}
\mathcal{Q}_{e \to e}(t) & = &
|\mathcal{P}^{>}_{ee}(\G k, t)|^2 \ + \ |\mathcal{P}^{>}_{\bar{e} e}(\G k, t)|^2 \\[2mm]
\mathcal{Q}_{e \to \mu} (t) & = &
|\mathcal{P}^{>}_{\mu e}(\G k, t)|^2 \ + \ |\mathcal{P}^{>}_{\bar{\mu} e}(\G k, t)|^2  \, .
\eea
This permits to give a more intuitive interpretation of the formulas \eqref{oscfor},\eqref{oscfor2}:  $\mathcal{Q}_{e \to e}(t)$, e.g., is the sum of the probabilities of different processes where an original electron-neutrino state is produced at $t=0$ and it is later detected at time $t$ - see also Section~\ref{neutsec} for a diagrammatic representation of such processes.

Let us notice that, if we consider the retarded propagator for the flavor fields\footnote{Note that for free fields this quantity is a c-number, thus there is no ambiguity in the choice of the Hilbert space.}:
\be
 \label{retGreen}
\mathcal{G}^{ret}(t,\G x;0,y) \, = \, \theta(t) \,   \lf\{\nu_\rho(t,\G x),\overline{\nu}_\si(0,\G y)\ri\} \, , \quad \rho,\si=e, \mu \, ,
\ee
and define the oscillation probability as \cite{PhysRevD.64.013011}
\be \label{yabprob}
\mathcal{Q}_{\nu_\rho \rightarrow \nu_\si}(\G k, t) \ = \ \mathrm{Tr}\lf[\mathcal{G}^{ret}_{\si \rho}(\G k, t)\mathcal{G}^{ret\dag}_{\si \rho}(\G k, t)\ri] \, ,
\ee
we re-obtain Eqs.~\eqref{oscfor},\eqref{oscfor2}. Once more, this fact confirms that the oscillation formula \eqref{oscfor}, \eqref{oscfor2} is a  solid result: such equation is a natural consequence of the relativistic formulation of neutrino oscillations and, in fact, it was also derived in the QM approach of Section \ref{DiracEq}. In Section \ref{perneu}, we will re-derive such result in perturbative QFT, confirming the above conclusions. 

\subsection{Neutrino oscillations from flavor currents} \label{seccur}

The derivation of neutrino oscillations in QFT presented  in Sections \ref{bfmixing} and \ref{greensec} is based on the plane-wave approximation. We have commented in Section \ref{stder} on the fact that a more realistic treatment based on wave-packets leads to a space-dependent oscillation formula, which includes damping of oscillations.  Here we briefly review how a wave-packet approach can be realize in a formalism based on QFT flavor states by means of the flavor currents\cite{Blasone:2002wp}.

A realistic description of neutrino oscillations needs to take
into account the finite size of the source and the detector. Moreover, in current experiments what is measured is the distance between the source and the detector
rather than the time of flight of neutrinos.
 We thus consider the flux of (electron) neutrinos through a detector surface $\Om$:
\bea
\Phi_{\nu_e\to \nu_e}(L)=\int_{t_0}^{T}
d t\int_{\Omega} \langle \nu_e| J_{\nu_e}^{i}(\bx,t)|
\nu_e\rangle\,\, d {\bf S}^i 
\eea
The  (localized) neutrino state is described by a wave packet:
\bea
|\nu_e\ran\equiv| \nu_e(\bx_0,t_0)\rangle=A\int d^3
\bk \,  {e}^{-i(\om_{k,1}t_0-\bk\cdot\bx_0)}
f(\bk)\,\alpha_{\bk,e}^{r\dag}(t_0) \,|0(t_0)\rangle_{e,\mu},,
\eea
where $A$ is a normalization constant. The flavor currents are defined as\cite{Blasone:2002wp}
\be
J^\al_{\nu_\si} (x) = \bar{\nu}_\si(x) \gamma^{\al} \nu_\si(x) \, , 
\ee
so that $Q_{\nu_\si}(t) \ = \ \intx J^0_{\nu_\si}(x)$.

One can check that
$_{e,\mu}\langle0|J^{\al}_{\nu_\si}(\bx,t)|0\rangle_{e,\mu}=0$ and
\bea\label{J1}
&&\langle \nu_e |J_e^\al(\bx,t)| \nu_e\rangle
\,=\, {\bar \Psi}(\bx,t)\, \Ga^\al \lf(\begin{array}{cc}1&1\\1&1\end{array}\ri)
\,\Psi(\bx,t)
\eea
with
\bea
&&{}\hspace{-.5cm}
\Psi(\bx,t) \equiv A \int\frac{d^3 \bk}{(2\pi)^\frac{3}{2}}\,
e^{i\bk \cdot \bx}\,f(\bk)\,
\lf(\begin{array}{l} u_{\bk,1}^r \, X_{\bk,e}(t)
\\ [3mm]
 \sum_{s} v_{-\bk,1}^s \,(\vec{\si}\cdot \bk)^{sr}\, Y_{\bk,e}(t)
 \end{array} \ri)\,,
\\ [2mm] \non
&&{}\hspace{-.5cm}
X_{\bk,e}(t)\,=\,\cos^2\theta e^{-i\om_{k,1}t}+\sin^2\theta
\left[e^{-i\om_{k,2} t}|U_{\bk}|^2+
e^{i \om_{k,2}t}|V_{\bk}|^2\right]
\,,
\\[2mm] \non
&&{}\hspace{-.5cm}
Y_{\bk,e}(t) \,=\,\sin^2\theta|U_{\bk}|\chi_1\chi_2
\left[\frac{1}{\om_{k,2}+m_2}  -
\frac{1}{\om_{k,1}+m_1}\right]\left[e^{-i \om_{k,2} t}-
e^{i \om_{k,2}t}\right]\,,
\eea
where  $\sigma_j$ are the Pauli matrices
and $\chi\,_{i}\equiv
\left(\frac{\omega_{k,i}+m_{i}}{4 \omega_{k,i}}\right)^{\frac{1}{2}}$.

Eq.(\ref{J1}) contains the most general
information about neutrino oscillations and can be explicitly
evaluated when the form of the wave-packet is specified. A
similar expression can be obtained also for $J_{\nu_\mu}^{\,{\al}}(x)$.

An oscillation formula in space is then obtained in Ref.\cite{Blasone:2002wp} in
the case of spherical symmetry and by assuming a
gaussian wave packet for the flavor state:
\bea
f(k)=\frac{1}{(2\pi\si_k^2)^{\frac{1}{4}}}\,\,
\exp\left[-\frac{(k-Q)^2}{4\sigma^2_{k}}\right]\,.
\eea
The resulting expression is studied  numerically in Ref.\cite{Blasone:2002wp}  and
it reduces  to the standard
formula \cite{PhysRevD.58.017301,Beuthe:2001rc} in the relativistic limit:
\bea\non
&&{}\hspace{-12mm}
\Phi_{\nu_e\to \nu_e}(z)\simeq  1-\frac{1}{2}\sin^2(2\theta)
\\ 
&& {}\hspace{8mm}+\frac{1}{2}\sin^2(2\theta)
\cos\left(2\pi\frac{z}{L^{osc}}\right)
\,\exp\lf[-\left(\frac{z}{L^{coh}}\right)^2  -2\pi^2
\left(\frac{\sigma_x}{L^{osc}}\right)^2\ri]\,,
\eea
with $L_{osc}=\frac{4\pi Q}{\De m^2}$ and
$ L_{coh}=\frac{L_{osc}Q}{\sqrt{2}\pi\sigma_{k}}$ being the
usual oscillation
length and  coherence length \cite{giunti2007fundamentals}.

\subsection{Lepton number conservation in the vertex} \label{lepnum}

In Ref.~\cite{PhysRevD.45.2414}, it was pointed out that the amplitude for the neutrino detection process \(\nu_\sigma + X_i \rightarrow e^- + X_f\) — where \(X_i\) and \(X_f\) denote the initial and final hadronic states, respectively, and \(e^-\) is the electron, is generally non-vanishing even for \(\sigma \neq e\), when Pontecorvo states are used. 
Indeed, for low-energy weak processes (where the four-fermion Fermi interaction is applicable), one has:
\begin{equation} \label{wae}
\langle e_{-}^{s} | \bar{e}(x)\, \gamma^\mu (1 - \gamma^5)\, \nu_e(x) | \nu_{\sigma}^{r} \rangle_{{}_P} \, h_\mu(x)
= \sum_j U_{e j}\, U^*_{\sigma j} \langle e_{-}^{s} | \bar{e}(x)\, \gamma^\mu (1 - \gamma^5)\, \nu_j(x) | \nu_{j}^{r} \rangle \, h_\mu(x),
\end{equation}
where \(h_\mu(x)\) represents the matrix element associated with the hadronic current. This expression is, in general, different from \(\delta_{\sigma e}\). Such a result appears to be inconsistent, given that the flavor of the neutrino is \emph{defined} through the flavor of the associated charged lepton in the lepton-neutrino doublet (see also Ref.~\cite{Bilenky:2001yh}).
To resolve this inconsistency, the concept of \emph{weak process states} was introduced in Ref.~\cite{PhysRevD.45.2414}. However these will not be discussed in this review.

An important remark is in order. The \(S\)-matrix is defined to connect asymptotic \emph{in} and \emph{out} states~\cite{itzykson2012quantum}:
\begin{equation} \label{sexp}
S_{AB} \ \equiv \ \langle A; \text{out} | B; \text{in} \rangle \, .
\end{equation}
Strictly speaking, flavor states, like unstable states, cannot be treated as asymptotically stable states. Therefore, the application of the \(S\)-matrix formalism to flavor transitions requires particular care. As will be discussed in Section~\ref{perneu}, even in the interaction picture, where mixing is treated as an interaction among flavor fields, the computation of flavor transition amplitudes and probabilities necessitates a finite-time approach, based on the time-evolution operator rather than on the \(S\)-matrix. 

This requirement is intimately related to TEUR, with the oscillating flavor playing the role of a clock observable, a point that will be explored in Section~\ref{TEUR}. Nonetheless, a precise meaning can still be assigned to Eq.~\eqref{sexp}, when interpreted as the asymptotic limit of a corresponding finite-time calculation~\cite{Lee:2017cqf}. In that case a lepton number violation due to neutrino oscillations is actually expected.

Since we were considering the asymptotic regime, the earlier example is not pathological in this context. However, it is essential to emphasize that lepton number must be conserved at short times (at tree level), where flavor oscillations are negligible\footnote{Loop-level processes may induce lepton number violation at production or detection vertices, but such effects are negligible for the present discussion.}.

In order to clarify the last point, let us consider the amplitude of the process $W^+\rightarrow e^+ + \nu_e$, where employ the Pontecorvo states:
\be
\mathcal{A}^P_{W^+ \rightarrow e^+ \, \nu_e} \ = \ {}_{_P}\lan \nu^r_{\G k, e}| \otimes \lan e^s_\G q |\lf[-i \int^{x^0_{out}}_{x^0_{in}} \, \dr^4 x \, \mathcal{H}^e_{int}(x) \ri]|W^+_{\G p, \la} \ran \, .
\ee
The interaction Hamiltonian density is obtained from Eq.\eqref{Linteract}:
\be
\mathcal{H}^e_{int}(x) \ = \  -\frac{g}{2\sqrt{2}} \, W^+_\mu(x) \, J^\mu_e(x) + h.c. \, ,
\ee
where the superscript index means that we are considering the electron-flavor piece, and the leptonic current is
\be
J^\mu_e(x) \ = \ \bar{\nu}_e(x) \, \ga^\mu \, (1-\ga^5)e(x) \, ,
\ee
The usual amplitude is obtained by taking the asymptotic limit \( x_{\text{out}}^0 \rightarrow +\infty \), \( x_{\text{in}}^0 \rightarrow -\infty \). However, as previously mentioned, flavor states are not asymptotically stable states, and our aim is to investigate the short-time behavior of the amplitude, specifically around the interaction time \( x^0 = 0 \). 
Explicit calculations (see Ref.~\cite{Blasone:2006jx}) yield:
\bea 
\mathcal{A}^P_{W^+ \rightarrow e^+ \, \nu_e} & = & \frac{i g }{2 \sqrt{4 \pi}}\frac{\varepsilon_{\G p, \mu, \la}}{\sqrt{2 E^W_\G p}} \de^3(\G p-\G q -\G k)  \\[2mm] \non
& \times & \sum^2_{j=1} \, U^2_{e j}  \int^{x^0_{out}}_{x^0_{in}} \!\! \dr x^0 \, e^{-i \om_{\G k,j} \, x^0_{out}} \, \bar{u}^r_{\G k,j} \, \ga^\mu (1-\ga^5) \, v^s_{\G q,e} \, e^{-i (E^W_\G p-E_\G q^e-\om_{\G k,j}) x^0}  
\eea
where \( E^W_{\boldsymbol{p}} \) and \( \varepsilon_{\boldsymbol{p}, \mu, \lambda} \) are the energy and polarization vector of the \( W^+ \) boson, respectively, and \( v^s_{\boldsymbol{q}, e} \) is the positron wave function. 
We choose \( x^0_{\text{in}} = -\Delta t / 2 \) and \( x^0_{\text{out}} = \Delta t / 2 \), with the hierarchy \( \tau_W \ll \Delta t \ll t_{\text{osc}} \), where \( \tau_W \) is the \( W^+ \) lifetime and \( t_{\text{osc}} \) is the neutrino oscillation time scale. 

Under such conditions, the amplitude can be expanded to leading order in \( \Delta t \), yielding:

\bea
\mathcal{A}^P_{W^+ \rightarrow e^+ \, \nu_e} & \approx & \frac{i g }{2 \sqrt{4 \pi}}\frac{\varepsilon_{\G p, \mu, \la}}{\sqrt{2 E^W_\G p}}\, \de^3(\G p-\G q -\G k) \De t \, \sum^2_{j=1} \, U^2_{e j} \ \, \bar{u}^r_{\G k,j} \, \ga^\mu (1-\ga^5) \, v^s_{\G q,e}  \, .
\eea
In the same manner, one can compute the flavor violating amplitude
\be
\mathcal{A}^P_{W^+ \rightarrow e^+ \, \nu_\mu} \ = \ {}_{_P}\lan \nu^r_{\G k, \mu}| \otimes \lan e^s_\G q |\lf[-i \int^{x^0_{out}}_{x^0_{in}} \, \dr^4 x \, \mathcal{H}^e_{int}(x) \ri]|W^+_{\G p, \la} \ran \, .
\ee
We get
\bea
\hspace{-4mm}\mathcal{A}^P_{W^+ \rightarrow e^+ \, \nu_\mu} & = & \frac{i g }{2 \sqrt{4 \pi}}\frac{\varepsilon_{\G p, \mu, \la}}{\sqrt{2 E^W_\G p}}\, \de^3(\G p-\G q -\G k)  \\[2mm] \non
& \times & \sum^2_{j=1} \, U_{\mu j} \, U_{e j} \, \int^{x^0_{out}}_{x^0_{in}} \!\! \dr x^0 \, e^{-i \om_{\G k,j} \, x^0_{out}} \, \bar{u}^r_{\G k,j} \, \ga^\mu (1-\ga^5) \, v^s_{\G q,e} \, e^{-i (E^W_\G p-E_\G q^e-\om_{\G k,j}) x^0} \, ,
\eea
which, in the short-time limit, becomes
\bea
\hspace{-6mm}\mathcal{A}^P_{W^+ \rightarrow e^+ \, \nu_\mu}  \approx  \frac{i g }{2 \sqrt{4 \pi}}\frac{\varepsilon_{\G p, \mu, \la}}{\sqrt{2 E^W_\G p}}\, \de^3(\G p-\G q -\G k) \De t \, \sum^2_{j=1} \, U_{\mu j} \, U_{e j} \ \, \bar{u}^r_{\G k,j} \, \ga^\mu (1-\ga^5) \, v^s_{\G q,e}  \, .
\eea
The fact we get a result different from zero indicates a clear inconsistency, because in the short-time limit the effect of flavor oscillations should be negligible.

The inconsistency disappears we employ the complete flavor states \eqref{bvflavstate}. Let us start from the decay $W^+ \rightarrow e^+ \, \nu_e$ \cite{Blasone:2006jx}:
\bea \non
\mathcal{A}_{W^+ \rightarrow e^+ \, \nu_e} & = & \frac{i g }{2 \sqrt{2}(2\pi)^\frac{3}{2}} \, \de^3(\G p-\G q -\G k) \, \int^{x^0_{out}}_{x^0_{in}} \!\! \dr x^0 \frac{\varepsilon_{\G p, \mu, \la}}{\sqrt{2 E^W_\G p}}\,  \\[2mm] \non
& \times &  \lf\{ \cos^2 \theta\, e^{-i \om_{\G k,1} \, x^0_{in}} \, \bar{u}^r_{\G k,1} \, \ga^\mu (1-\ga^5) \, v^s_{\G q,e} \, e^{-i (E^W_\G p-E_\G q^e-\om_{\G k,1}) x^0} \ri.	\\[2mm]
& + &\sin^2 \theta 	\lf[|U_\G k| \, e^{-i \om_{\G k,2} \, x^0_{in}} \, \bar{u}^r_{\G k,2} \, \ga^\mu (1-\ga^5) \, v^s_{\G q,e} \, e^{-i (E^W_\G p-E_\G q^e-\om_{\G k,2}) x^0}\ri. \non\\[2mm]
 & + & \lf.\lf. \epsilon^r |V_\G k| \, e^{i \om_{\G k,2} \, x^0_{in}} \, \bar{v}^r_{-\G k,2} \, \ga^\mu (1-\ga^5) \, v^s_{\G q,e} \, e^{-i (E^W_\G p-E_\G q^e+\om_{\G k,2}) x^0}\ri]\ri\}\, .
\eea
The amplitude of the flavor-changing process $W^+ \rightarrow e^+ \, \nu_\mu$ reads
\bea \non
\mathcal{A}_{W^+ \rightarrow e^+ \, \nu_\mu} & = & \sin \theta \, \cos \theta \, \frac{i g }{2 \sqrt{2}(2\pi)^\frac{3}{2}} \, \de^3(\G p-\G q -\G k) \, \int^{x^0_{out}}_{x^0_{in}} \!\! \dr x^0 \frac{\varepsilon_{\G p, \mu, \la}}{\sqrt{2 E^W_\G p}}\,  \\[2mm] \non
& \times &  \lf\{ e^{-i \om_{\G k,2} \, x^0_{in}} \, \bar{u}^r_{\G k,2} \, \ga^\mu (1-\ga^5) \, v^s_{\G q,e} \, e^{-i (E^W_\G p-E_\G q^e-\om_{\G k,2}) x^0} \ri.	 \\[2mm]
& - & 	\lf[|U_\G k| \, e^{-i \om_{\G k,1} \, x^0_{in}} \, \bar{u}^r_{\G k,1} \, \ga^\mu (1-\ga^5) \, v^s_{\G q,e} \, e^{-i (E^W_\G p-E_\G q^e-\om_{\G k,1}) x^0}\ri. \non\\[2mm]
 & + & \lf.\lf. \epsilon^r |V_\G k| \, e^{i \om_{\G k,1} \, x^0_{in}} \, \bar{v}^r_{-\G k,1} \, \ga^\mu (1-\ga^5) \, v^s_{\G q,e} \, e^{-i (E^W_\G p-E_\G q^e+\om_{\G k,1}) x^0}\ri]\ri\}\, .
\eea
Under the condition $\tau_W \ll \De t \ll t_{osc}$ we find, for the short-time result~\cite{Blasone:2006jx}
\bea \non
\hspace{-4mm}\mathcal{A}_{W^+ \rightarrow e^+ \, \nu_e} & \approx & \frac{i g }{2 \sqrt{2}(2\pi)^\frac{3}{2}} \, \de^3(\G p-\G q -\G k)  \frac{\varepsilon_{\G p, \mu, \la}}{\sqrt{2 E^W_\G p}}\,  \De t\\[2mm]
& \times &  \lf\{ \cos^2 \theta \, \bar{u}^r_{\G k,2}+\sin^2 \theta 	 \lf[|U_\G k| \bar{u}^r_{\G k,2}+ \epsilon^r |V_\G k| \, \bar{v}^r_{-\G k,2} \ri]\ri\} \, \ga^\mu (1-\ga^5) \, v^s_{\G q,e}\, , 	\label{aee}
\eea
and
\bea 	\label{aemu}
\mathcal{A}_{W^+ \rightarrow e^+ \, \nu_\mu} & \approx & 0\, .
\eea
i.e., as anticipated, the calculation performed with the complete flavor states leads to the expected result. This is a consequence of the fact that the complete flavor states are eigenstates of the lepton charges and that such charges commute with the interaction Hamiltonian.

\subsection{Entanglement in neutrino mixing} \label{Entangl}

The phenomenon of entanglement is a fundamental aspect of QM and QFT \cite{nielsen2000quantum}. Over the past few decades, it has been experimentally verified, and significant efforts have been dedicated to its investigation, particularly in quantum optics and quantum computing. Notably, entanglement also plays a role in neutrino mixing and oscillations \cite{ill1,ill2,ill3,BLASONE2013320,Blasone:2014jea, Bittencourt:2014pda,ill4,ill5}, as we will now briefly review. 

Let us start from the standard QM treatment of flavor states presented in Section \ref{stder}. We rewrite the flavor states
 \eqref{stflavstates} in the form
\bea
|\nu_{e}\ran & = &\cos \theta \, |1\ran_1  |0\ran_2+\sin \theta \, |0\ran_1  |1\ran_2 \, ,  \\[2mm]
|\nu_{\mu}\ran & = & -\sin \theta \, |1\ran_1  |0\ran_2+\cos \theta \, |0\ran_1  |1\ran_2\, ,
\eea
where we introduced the notation
\bea
|\nu_{1}\ran & = & |1\ran_1 |0\ran_2  \ \equiv \ |1\ran_1 \otimes |0\ran_2 \, ,  \\[2mm]
|\nu_{2}\ran & = & |0\ran_1 |1\ran_2 \ \equiv \  |0\ran_1 \otimes |1\ran_2 \, .
\eea
Let us remark that the necessity of the tensor product structure directly comes from QFT. where Hilbert spaces for fields with different masses are orthogonal (see \ref{ineqcar}). Because the QM treatment should be viewed as a limiting case of the QFT treatment, we should maintain the tensor product structure. The above state is clearly an entangled state of the massive neutrino qubits $|\nu_{j}\ran$, $j=1,2$. This phenomenon is known as \emph{static entanglement} \cite{Blasone:2010ta,blasone2011quantum}. We can now write the density matrix, e.g. for a neutrino with a given flavor $\si$
\be
\rho^{(\si)} \ = \ |\nu_\si\ran \lan \nu_\si| \, .
\ee
The \emph{reduced density operators} are defined tracing out one of the two mass components
\be
\rho^{(\si)}_i \ = \ \mathrm{Tr}_j \lf[\rho^{(\si)}\ri] \, , \qquad i \ \neq \  j \, .
\ee
Explicitly, e.g. for $\rho^{(e)}$, we find
\bea
\rho^{(e)}_1 & = & \cos^2 \theta \, |1\ran_1 \, {}_1\lan 1 | \, + \, \sin^2 \theta \, |0\ran_1 \,  {}_1\lan 0| \, , \\[2mm]
\rho^{(e)}_2 & = & \cos^2 \theta \, |0\ran_2 \, {}_2\lan 0 | \, + \, \sin^2 \theta \, |1\ran_2 \, {}_2\lan 1| \, .
\eea
In order to quantify the entanglement, we compute the corresponding \emph{linear entropies} (also known as \emph{impurity})
\bea \label{sl1}
S^{1;2}_L & = & 2\, \lf(1-\mathrm{Tr}_1\lf[\lf(\rho^{(e)}_1\ri)^2\ri]\ri)   \, , \\[2mm]
S^{2;1}_L & = & 2\, \lf(1-\mathrm{Tr}_2\lf[\lf(\rho^{(e)}_2\ri)^2\ri]\ri)  \, . \label{sl2}
\eea
By means of the above expressions it is easy to compute\cite{ill2}
\be \label{sten}
S^{1;2}_L \ = \ S^{2;1}_L \ = \ \sin^2(2\theta)\, .
\ee

Let us now consider the time evolution of the flavor states
\be
 |\nu_\si(t)\ran \ = \ \sum_{\rho=e,\mu} \, \tilde{U}_{\si \rho}(t)  |\nu_\rho\ran \, , \qquad \si=e,\mu \, .
\ee
where
\be
{U}_{\si \rho}(t) \ \equiv \ U \, \begin{bmatrix} e^{-i \om_1 t} & 0 \\ 0 & e^{-i \om_2 t} \end{bmatrix} \, U^{-1} \, , 
\ee
being $U$ the mixing matrix.
Defining
\bea
|\nu_{e}\ran & = & |1\ran_e \otimes |0\ran_\mu \ = \ |1\ran_e |0\ran_\mu \, ,  \\[2mm]
|\nu_{\mu}\ran & = & |0\ran_e \otimes |1\ran_\mu \ = \ |0\ran_e |1\ran_\mu\, ,
\eea
We can re-interpret the flavor states at time $t$, as entangled states of neutrinos with a definite flavor at a reference time, identified as flavor neutrino qubits. This phenomenon is known as \emph{dynamical entanglement}.

Let us then consider the density matrix at time $t$:
\be
\rho^{(\si)}(t) \ = \ |\nu_\si(t)\ran \lan \nu_\si(t)| \, .
\ee
The reduced density operators are defined tracing out one of the two flavors (at $t=0$)
\be
\rho^{(\si)}_\al(t) \ = \ \mathrm{Tr}_\bt \lf[\rho^{(\si)}(t)\ri] \, , \qquad \al \ \neq \  \bt \, ,
\ee
The corresponding \emph{linear entropies} are~\cite{ill2}
\bea \label{sle}
S^{e;\mu}_L (t)& = & 2\, \lf(1-\mathrm{Tr}_e\lf[\lf(\rho^{(e)}_\mu(t)\ri)^2\ri]\ri)   \, , \\[2mm]
S^{\mu;e}_L(t) & = & 2\, \lf(1-\mathrm{Tr}_\mu\lf[\lf(\rho^{(e)}_e(t)\ri)^2\ri]\ri)  \, . \label{slmu}
\eea
An explicit computation leads to
\be \label{leemu}
S^{e;\mu}_L (t)\ = \ S^{\mu;e}_L (t)\ = \  \, 4\,\mathcal{P}_{e \rightarrow \mu} (t) \, (1-\mathcal{P}_{e \rightarrow \mu} (t)) \, .
\ee
Thus, dynamical entanglement in terms of the flavor qubits develops in time and it is maximal when the oscillation probabilities are equal, which means maximal ignorance about the flavor of the neutrino state\footnote{It is important to note that a geometric phase contribution is contained in the phase of flavor neutrino state at time $t\neq 0$, see Ref.\cite{Blasone:1999tq}}.

Notice also that when there is no-mixing, both static and dynamic linear entropies correctly go to zero.

How can one quantify static and dynamic entanglement in QFT? Interestingly, it turns out that the variance of conserved charges evaluated on a given flavor state, whose entanglement we wish to quantify, coincides with the previously computed impurity~\cite{Blasone:2014jea}. This observation aligns with the more general discussion presented in Ref.~\cite{PhysRevA.75.032315}, where the variance is proposed as a measure of entanglement.

As an example, let us consider the Noether charges
\be
Q_{\nu_j} \ = \ \intx \, \nu^\dag_j(x) \nu_j(x) \, , \qquad j=1,2 \, , 
\ee
which are conserved because of the phase invariance of the Dirac equations \eqref{direq}, we obtain a measure of the static entanglement on the exact flavor state $|\nu^r_{\G k,\rho}\ran$ \cite{Blasone:2014jea}:
\bea \label{DEQn1}
\si^2_{Q_{j}} \ \equiv \ \lf(\De Q_{\nu_j}\ri)^2  \ = \ \lan Q_{\nu_j}^2\ran_\rho \,-\,
\lan Q_{\nu_j}\rangle_\rho^2
\ = \  \frac{1}{4} \sin^2(2\theta)\, .
\eea
Remarkably this result coincides with the QM one (see Eq.\eqref{sten}).
On the other hand, the \emph{dynamic} entanglement can be measured by the variances of the flavor
charges:
\bea \label{varq}
\sigma^2_Q \ \equiv \ \lf(\De Q_{\nu_\rho}\ri)^2  \ = \ \lan Q^2_{\nu_\rho}(t)\ran_\rho \ - \ \lan Q_{\nu_\rho}(t)\ran_\rho^2 \  = \ \ \mathcal{Q}_{\rho \rightarrow \rho}(t)\lf(1-\mathcal{Q}_{\rho\rightarrow \rho}(t)\ri) \, .
\eea
In this case, the result formally looks the same (up to a  constant)  as Eq.~\eqref{leemu}. However, the two expressions coincide only in the relativistic limit, where the exact oscillation formula~\eqref{oscfor} reduces approximately to the standard one~\eqref{sform}.

\subsection{Flavor--Energy uncertainty relations} \label{TEUR}
In Section \ref{bfmixing} we remarked that
flavor charges do not commute with the free part of the Lagrangian
and then with the corresponding Hamiltonian $H$. This leads to a \emph{flavor-energy uncertainty relation} \cite{Blasone2019},  that can be formalized via the Mandelstam--Tamm \emph{time-energy uncertainty relation} (TEUR) \cite{ManTam}, where flavor charges play the role of clock observables for the oscillating neutrino systems. Note that, because only flavor can be detected in weak processes, such uncertainty relations put a fundamental bound on neutrino energy/mass precision. As remarked in  Refs. \cite{Bilenky2008,Akhmedov:2008zz,Bilenky:2009zz}, Mossbauer neutrinos could furnish an example of neutrinos produced with a definite energy and then this could spoil the oscillation phenomenon. Here we do not discuss such a problem.

Mandelstam--Tamm version of TEUR is formulated as~\cite{ManTam}
\be \label{teunc}
\Delta E \, \Delta t \, \geq \frac{1}{2} \, .
\ee
We put
\be
\Delta E \equiv \si_H \, \qquad \Delta t \equiv \si_O/\lf|\frac{\dr \lan O(t) \ran}{\dr t}\ri| \, .
\label{teunc1}
\ee
Here $O(t)$ represents the ``clock observable'' whose dynamics quantifies temporal changes in a system
and $\Delta t$ is the characteristic time interval over which the mean value of $O$ changes by a standard deviation.

TEUR for neutrino oscillations in flat spacetime have been extensively studied in Refs. \cite{Bilenky:2005hv,Bilenky2008,Akhmedov:2008zz,Bilenky:2009zz} and later extended to stationary curved spacetimes \cite{Blasone2020}.

\subsubsection{Time-energy uncertainty relations for neutrino oscillations in quantum mechanics} \label{bilteur}
We start reviewing how TEUR for neutrino oscillations was treated in the standard QM approach  presented in Section \ref{stder}. 

Following~\cite{Bilenky:2005hv}, we set $O(t)=P_\si=|\nu_\si(t)\rangle\langle\nu_\si(t)|$ in Eq.\eqref{teunc}. Projection operators are idempotent: $P^2=P$. From this condition, it follows that the standard deviation $\Delta P_\si$ obeys
\be
\Delta P_\si(t)=\sqrt{P_{\si\rightarrow \si}(t)-P^2_{\si\rightarrow \si}(t)}\,,
\ee
where the survival probability is generally given by the expression Eq.\eqref{psirho}.
By substitution into Eq.\eqref{teunc}, we derive
\be
\label{BilTEUR}
\Delta E\ge\frac{1}{2}\hspace{0.2mm}\frac{|\frac{d}{dt}P_{\si\rightarrow\si}(t)|}
{\sqrt{P_{\si\rightarrow\si}(t)-P^2_{\si\rightarrow\si}(t)}}\,.
\ee

Consider the survival probability $P_{\si\rightarrow\si}(t)$ in the time interval $0 \leq t \leq t_{1min}$,
where $t_{1min}$ is the time when $P_{\si\rightarrow\si}(t)$ reaches the first minimum. In this range,  $P_{\si\rightarrow\si}(t)$ is a monotonically decreasing function~\cite{Bilenky:2005hv,Bilenky2008,Akhmedov:2008zz,Bilenky:2008dk,Bilenky:2009zz,Bilenky:2011pk}. By integration of Eq.~\eqref{BilTEUR} from $0$ to $t$, we finally get
\be 
\label{final}
\Delta E\, t\ge \frac{1}{2}\left[\frac{\pi}{2}-\sin^{-1}\left(2P_{\si\rightarrow\si}(t)-1\right)\right].
\ee

It is instructive to apply the above inequality to specific cases of experimental relevance. A notable example involves atmospheric neutrinos, which are produced when cosmic rays interact with molecules in the Earth's upper atmosphere. These interactions generate pions and kaons, which subsequently decay into muon neutrinos and muons; many of the resulting muons then decay into electrons, along with a pair consisting of a muon neutrino and an electron neutrino. A rough estimate suggests that muon neutrinos are approximately twice as abundant as electron neutrinos~\cite{K2K:2004iot,Super-Kamiokande:2004orf}.
In the atmospheric long-baseline regime, the survival probability $ P_{\mu \rightarrow \mu}(t) $ is primarily governed by the mass-squared difference $ \Delta m_{23}^2 $. Noting that 
$
P_{\mu \rightarrow \mu}(t^{(23)}_{1\text{min}}) \simeq 0\,,
\quad \text{with} \quad
t^{(23)}_{1\text{min}} = \frac{2\pi E}{\Delta m_{23}^2}\,,
$
we obtain the following time-energy uncertainty relation (TEUR) for ultrarelativistic neutrinos (where \( L \simeq t \))~\cite{Bilenky:2005hv}:

\be
\label{atm}
\Delta E\, t_{osc}^{(23)}\ge\pi\,,
\ee
where $ t_{\text{osc}}^{(23)} = 2t^{(23)}_{1\text{min}} $ denotes the oscillation period in the atmospheric long-baseline regime. As explained in Ref.~\cite{Bilenky:2005hv}, the above relation provides a necessary condition for atmospheric neutrino oscillations to be observable.


An alternative way to simplify the inequality \eqref{BilTEUR} is to notice that maximum value of the square root in the denominator of the r.h.s. is $\ha$. Then, the simpler inequality
\be
\Delta E  \ \geq  \ \lf|\frac{\dr P_{\si\rightarrow \si}(t)}{\dr t} \ri| \, ,
\ee
is also true. By means of triangular inequality and integrating both
sides from $0$ to a generic $T$, which is not forced to be within the interval $[0,t_{1 \text{min}}]$. We thus get
\be
\mbox{\hspace{-2mm}}\Delta E \, T  \ \geq  \ \int^T_{0} \!\! \dr t \, \lf|\frac{\dr P_{\si\rightarrow \si}(t)}{\dr t} \ri| \, \geq  \,\lf|\int^T_{0} \!\! \dr t \, \frac{\dr P_{\si\rightarrow \si}(t)}{\dr t} \ri| \, ,
\ee
where, in the last passage, we used the triangular inequality.
Therefore, one finds
\be \label{etrel}
\Delta E\,  T \geq P_{\si\rightarrow \rho}(T)  \, ,  \quad \si \neq \rho  \, ,
\ee
with $
P_{\si\rightarrow \rho}(t) \ = \  1-P_{\si\rightarrow \si}(t) $, and where we noticed the the probability is always non-negative.
For $T=T_h$, so that $P_{\si\rightarrow \rho}(T_h)=\ha$, we find the suggestive form
\be
\Delta E\, T_h \geq \ha \, ,
\ee
which looks like the standard Heisenberg inequality of the TEUR.

\subsubsection{The quantum field theory case: flavor vs energy} \label{qftfeur}
Let us now consider the QFT treatment of TEUR. In the previous subsection we have seen that TEUR arises from the the incompatibility of the Hamiltonian with the projector on a definite-flavor state. In other words, we are dealing with a \emph{flavor-energy uncertainty relation}.

In the QFT treatment, the neutrino lepton charges, introduced in Eq.\eqref{QflavLept}, are regarded as the relevant flavor operators: we thus choose them as {\em clock observables}.

Then, our starting point is
\be
\lf[Q_{\nu_\si}(t) \ , \, H\ri] \ = \ i \, \frac{\dr Q_{\nu_\si}(t)}{\dr t} \ \neq \ 0 \, ,
\ee
which leads to
\be \label{neutun}
\sigma_H \, \sigma_Q \ \geq \ \frac{1}{2}\lf|\frac{\dr \mathcal{Q}_{\si\rightarrow \si}(t)}{\dr t}\ri|.
\ee
The flavor variance was computed in Eq.\eqref{varq} and quantifies the dynamic entanglement for neutrino states in QFT (cf. Section \ref{Entangl}). Proceeding as before one finds the analogous of the inequality Eq.\eqref{BilTEUR}:
\be
\label{ourTEUR}
\Delta E\ge\frac{1}{2}\hspace{0.2mm}\frac{|\frac{d}{dt}\mathcal{Q}_{\si\rightarrow\si}(t)|}
{\sqrt{\mathcal{Q}_{\si\rightarrow\si}(t)-\mathcal{Q}^2_{\si\rightarrow\si}(t)}}\,.
\ee
As noticed above, the square root of the charge variance is always smaller or equal than $\ha$. Then, we arrive at the simpler inequality:
\be
\Delta E \ \geq \ \lf|\frac{\dr \mathcal{Q}_{\si\rightarrow \si}(t)}{\dr t}\ri| \, .
\ee
Integrating by parts and employing the triangular inequality
\be
\Delta E \, T  \ \geq  \ \int^T_{0} \!\! \dr t \, \lf|\frac{\dr \mathcal{Q}_{\si\rightarrow \si}(t)}{\dr t} \ri| \, \geq  \,\lf|\int^T_{0} \!\! \dr t \, \frac{\dr \mathcal{Q}_{\si\rightarrow \si}(t)}{\dr t} \ri| \, ,
\ee
and we arrive at the Mandelstam--Tamm TEUR in the form
\be \label{etq}
\Delta E \,T \ \geq\  \mathcal{Q}_{\si\rightarrow \rho}(T)  \, ,  \quad \si \neq \rho .
\ee

Let us look at this relation in some relevant cases.
When $m_i/|\G k|\rightarrow 0$, i.e. in the relativistic case, we have
\bea \label{firstapprox1}
|U_\G k|^2  \approx  1 \  -  \ \varepsilon(\G k)  \, , \;\;\;\;\;
|V_\G k|^2  \approx  \varepsilon(\G k)  \, ,
\eea
with $\varepsilon(\G k)   \equiv {(m_1-m_2)^2}/{4 |\G k|^2}$.
In the same limit
\be
\Om_{\G k}^{_-} \ \approx \ \frac{\delta m^2}{4 |\G k|}\ = \ \frac{\pi}{L_{osc}} \, , \qquad \Om_{\G k}^{_+} \ \approx \ |\G k| \, .
\ee
Therefore, as previously remarked, at the leading order  $|U_\G k|^2 \rightarrow 1$, $|V_\G k|^2 \rightarrow 0$ and the standard oscillation formula \eqref{stoscfor} is recovered. The r.h.s. of (\ref{stoscfor}) reaches its maximum at $L=L_{osc}/2$ and the inequality (\ref{etq}) reads
\be \label{condne1}
\De E \ \geq \ \frac{2 \sin^2(2\theta)}{L_{osc}} \, .
\ee
Note that $\De E$ is time independent and then~(\ref{condne1}) applies in the interaction vertex.

Let us now consider the exact oscillation formula (\ref{oscfor}) in the next-to-the-leading relativistic order in $\varepsilon(\G k)$:
\begin{eqnarray}
\mathcal{Q}_{\si\rightarrow \rho}(t) \approx \sin^2 (2 \theta) \lf[\sin^2\lf(\frac{\pi t}{L_{osc}}\ri) \lf(1 - \varepsilon(\G k) \ri)  +  \varepsilon(\G k)  \sin^2\lf(|\G k|t\ri)\ri] \, , \quad \si \neq \rho \, .
\end{eqnarray}
By setting $T= L_{osc}/2$, the relation~(\ref{etq}), can be  written as
\be
\De E  \ \geq \ \frac{2 \, \sin^2 2 \theta}{L_{osc}} \, \lf[1  -  \varepsilon(\G k) \,  \cos^2\lf(\frac{|\G k|L_{osc}}{2}\ri)\ri] \, ,
\ee
i.e. the bound on the energy is lowered with respect to \eqref{condne1}.
For neutrino masses~\footnote{The values for neutrino masses are taken from Ref.~\cite{PhysRevD.98.030001}.}: $m_1=0.0497 \, {\rm eV}$, $m_2=0.0504 \, {\rm eV}$,
and $|\G k|= 1 \, {\rm MeV}$, then $\varepsilon(\G k) = 2 \times 10^{-19}$, which is negligible in any respect.

On the other hand, in the non-relativistic regime consider, e.g., $|\G k|= \sqrt{m_1 m_2}$. In this case,
\bea
|U_\G k|^2 & = & \ha \ +\ \frac{\xi}{2} \ = \  1-|V_\G k|^2 \, , \\[2mm]
\xi & = & \frac{2\sqrt{m_1 m_2}}{m_1+m_2} \, ,
\eea
and we can rewrite~(\ref{etq}) as
\bea
\De E \, T \geq  \frac{\sin^2 2 \theta}{2} \, \Big[1 - \,  \cos \lf(\tilde{\om}_{1}T\ri)\cos \lf(\tilde{\om}_{2}T\ri)  - \, \xi  \sin \lf(\tilde{\om}_{1}T\ri)\sin \lf(\tilde{\om}_{2}T\ri)\Big] \, ,
\eea
with $\tilde{\om}_j = \sqrt{m_j(m_1+m_2)}$. To compare it with the relativistic case,
we take $T=\tilde{L}_{osc}/4$, with $\tilde{L}_{osc}=4\pi\sqrt{m_1 m_2}/\de m^2$, obtaining
\bea
\De E  & \geq & \frac{2\sin^2 2 \theta}{\tilde{L}_{osc}} \ \lf(1-\chi\ri) \, .
\eea
Here
\begin{eqnarray}
\hspace{-3mm}\chi \ = \ \xi \, \sin \lf(\tilde{\om}_{1}\tilde{L}_{osc}/4\ri)\sin\lf(\tilde{\om}_{2} \tilde{L}_{osc}/4\ri) \ + \ \cos \lf(\tilde{\om}_{1}\tilde{L}_{osc}/4\ri)\cos\lf(\tilde{\om}_{2} \tilde{L}_{osc}/4\ri) \, .
\end{eqnarray}
Substituting the same values for neutrino masses, we obtain $\chi=0.1$, i.e. the original bound on energy decreased by $10\%$.

Let us stress that the TEUR has traditionally been interpreted as a condition for neutrino oscillations: if one were able to measure the neutrino energy with extremely high precision, oscillations would be suppressed, as the oscillation length would diverge~\cite{PhysRevD.24.110,Bilenky:2005hv}.
However, the inequivalence between mass and flavor representations suggests a new interpretation. From the above  inequalities we infer that neutrinos, being always produced and detected in definite flavor states, possess an intrinsic energy uncertainty. This uncertainty sets a fundamental limit on the experimental precision achievable in energy or mass measurements.

In conclusion, the fact that in the presence of mixing, neutrino flavor charges are not conserved, i.e. they do not commute with the Hamiltonian, poses a fundamental dichotomy: the eigenstates of the Hamiltonian and those of the flavor charges belong to different (orthogonal) Hilbert spaces. Therefore, the problem of the choice arise, which ones are the physical states. Here we argue in favor of the flavor states - excitations of the flavor vacuum - as the  states produced in weak interaction processes.

\subsection{Three-flavor neutrino mixing in quantum field theory} \label{3flavorAppA}

The mixing transformation in the three-flavor case can be rewritten as\cite{Blasone:1995zc,PhysRevD.66.025033}:
\be \label{3flavmix}
\nu_{\si}(x)=G_\theta ^{-1}(t)\,\nu_{j}(x)
\,G_\theta (t),  
\ee
where $(\si,j)=(e,1), (\mu,2), (\tau,3),$  and the generator of the mixing transformation is
\be
G_\theta (t) =G_{23}(t)G_{13}(t)G_{12}(t)
\ee
with
\bea  
G_{12}(t)&=& \exp\lf[\theta_{12}
\int d^{3}\bx(\nu_{1}^{\dag}(x)\nu_{2}(x)-\nu_{2}^{\dag}(x)\nu_{1}(x))\ri],
\\
G_{13}(t)&=& \exp\lf[\theta_{13}\int d^{3}\bx(\nu_{1}^{\dag}(x)\nu_{3}(x)
e^{-i\delta}-\nu_{3}^{\dag}(x)\nu_{1}(x)e^{i\delta})\ri],
\\ 
G_{23}(t)&=& \exp\lf[\theta_{23}\int d^{3}\bx(\nu_{2}^{\dag}(x)\nu_{3}(x)-
\nu_{3}^{\dag}(x)\nu_{2}(x))\ri].
\eea
The proof of Eq.\eqref{3flavmix} is similar to the one given for Eq.\eqref{Gmix2}.

As for the two flavor mixing, the flavor vacuum is defined by
\bea |0(t)\rangle_{f} \; = \;G_\theta^{-1}(t) \;|0\rangle_{m} \, ,  
\eea 
where $|0\rangle_{m}$ is the mass vacuum. This is annihilated by the operators
\bea
\alpha_{{\bf k},e}^{r}&=&c_{12}c_{13}\;\alpha_{{\bf k},1}^{r} +
s_{12} c_{13} \left( U_{12}^{{\bf k} *}\; \alpha_{{\bf k},2}^{r}
+\epsilon^{r} V^{{\bf k}}_{12} \; \beta_{- {\bf k},2}^{r\dag}\right) \non \\[2mm]
&& + e^{-i\delta} \; s_{13} \left(U^{{\bf k}*}_{13}\;\alpha_{{\bf k},3}^{r}
+\epsilon^{r} V^{{\bf k}}_{13}\;\beta_{-{\bf
k},3}^{r\dag}\right)\, , \\[2mm]  \non
\alpha_{{\bf k},\mu}^{r} &=&\left(c_{12}c_{23}- e^{i\delta}
\;s_{12}s_{23}s_{13}\right)\;\alpha_{{\bf k},2}^{r}
- \left(s_{12}c_{23}+e^{i\delta}\;c_{12}s_{23}s_{13}\right)
\left(U^{{\bf k}}_{12}\;\alpha_{{\bf k},1}^{r}
-\epsilon^{r} V^{{\bf k}}_{12}\;\beta_{-{\bf k},1}^{r\dag}\right)  \\[2mm]
&&+\;s_{23}c_{13}\left(U^{{\bf k}*}_{23}\;\alpha_{{\bf k},3}^{r}
+ \epsilon^{r} V^{{\bf k}}_{23}\;\beta_{-{\bf k},3}^{r\dag} \right)\, , \\[2mm]
{}\hspace{-.8cm}
\alpha_{{\bf k},\tau}^{r} &=& c_{23}c_{13}\;\alpha_{{\bf k},3}^{r}
- \left(c_{12}s_{23}+e^{i\delta}\;s_{12}c_{23}s_{13}\right)
\left(U^{{\bf k}}_{23}\;\alpha_{{\bf k},2}^{r}
 -\epsilon^{r} V^{{\bf k}}_{23}\;\beta_{-{\bf k},2}^{r\dag}\right)  \non
\\[2mm]
&&+\;\left(s_{12}s_{23}- e^{i\delta}\;c_{12}c_{23}s_{13}\right)
\left(U^{{\bf k}}_{13}\;\alpha_{{\bf k},1}^{r}
-\epsilon^{r} V^{{\bf k}}_{13}\;\beta_{-{\bf k},1}^{r\dag}\right) \, .
\eea

Similar expressions are found for antiparticles, with $\de\rar -\de$. The  Bogoliubov coefficients read
\bea
V^{{\bf k}}_{ij}=|V^{{\bf
k}}_{ij}|\;e^{i(\omega_{\G k,j}+\omega_{\G k,i})t}\;\;,\;\;
U^{{\bf k}}_{ij}=|U^{{\bf k}}_{ij}|\;e^{i(\omega_{\G k,j}-\omega_{\G k,i})t}, ~~i,j =1,2,3, ~~i>j,
\eea
\bea |U^{{\bf
k}}_{ij}|=\left(\frac{\omega_{\G k,i}+m_{i}}{2\omega_{\G k,i}}\right)
^{\frac{1}{2}}
\left(\frac{\omega_{\G k,j}+m_{j}}{2\omega_{\G k,j}}\right)^{\frac{1}{2}}
\left(1+\frac{|{\bf
k}|^{2}}{(\omega_{\G k,i}+m_{i})(\omega_{\G k,j}+m_{j})}\right) ,
\eea
\bea
|V^{{\bf k}}_{ij}|=
\left(\frac{\omega_{\G k,i}+m_{i}}{2\omega_{\G k,i}}\right)
^{\frac{1}{2}}
\left(\frac{\omega_{\G k,j}+m_{j}}{2\omega_{\G k,j}}\right)^{\frac{1}{2}}
\left(\frac{|{\bf k}|}{(\omega_{\G k,j}+m_{j})}-\frac{|{\bf
k}|}{(\omega_{\G k,i}+m_{i})}\right) ,
\eea
\bea
|U^{{\bf k}}_{ij}|^{2}+|V^{{\bf k}}_{ij}|^{2}=1 \quad, \quad i,j=1,2,3 \;\;,
\;\; i>j.
\eea


Identities like the following one are found to be valid\cite{PhysRevD.66.025033}: 
\bea
V^{{\bf k}}_{23}(t)V^{{\bf
k}*}_{13}(t)+U^{{\bf k}*}_{23}(t)U^{{\bf k}}_{13}(t) = U^{{\bf
k}}_{12}(t).
\eea

Therefore, the neutrino flavor-fields can be expanded exactly as in Eqs.\eqref{nue},\eqref{numu}, with an analogous expansion for $\nu_\tau$. The same is true for the flavor/lepton charges: one just has to add the tau-neutrino charge $Q_{\nu_\tau}(t)$. Notice that the flavor charges in the three-flavor case, are obtained by means of $SU(3)$ transformations, see Ref.\cite{PhysRevD.66.025033}.

It is interesting to report that the various parametrizations of the PMNS
matrix can be obtained by introducing different phases and defining the more general generators:
\bea 
&&G_{12}\equiv\exp\Big[\theta_{12}\int
d^{3}x\lf(\nu_{1}^{\dag}\nu_{2}e^{-i\de_{2}}-
\nu_{2}^{\dag}\nu_{1}e^{i\de_{2}}\ri)\Big]
\\
&&G_{13}\equiv\exp\Big[\theta_{13}\int
d^{3}x\lf(\nu_{1}^{\dag}\nu_{3}e^{-i\de_{5}}-
\nu_{3}^{\dag}\nu_{1}e^{i\de_{5}}\ri)\Big]
\\ 
&&G_{23}\equiv \exp\Big[\theta_{23}\int
d^{3}x\lf(\nu_{2}^{\dag}\nu_{3}e^{-i\de_{7}}-
\nu_{3}^{\dag}\nu_{2}e^{i\de_{7}}\ri)\Big]
\eea
Permutating the above generators, we get six different matrices, so that we can obtain all possible parametrizations of the PMNS matrix setting to zero
two of the phases and permuting rows/columns.

The oscillations formula can be obtained, as above, taking the expectation value of the flavor charges on a reference-time flavor state. For example, considering an initial electron-neutrino state, one gets

\begin{footnotesize}
\bea \non {\cal Q}_{e \to e}(t) \, &=& \,1 \,-\, \sin^{2}( 2
\theta_{12})\cos^{4}\theta_{13} \, \Big[|U_{12}^{\bf k}|^2\, \sin^{2}
\lf(  \De_{12}^{\bf k}  t \ri)
 +\,|V_{12}^{\bf k}|^2 \, \sin^{2}
\lf(\Om_{12}^{\bf k}t \ri)\Big] \
\\ \non && -\, \sin^{2}(2 \theta_{13})\cos^{2}\theta_{12}
\, \Big[|U_{13}^{\bf k}|^2\,
 \sin^{2} \lf( \De_{13}^{\bf k}  t \ri)
 +\,|V_{13}^{\bf k}|^2\, \sin^{2} \lf( \Om_{13}^{\bf k} t \ri)\Big] \
\\ &&
-\, \sin^{2}(2 \theta_{13})\sin^{2}\theta_{12} \, \Big[|U_{23}^{\bf
k}|^2 \, \sin^{2} \lf( \De_{23}^{\bf k} t \ri) +\,|V_{23}^{\bf
k}|^2\, \sin^{2} \lf( \Om_{23}^{\bf k} t \ri)\Big]\,, \eea
\bea\non 
\hspace{-4mm}
{\cal Q}_{e \to \mu}(t)
 &=& 2 J_{\CP}
 \Big[|U_{12}^{\bf k}|^2\, \sin(2\De_{12}^{\bf k}t)
- |V_{12}^{\bf k}|^2\, \sin(2\Om_{12}^{\bf k} t) + (|U_{12}^{\bf
k}|^2 - |V_{13}^{\bf k}|^2 ) \sin(2\De_{23}^{\bf k}t)
\\ \non
&+& (|V_{12}^{\bf k}|^2 - |V_{13}^{\bf k}|^2 ) \sin(2\Om_{23}^{\bf
k}t)
  - |U_{13}^{\bf k}|^2\,
\sin(2\De_{13}^{\bf k}t)+ |V_{13}^{\bf k}|^2\, \sin(2\Om_{13}^{\bf
k}t)\Big]
\\ \non
&+&\, \cos^{2}\theta_{13} \sin\theta_{13}
\Big[\cos\de\sin(2\theta_{12})\sin(2\theta_{23}) + 4
\cos^2\theta_{12}\sin\theta_{13}\sin^2\theta_{23}\Big]
\times
\\ \non
&\times &\Big[|U_{13}^{\bf
k}|^2\sin^{2} \lf(\De_{13}^{\bf k} t \ri) + |V_{13}^{\bf k}|^2\
\sin^{2} \lf( \Om_{13}^{\bf k} t \ri)\Big]
\\ \non
& -& \cos^{2}\theta_{13}\sin\theta_{13}
 \Big[\cos\de\sin(2\theta_{12})\sin(2\theta_{23}) -
4 \sin^2\theta_{12}\sin\theta_{13}\sin^2\theta_{23}\Big] \times
\\ \non
&\times&\Big[|U_{23}^{\bf
k}|^2\ \sin^{2} \lf( \De_{23}^{\bf k} t \ri)
 + |V_{23}^{\bf k}|^2\
\sin^{2} \lf( \Om_{23}^{\bf k} t \ri)\Big]
\\ \non
& +&\cos^{2}\theta_{13} \sin(2\theta_{12}) \Big[ (\cos^2\theta_{23} -
\sin^2\theta_{23}\sin^2\theta_{13})\sin(2\theta_{12})
+\cos\de\cos(2\theta_{12})\sin\theta_{13}\sin(2\theta_{23})\Big] \times
\\
&\times& \Big[|U_{12}^{\bf k}|^2\ \sin^{2} \lf(\De_{12}^{\bf k} t
\ri) + |V_{12}^{\bf k}|^2\ \sin^{2} \lf( \Om_{12}^{\bf k} t
\ri)\Big]\,, \eea
\bea\non
\hspace{-4mm}{\cal Q}_{e \to \tau}(t)&=& - 2 J_{\CP}
 \Big[|U_{12}^{\bf k}|^2\ \sin(2\De_{12}^{\bf k}t)
- |V_{12}^{\bf k}|^2\, \sin(2\Om_{12}^{\bf k} t) + (|U_{12}^{\bf
k}|^2\, - |V_{13}^{\bf k}|^2 ) \sin(2\De_{23}^{\bf k}t)
\\ \non
&+& (|V_{12}^{\bf k}|^2\, - |V_{13}^{\bf k}|^2 )
\sin(2\Om_{23}^{\bf k}t)
 \, - |U_{13}^{\bf k}|^2\,
\sin(2\De_{13}^{\bf k}t)+ |V_{13}^{\bf k}|^2\, \sin(2\Om_{13}^{\bf
k}t)\Big]
\\ \non
&-& \cos^{2}\theta_{13}\sin\theta_{13}
\Big[\cos\de\sin(2\theta_{12})\sin(2\theta_{23}) -4
\cos^2\theta_{12}\sin\theta_{13}\cos^2\theta_{23}\Big] \times
\\ \non
&\times&\Big[|U_{13}^{\bf
k}|^2\, \sin^{2} \lf( \De_{13}^{\bf k} t \ri) + |V_{13}^{\bf
k}|^2\, \sin^{2}\lf(\Om_{13}^{\bf k} t\ri)\Big]
\\ \non
& +&\cos^{2}\theta_{13}\sin\theta_{13}
 \Big[\cos\de\sin(2\theta_{12})\sin(2\theta_{23}) +
  4 \sin^2\theta_{12}\sin\theta_{13}\cos^2\theta_{23}\Big]\times
\\ \non
&\times&\Big[|U_{23}^{\bf k}|^2 \,
\sin^{2} \lf(\De_{23}^{\bf k} t \ri) +
 |V_{23}^{\bf k}|^2\,\sin^{2} \lf(\Om_{23}^{\bf k}  t \ri)\Big]
\\ \non
&+& \cos^{2}\theta_{13} \sin(2\theta_{12}) \Big[ (\sin^2\theta_{23} -
\sin^2\theta_{13}\cos^2\theta_{23})\sin(2\theta_{12})
-\cos\de\cos(2\theta_{12})\sin\theta_{13}\sin(2\theta_{23})\Big] \times
\\
&\times & \Big[|U_{12}^{\bf k}|^2\, \sin^{2} \lf( \De_{12}^{\bf k}
t \ri) + |V_{12}^{\bf k}|^2\, \sin^{2} \lf(\Om_{12}^{\bf k} t
\ri)\Big]\,, \eea
\end{footnotesize}
where $ \De_{ij}^{\bf k}  \equiv (\om_{\G k,i}-\om_{\G k,j})/2$, $ \Om_{ij}^{\bf k}  \equiv (\om_{\G k,i}+\om_{\G k,j})/2$, and we introduced the Jarlskog invariant factor $J_{\CP}$ defined
as\cite{Jarlskog}
\bea J_{\CP} \equiv \Im \text{m} (U_{i\alpha}U_{j \beta}U^{*}_{i
\beta}U^{*}_{j\alpha}), \eea
where the $U_{i\al}$ are the elements of PMNS matrix $U$
and $i \neq j, \,\ \alpha \neq \beta$. $J_{\CP}$ vanishes when there is no $CP$--violation.

From the above oscillation formulas, the CP asymmetries can be computed, see Ref.\cite{PhysRevD.66.025033}.

\section{Perturbation theory of flavor oscillations} \label{perneu}
As discussed in Section~\ref{bfmixing}, the computation of the time-evolution operator amplitude cannot, in general, be used to derive the oscillation formula within the QFT framework. However, this statement is not entirely accurate. In what follows, we show that such a computation is indeed possible within a perturbative framework, where the non-perturbative flavor vacuum is not defined and, consequently, inequivalent representations do not play apparently a significant role \cite{Blasone:2023brf}.

Although this approach may appear disconnected from the flavor Fock space construction reviewed in the previous section, we will show that, within the limits of the employed approximations, it leads to the same oscillation formula Eq.~\eqref{oscfor}.

\subsection{Mixing as an interaction} \label{gencon}
In the interaction (Dirac) picture, it is fundamental to decompose the Lagrangian into a free and an interacting part  $\mathcal{L}=\mathcal{L}_0+\mathcal{L}_{int}$.

In Eq.~\eqref{Lagrangian}, we expressed the lepton sector of the (charged-current) weak-interaction Lagrangian as $\mathcal{L} = \mathcal{L}_0 + \mathcal{L}_{CC}$, where $\mathcal{L}_0$ denotes the Dirac Lagrangian for both charged leptons and neutrinos, including the non-diagonal mass matrix $M_\nu$ for the neutrinos. A natural choice is thus to identify the interaction term as $\mathcal{L}_{int} = \mathcal{L}_{CC}$.
Then, when working in the flavor basis, the fields should be expanded as described in the previous section. The flavor states introduced in Eq.~\eqref{bvflavstate} can then be used to compute weak-interaction amplitudes via perturbative expansion in the weak coupling $g$. However, as discussed in Section~\ref{lepnum}, such states cannot be interpreted as asymptotic $in$ or $out$ states. Therefore, a finite-time approach — based on the time-evolution operator rather than the conventional $S$-matrix formalism — should be adopted. The asymptotic limit may still be considered, but only with appropriate care~\cite{Lee:2017cqf}. Using these finite-time flavor states directly within an $S$-matrix framework may lead to incorrect conclusions~\cite{Giunti:2003dg,Li:2006qt}. 

An alternative approach is to work in the mass basis, where the effects of mixing are incorporated into the charged-current interaction term $\mathcal{L}_{CC}$ (see Eq.~\eqref{Lcc}), while the free part $\mathcal{L}_0$ reduces to the standard Dirac Lagrangian for both neutrinos and charged leptons. In this framework, flavor states are typically defined as linear combinations of mass eigenstates, like in the QM treatment originally proposed by Pontecorvo~\cite{PhysRevD.45.2414}. However, such flavor states are not eigenstates of the lepton flavor charges $Q_{\nu_\sigma}$, which can lead to violations of lepton number even on time scales where oscillation effects are negligible~\cite{Blasone:2019rxl}.

In Ref.~\cite{Blasone:2023brf}, a different approach has been proposed: to include the mixing term, originating from the off-diagonal elements of the neutrino mass matrix $M_\nu$, in the interaction Lagrangian $\mathcal{L}_{int}$. In other words, in this framework, mixing is interpreted as an interaction between different flavor fields. Consequently, $\mathcal{L}_0$ takes the form of a standard Dirac Lagrangian, where the masses of the perturbative neutrino fields correspond to the diagonal elements of $M_\nu$.

Because our aim is to compute the oscillations formula, we will only focus on the neutrino Lagrangian, neglecting $\mathcal{L}_{CC}$. Then, we decompose \eqref{Lnu} as
\be \label{pneutlag}
{\cal L}_{\nu}  \ = \  \mathcal{L}_{\nu,0}+ \mathcal{L}_{int}
\ee
where
\bea
{\cal L}_{\nu,0} & =  & \sum_\si \, \overline{\nu}_\si \lf( i \ga_\mu \pa^\mu - m_\si \ri)\nu_\si  \, \\[2mm]
 \mathcal{L}_{int} & = & -m_{e \mu} \lf(\overline{\nu}_e \nu_\mu+\overline{\nu}_\mu \nu_e\ri) \, .
\eea

Then, we can compute the transition amplitudes among different flavors by means of the usual Dyson formula for the time evolution operator
\be \label{dyfor}
U(t_i,t_f) \ = \ \mathcal{T} \exp \lf[i \int^{t_f}_{t_i} \!\! \dr^4 x \, :\mathcal{L}_{int}(x): \ri] \ = \ \mathcal{T}\exp  \lf[-i \int^{t_f}_{t_i} \!\! \dr^4 x \, :\mathcal{H}_{int}(x): \ri] \, ,
\ee
where $\mathcal{H}_{int}(x)=-\mathcal{L}_{int}(x)$ is the interaction Hamiltonian density and $\mathcal{T}$ is the chronological product. In the following we will only need the expression of the operator up to the second order
\be \label{dyfor2}
U(t_i,t_f) 
\ = \ 1-i\int_{t_{i}}^{t_{f}}\!\!\dr t_{1} \,H_{int}(t_{1})+(-i)^{2}%
\int_{t_{i}}^{t_{f}}\!\!\dr t_{1} \,H_{int}(t_{1})%
\int_{t_{i}}^{t_{1}}\dr t_{2} \, H_{int}(t_{2})+...
\ee
where $H_{int}=\intx \,\mathcal{H}_{int}(x)$ is the interaction Hamiltonian in interaction picture.

We emphasize that, even in the present case, our analysis focuses on the time evolution operator rather than the $S$-matrix. This is because, as already stressed, the phenomenon of flavor oscillations can only be properly described at finite time. In other words, flavor neutrino states do not exist as asymptotically stable states. As will become evident from the examples discussed below, taking the limits \( t_i \to -\infty \) and \( t_f \to +\infty \) effectively suppresses the flavor-changing processes under consideration. At the same time, this asymptotic limit ensures exact energy conservation.
This feature is intimately related to the flavor-energy uncertainty relation discussed in Section~\ref{qftfeur}. Moreover, it is analogous to the case of unstable particles~\cite{Bernardini:1993,FacchiPascazio1999,Giacosa:2010br,Giacosa:2011xa,Giacosa:2018dzm,Giacosa:2021hgl} (see also~\cite{Anselmi:2023wjx,Anselmi:2023phm}, where the importance of finite-time quantum field theory in the study of decays has been emphasized). Indeed, both the decay of unstable particles~\cite{Bhattacharyya_1983} and neutrino oscillations~\cite{Bilenky:2009zz} can be interpreted within the framework of TEUR.

In the following we will first study the cases of 0+1D QFT (that is, QM), and a 3+1D scalar model. This preliminary analysis permits to grasp the main features of the problem, without the complication of dealing with spinors, which we consider in Section \ref{neutsec}.
\subsection{A quantum mechanics toy model of flavor mixing} \label{childsec}

Let us consider the QM problem of two coupled harmonic
oscillators $A$ and $B$. 
We can see this problem as a $0+1D$ field theory described by the Lagrangian 
\begin{equation}
L=\frac{1}{2}\left( \frac{dx_{A}}{dt}\right) ^{2}-\frac{\om_{A}^{2}}{%
2}x_{A}^{2}+\frac{1}{2}\left( \frac{dx_{B}}{dt}\right) ^{2}-\frac{\om_{B}^{2}}{%
2}y^{2}-\om_{AB}^{2}x_{A}x_{B}  \, . \label{toylagrangian}
\end{equation}%
According with the previous discussion, we regard the mixing term $L%
_{int}=\om_{AB}^{2}x_{A}x_{B}$ as an interaction, where $\om_{AB}^{2}$ plays the role of the coupling constant. 

In the interaction picture,  the
fields have the form
\begin{eqnarray}
x_{A}(t) &=&\frac{1}{\sqrt{2\om_{A}}}\left( a_{A}e^{-i\om_{A}t}+a_{A}^{\dagger
}e^{i\om_{A}t}\right) \text{ ,}  \label{xex} \\[2mm]
x_{B}(t) &=&\frac{1}{\sqrt{2\om_{B}}}\left( a_{B}e^{-i\om_{B}t}+a_{B}^{\dagger
}e^{i\om_{B}t}\right) \text{ ,}  \label{yex}
\end{eqnarray}%
where the creation and annihilation operators follow the usual commutation
relations 
\be
[a_{A},a_{A}^{\dagger }]=[a_{B},a_{B}^{\dagger }] \ = \ 1
\ee
and zero otherwise.

To proceed we employ the formula \eqref{dyfor2} with
\be
H_{int}(t)=\om_{AB}^{2}x_{A}(t)x_{B}(t) \, .
\ee
As initial state at $t=t_{i}$, we take
\be
\vert A\rangle =a_{A}^{\dagger }\vert 0\rangle
\ee
We then evaluate the probability that the state has changed at time $t_{f} > t_{i}$, a situation that, roughly speaking, corresponds to a decay of the initial state. The first possible transition is the mixing $\vert A\rangle = a_{A}^{\dagger} \vert 0\rangle \rightarrow a_{B}^{\dagger} \vert 0\rangle = \vert B\rangle$, driven by the interaction term. The corresponding amplitude reads:
\begin{eqnarray}
\langle B\vert U(t_{f},t_{i})\vert A\rangle 
&=&\langle 0\vert a_{B}U(t_{f},t_{i})a_{A}^{\dagger }\vert
0\rangle =-i\frac{\om_{AB}^{2}}{\sqrt{2\om_{A}}\sqrt{2\om_{B}}}%
\int_{t_{i}}^{t_{f}}dt_{1}e^{-i(\om_{A}-\om_{B})t_{1}}  \notag \\
&=&\frac{\om_{AB}^{2}}{\sqrt{2\om_{A}}\sqrt{2\om_{B}}}\frac{%
e^{-i(\om_{A}-\om_{B})t_{f}}-e^{-i(\om_{A}-\om_{B})t_{i}}}{(\om_{A}-\om_{B})}\text{ .}
\end{eqnarray}%
Therefore, the probability of this process is
\begin{equation}
\mathcal{P}_{A\rightarrow B}(\De t)=\frac{\om_{AB}^{4}}{\om_{A}\om_{B}}\frac{\sin^{2}\left[ (\om_{A}-\om_{B})\Delta t/2\right] }{(\om_{A}-\om_{B})^{2}}\text{ } \, , \qquad
\Delta t=t_{f}-t_{i} \, .
\end{equation}%
This formula describes an oscillation between $A$ and $B$ states, whose frequency is proportional to the
frequencies difference. We will refer to the above expression as the ``low frequency'' term. Note, for
short times $\mathcal{P}_{A\rightarrow B}(\De t)\simeq \frac{\om_{AB}^{4}\Delta t^{2}}{%
4\om_{A}\om_{B}}$. This is the same behavior encountered in the study of particles decay and which is responsible for the \emph{quantum Zeno effect} \cite{Bernardini:1993,FacchiPascazio1999}: if one performs measurements at very short time there is no decay/oscillation (the first non-trivial term is quadratic).

At first-order there is another possible transition: 
\be
a_{A}^{\dagger }\vert 0\rangle \rightarrow \frac{\left(
a_{A}^{\dagger }\right) ^{2}}{\sqrt{2}}a_{B}^{\dagger }\left\vert
0\right\rangle
\ee
that is a single excitation along $A$ converts into $AAB.$
The amplitude of such process reads:%
\begin{eqnarray}
\frac{1}{\sqrt{2}}\langle 0\vert
a_{B}a_{A}^{2} \, U(t_{f},t_{i})\, a_{A}^{\dagger }\vert 0\rangle  &=&-i%
\frac{\sqrt{2}\om_{AB}^{2}}{\sqrt{2\om_{A}}\sqrt{2\om_{B}}}%
\int_{t_{i}}^{t_{f}}dt_{1}e^{-i(\om_{A}+\om_{B})t}  \notag \\[2mm]
&=&\frac{\sqrt{2}\om_{AB}^{2}}{\sqrt{2\om_{1}}\sqrt{2\om_{2}}}\frac{%
e^{-i(\om_{A}+\om_{B})t_{f}}-e^{-i(\om_{A}+\om_{B})t_{i}}}{(\om_{A}+\om_{B})}\text{ ,}
\end{eqnarray}%
hence%
\begin{equation}
\mathcal{P}_{A\rightarrow AAB}(\De t)=\frac{2\om_{AB}^{4}}{\om_{A}\om_{B}}\frac{\sin ^{2}\left[ 
 (\om_{A}+\om_{B})\Delta t/2 \right] }{(\om_{A}+\om_{B})^{2}}\text{ ,}
\end{equation}%
which depends on the sum of the frequencies and is named the `high
frequency' term. For short times, $P^{A\rightarrow AAB}(\De t)\simeq \frac{%
\om_{AB}^{4}t^{2}}{2\om_{A}\om_{B}}$. This confirms the above analogy with unstable particles.

Therefore, the total transition probability is the sum of the above terms
\bea \non
\mathcal{P}_{D}^{A}(\De t) & =& \mathcal{P}_{A\rightarrow B}(\De t)+\mathcal{P}_{A\rightarrow AAB}(\De t)
\\ [2mm]
& =&\frac{\om_{AB}^{4}}{%
\om_{A}\om_{B}}\left[ \frac{\sin ^{2}\left[ (\om_{A}-\om_{B})\Delta t/2 
\right] }{(\om_{A}-\om_{B})^{2}}+2\frac{\sin ^{2}\left[  
(\om_{A}+\om_{B})\Delta t/2 \right] }{(\om_{A}+\om_{B})^{2}}\right] \, .
\eea
For short times, $\mathcal{P}_{D}^{A}(\De t)\simeq \frac{3\om_{AB}^{4}\Delta
t^{2}}{4\om_{A}\om_{B}}.$

In the same way one can easily calculate the survival
probability. The corresponding amplitude is given by 
\begin{equation}
\langle A\vert U(t_{f},t_{i})\vert A\rangle
=\langle 0\vert a_{A}U(t_{f},t_{i})a_{A}^{\dagger }\vert
0\rangle \, .
\end{equation}%
Up to the second order we obtain

\begin{equation}
\langle A\vert U(t_{f},t_{i})\vert A\rangle =1-i \, %
\mathcal{T}\langle 0\vert
a_{A}\int_{t_{i}}^{t_{f}}dt_{1}H_{int}(t_{1})%
\int_{t_{i}}^{t_{1}}dt_{2}H_{int}(t_{2})a_{A}^{\dagger }\vert
0\rangle \, .
\end{equation}%
The survival probability reads
\begin{align}\non
\mathcal{P}_{A \rightarrow A}(\De t)& =\left\vert 1-\frac{\om_{AB}^{4}}{4\om_{A}\om_{B}}\left[ 2\frac{t}{%
i(\om_{A}+\om_{B})}-2\frac{e^{-i(\om_{A}+\om_{B})\Delta t}-1}{(\om_{A}+\om_{B})^{2}}\ri.\ri.
\\ \non
&\qquad \quad\lf.\lf.+
\frac{t}{-i(\om_{A}-\om_{B})}-\frac{e^{i(\om_{A}-\om_{B})\Delta t}-1}{%
(\om_{A}-\om_{B})^{2}}\right] \right\vert ^{2} 
\\
& =1-2R+... \ ,
\end{align}%
where 
\begin{eqnarray}
R&=&\frac{\om_{AB}^{4}}{2\om_{A}\om_{B}}\left( \frac{\sin ^{2}\left[  
(\om_{A}-\om_{B})\Delta t/2\right] }{(\om_{A}-\om_{B})^{2}}+2\frac{\sin ^{2}\left[ 
(\om_{A}+\om_{B})\Delta t/2\right] }{(\om_{A}+\om_{B})^{2}}\right) \, .
\end{eqnarray}%
Then
\begin{equation}
\mathcal{P}_{S}^{A}(\De t) = \mathcal{P}_{A \rightarrow A}(\De t) \ = \ 1-\frac{\om_{AB}^{4}}{\om_{A}\om_{B}}\left( \frac{\sin ^{2}\left[ 
(\om_{A}-\om_{B})\Delta t/2\right] }{(\om_{A}-\om_{B})^{2}}+2\frac{\sin ^{2}%
\left[(\om_{A}+\om_{B})\Delta t/2\right] }{(\om_{A}+\om_{B})^{2}}\right) \, .  
\end{equation}%
The unitarity is verified at order $g^{2}$:
\begin{equation}
\mathcal{P}_{S}^{A}(\De t)+\mathcal{P}_{D}^{A}(\De t)\ =\ 1 \, ,
\end{equation}%
for each $t$. For small $t,$ $p_{S}^{A}(t)\simeq 1-\frac{%
\om_{AB}^{4}}{\om_{A}\om_{B}}\left( 2\frac{\Delta t^{2}}{4}+\frac{\Delta t^{2}}{4}%
\right) =1-\frac{3\om_{AB}^{4}\Delta t^{2}}{4\om_{A}\om_{B}}.$

\subsection{Scalar field mixing in the interaction picture} \label{bosesec}

Let us now move to the more complex example of two real scalar fields $\phi_{A}=\phi_{A}(t,\mathbf{x})$ and $\phi_{B}=\phi_{B}(t,%
\mathbf{x})$. 
Our Lagrangian density is
\begin{equation}
\mathcal{L}=\frac{1}{2}\left( \partial _{\alpha }\phi_{A}\right) ^{2}-\frac{%
m_{A}^{2}}{2}\phi_{A}^{2}+\frac{1}{2}\left( \partial _{\alpha }\phi_{B}\right)
^{2}-\frac{m_{B}^{2}}{2}\phi_{B}^{2}-m_{AB}^{2}\phi_{A}\phi_{B}\text{.}
\end{equation}%
The Hamiltonian density is
\begin{equation}
\mathcal{H}=\frac{\pi _{A}}{2}+\frac{\pi _{B}}{2}+\frac{1}{2}\left(
\nabla \phi_{A}\right) ^{2}+\frac{1}{2}\left( \nabla
\phi_{B}\right) ^{2}+\frac{m_{A}^{2}}{2}\phi_{A}^{2}+\frac{m_{A}^{2}}{2}%
\phi_{B}^{2}+m_{AB}^{2}\phi_{A}\phi_{B} \, ,
\end{equation}%
with $\pi _{A}=\partial _{t}\phi_{A}$ and $\pi _{B}=\partial _{t}\phi_{B}$. As in the previous example, we treat 
the mixing term as a perturbation:

\bea
\mathcal{H}_{0} & = & \frac{\pi _{A}}{2}+\frac{\pi _{B}}{2}+\frac{1}{2}\left(
\nabla \phi_{A}\right) ^{2}+\frac{1}{2}\left( \nabla
\phi_{B}\right) ^{2}+\frac{m_{A}^{2}}{2}\phi_{A}^{2}+\frac{m_{A}^{2}}{2}\phi_{B}^{2} \, 
 , \\[2mm]
 \mathcal{H}_{int} & = & m_{AB}^{2}\phi_{A}\phi_{B} \, .
\eea
In the interaction picture we can expand the field $\phi_A$ and its conjugate momentum as
\begin{align}
\phi_{A}(x)& =\phi_{A}(t,\mathbf{x})=\frac{1}{\sqrt{V}}\sum_{\textbf{k}=2\pi 
\mathbf{n}/L}\frac{1}{\sqrt{2\omega _{\G k,A}}}\left( a_{\G k,A%
}e^{-ikx}+a_{\G k,A}^{\dagger }e^{ikx}\right)  \, ,\\
\pi _{A}(x)& =\pi _{A}(t,\mathbf{x})=\frac{-i}{\sqrt{V}}\sum_{\textbf{k}%
=2\pi \mathbf{n}/L}\sqrt{\frac{\omega_{\G k,A}}{2}}\left( a_{\G k,%
A}e^{-ikx}-a_{\G k,A}^{\dagger }e^{ikx}\right) \, ,
\end{align}%
with $k^{0}=\omega_{\G k,A}=\sqrt{\textbf{k}^{2}+m_{A}^{2}}$. 

The
commutation relation is the usual
\be
\left[ \phi_{A}(t,\mathbf{x}),\pi _{A}(t,\mathbf{y})%
\right] =\frac{1}{V}\sum_{\textbf{k}}e^{i\mathbf{k\cdot (x-y)}}=i\delta_V(%
\mathbf{x-y}) \, , 
\ee
which implies that $\left[ a_{\G k,A},a_{\G p,A}^{\dagger }\right] =\delta_{\G k,\G p}$, zero
otherwise. Analogous expressions hold for $\phi_{B}(x)$ and $\pi _{B}(x)$. 

We can thus expand the interacting Hamiltonian as

\begin{align} \non 
&H_{int}(t) =\intx_{1}\mathcal{H}_{int}(x)=\sum_{\textbf{q}}\frac{%
m_{AB}^{2}}{\sqrt{2\omega_{\G q,A}}\sqrt{2\omega_{\G q,B}}}\left( a_{\G q,A} a_{\G q,B}^{\dagger }e^{-i\left( \omega_{\G q,A} 
 -\omega_{\G q,B}\ri)t} \right.
 \\
& \left. + a_{\G q,A}^{\dagger
}a_{\G q,B} e^{i\left( \omega_{\G q,A} -\omega_{\G q,B}%
\ri)t}  + a_{\G q,A}a_{-\G q,B}e^{-i\left( \omega_{\G q,A}%
 +\omega_{\G q,B}\ri)t} + a_{\G q,A}^{\dagger}a_{-\G q,B}%
^{\dagger }e^{i\left( \omega_{\G q,A} +\omega_{\G q,B}%
\ri)t}\right) \, .
\end{align}

The last ingredient is the ``flavor state'' $A$
\begin{equation}
\left\vert A,\textbf{p}\right\rangle =a_{\G p,A}^{\dagger }\left\vert
0\right\rangle \, .
\end{equation}

As in the previous example we want to compute the probability that such a state, created at $t=0$, is later transformed into a different state at the time $t>0$ or,
conversely, that it has not changed.

Firstly we
compute the probability amplitude for the transition 
\be
\left\vert A,\mathbf{%
p}\right\rangle \rightarrow \left\vert B,\textbf{k}\right\rangle \, .
\ee
which reads
\begin{align}\non
\mathcal{A}_{A\rightarrow B}\left( \textbf{p},\G{k}; t_i,t_f\right) &
=\langle B,\textbf{k}\vert U(t_{f},t_{i})\vert A,\textbf{p}%
\rangle =-i\int_{t_{i}}^{t_{f}}dt_{1}\langle 0\vert a_{\G k,B}%
\,H_{int}(t_{1})\,a_{\G p,A}^{\dagger }\vert 0\rangle
+... \\
& =\frac{m_{AB}^{2}}{\sqrt{2\omega_{\G p,A}}\sqrt{2\omega_{\G k,B}%
}}\delta _{\textbf{k},\textbf{p}}\frac{e^{-i\left( \omega_{\G p,A}%
 -\omega_{\G k,B}\ri)t_{f}}-e^{-i\left( \omega_{\G p,A}%
 -\omega_{\G k,B}\ri)t_{i}}}{ \om_{\G p,A}%
-\omega_{\G k,B}}\, .
\end{align}%
The corresponding probability is obtained upon taking the square and summing over the density of final states $%
\sum_{\textbf{k}}$ :%
\begin{equation} \label{probsum}
\mathcal{P}_{A\rightarrow B}(\G{p}; \De t)=\sum_{\textbf{k}}\vert
\mathcal{A}_{A\rightarrow B}\left( \textbf{p},\G{k};t_i,t_f\right) \vert ^{2}=%
\frac{m_{AB}^{4}}{\omega_{\G p,A}\omega_{\G p,B}}\frac{\sin
^{2}\left[ \left( \omega_{\G p,A} -\omega_{\G p,B} 
\ri)\Delta t /2 \right] }{\left( \omega_{\G p,A} -\omega_{\G p,B}%
\ri)^{2}}\, .
\end{equation}%
The result is finite and well behaved. Even in this case we have a behavior which resembles the one of unstable particles for short times, i.e. $\mathcal{P}_{A\rightarrow
B}(\G{p}; \De t)\simeq \frac{m_{AB}^{4}}{4\omega_{\G p,A}%
\omega_{\G p,B}}t^{2}$. 

Another possible process is
\begin{equation}
\left\vert A,\textbf{p}\right\rangle \rightarrow \left\vert A,\textbf{k}%
_{1}\right\rangle \left\vert A,\textbf{k}_{2}\right\rangle \left\vert B,%
\textbf{k}_{3}\right\rangle =a_{A,\textbf{k}_{1}}^{\dagger }a_{A,\textbf{k}%
_{2}}^{\dag }a_{B,\textbf{k}_{3}}^{\dag }\left\vert 0\right\rangle \text{ ,}
\end{equation}%
where the two emitted $A$ particles have different momentum, $\textbf{k}%
_{1}\neq \textbf{k}_{2}$. The corresponding amplitude of this process reads
\begin{equation}
\mathcal{A}_{A\rightarrow AAB}^{\textbf{k}_{1}\neq \textbf{k}_{2}}(\textbf{p}%
,\textbf{k}_{1},\textbf{k}_{2},\textbf{k}_{3};t_i,t_f)=-i\int_{t_{i}}^{t_{f}}dt_{1}\langle 0\vert a_{A,\textbf{k}%
_{1}}a_{A,\textbf{k}_{2}}a_{B,\textbf{k}_{3}}H_{int}(t_{1})a_{A,\textbf{p}%
}^{\dagger }\vert 0\rangle \text{ .}
\end{equation}%
After an explicit calculation up to first order, its squared modulus turns
out to be: 
\bea
&&{}\hspace{-4mm}\vert \mathcal{A}_{A\rightarrow AAB}^{\textbf{k}_{1}\neq \textbf{k}%
_{2}}(\textbf{p},\textbf{k}_{1},\textbf{k}_{2},\textbf{k}%
_{3},t_i,t_f)\vert ^{2} =
\\ \non &&
\qquad
=\frac{m_{AB}^{4}}{\omega_{\G k_3,A}%
\omega _{\G k_3,B}}\frac{\sin^{2}\lf[\left( \omega
_{\textbf{k}_{3},A} +\omega _{\G k_3,B}\ri)\Delta t/2 \ri] }{%
\left( \omega_{\textbf{k}_{3},A} +\omega _{\G k_3,B}\ri)^{2}}
\left( \delta _{\textbf{k}_{1},\textbf{p}}\delta _{\textbf{k}_{2},-\textbf{k}%
_{3}}+\delta _{\textbf{k}_{1},-\textbf{k}_{3}}\delta _{\textbf{k}_{2},%
\textbf{p}}\right) \text{ .}
\eea
In order to compute the probability one needs to square and sum over final states $\textbf{k}_{1},\textbf{k}_{2},%
\textbf{k}_{3}.$, getting:
\begin{eqnarray}
&&\mathcal{P}^{\G k_1 \neq \G k_2}_{A\rightarrow AAB}(\G{p};\De t) =\text{ }\frac{1}{2}\sum_{%
\textbf{k}_{1},\textbf{k}_{2},\textbf{k}_{3}}\left\vert 
\mathcal{A}_{A\rightarrow AAB}^{\textbf{k}_{1}\neq \textbf{k}_{2}}(\textbf{p}%
,\textbf{k}_{1},\textbf{k}_{2},\textbf{k}_{3},t_i,t_f)\right\vert ^{2} \\ \non
&&=\sum_{\textbf{k}_{3}}\frac{m_{AB}^{4}}{\omega _{\G k_3,A}\omega
_{\G k_3,B}}\frac{\sin^{2}\lf[  \left( \omega _{\G k_3,A}%
 +\omega _{\G k_3,B}\ri)\Delta t/2\ri] }{\left( \omega
_{\textbf{k}_{3},A} +\omega _{\G k_3,B}\ri)^{2}}-\frac{%
m_{AB}^{4}}{\omega _{\G p,A}\omega _{\G p,B}}\frac{\sin ^{2}%
\left[ \left( \omega _{\G p,A} +\omega _{\G p,B}\ri)%
\Delta t /2 \right] }{\left( \omega _{\G p,A} +\omega _{\G p,B}%
\ri)^{2}} \, ,
\end{eqnarray}%
where the factor $1/2$ takes into account that the two $A
$ in the final state are identical bosons. 

The sum term diverges so that a cutoff must be introduced to
keep the intermediate results finite. At the end of the calculation it will be removed. 
What is the meaning of such divergent term? In order to understand it, we consider the expression for large volumes
\bea \non
\mathcal{P}^{\G k_1 \neq \G k_2}_{A\rightarrow AAB}(\G p;\Delta t) &=&V\int \!\! \frac{\dr^{3}\G k_{3}}{(2\pi
)^{3}} \, \lf[\frac{m_{AB}^{4}}{\omega _{\G k_3,A}\omega
_{\G k_3,B}}\frac{\sin^{2}\lf[  \left( \omega _{\G k_3,A}%
 +\omega _{\G k_3,B}\ri)\Delta t/2 \ri] }{\left( \omega
_{\textbf{k}_{3},A} +\omega _{\G k_3,B}\ri)^{2}}
\ri.
\\ &&
\lf. 
-\frac{%
m_{AB}^{4}}{\omega _{\G p,A}\omega _{\G p,B}}\frac{\sin ^{2}%
\left[ \left( \omega_{\G p,A} +\omega _{\G p,B}\ri)%
\Delta t /2 \right] }{\left( \omega _{\G p,A} +\omega _{\G p,B}%
\ri)^{2}}\ri] \, , 
\label{AABdiff}
\eea
where, as remarked, a cutoff is implicit in the integral over $|\G k_{3}|$. The divergent term is
proportional to $V$ and is a typical vacuum term that needs to be subtracted.

The last possible transition is obtained in the case in which $\textbf{k}_{2}= 
\textbf{k}_{1}$:
\begin{equation}
\vert A,\textbf{p}\rangle \rightarrow \frac{1}{\sqrt{2}}a_{\G k_1,%
A}^{\dagger }a_{\textbf{k}_{1},A}^{\dag }a_{\textbf{k}_{3},B%
}^{\dag }\vert 0\rangle \text{ .}
\end{equation}%
The first-order amplitude is 
\begin{equation}
\mathcal{A}_{A\rightarrow AAB}^{\textbf{k}_{1}=\textbf{k}_{2}}(\textbf{p},%
\textbf{k}_{1},\textbf{k}_{3};t_i,t_f)=-\frac{i}{\sqrt{2}}%
\int_{t_{i}}^{t_{f}}dt_{1}\langle 0\vert a_{\textbf{k}_{1},A}a_{%
\textbf{k}_{1},A}a_{\textbf{k}_{3},B}H_{int}(t_{1})a_{\textbf{p},A}^{\dagger
}\vert 0\rangle \, . 
\end{equation}%
Taking the square and summing over the final momenta $\textbf{k}_{1},\textbf{k}_{3}$ we get
\bea\non
\mathcal{P}^{\G k_1 = \G k_2}_{A\rightarrow AAB}(\G{p},\De t) 
&=& \sum_{\G k_{1},\G k_{3}}\left| \mathcal{A}_{A\rightarrow AAB}^{\G k_{1}=
\G k_{2}}(\textbf{p},\G k_{1},\G k_{3};t_i,t_f)\right| ^{2}
\\
&=&2\frac{m_{AB}^{4}}{\omega _{\G p,A}\omega _{\G p,B}}\frac{\sin ^{2}%
\left[ \left( \omega _{\G p,A} +\omega _{\G p,B}\ri)%
\Delta t/2\right] }{\left( \omega _{\G p,A} +\omega _{\G p,B}%
\ri)^{2}} \, .
\eea
Note, the factor $2$ is the same appearing in the QM toy model. 

The total probability is thus
\begin{eqnarray}\non
\mathcal{P}_{D}^{A}(\G{p}; \De t) &=&\mathcal{P}_{A\rightarrow B}(\G{p}; \De t) +\mathcal{P}^{\G k_1 \neq \G k_2}_{A\rightarrow AAB}(\G{p}; \De t) +\mathcal{P}^{\G k_1 = \G k_2}_{A\rightarrow AAB}(\G{p}; \De t)  
\\ [2mm] \non
&=&\frac{%
m_{AB}^{4}}{\omega _{\G p,A}\omega _{\G p,B}}\frac{\sin ^{2}
\left[ \left( \omega _{\G p,A} -\omega _{\G p,B}\ri)%
\Delta t/2\right] }{\left( \omega _{\G p,A} -\omega _{\G p,B}%
\ri)^{2}}+\frac{%
m_{AB}^{4}}{\omega _{\G p,A}\omega _{\G p,B}}\frac{\sin ^{2}
\left[ \left( \omega _{\G p,A} +\omega _{\G p,B}\ri)%
\Delta t/2\right] }{\left( \omega _{\G p,A} +\omega _{\G p,B}%
\ri)^{2}}\\
&&+V\int \!\! \frac{\dr^{3}\G k_{3}}{(2\pi )^{3}}\frac{m_{AB}^{4}}{\omega _{\G k_3,A}%
\omega _{\G k_3,B}}\frac{\sin ^{2}\lf[ \left( \omega
_{\G k_3,A}+\omega _{\G k_3,B}\ri)\Delta t/2\ri] }{%
\left( \omega _{\G k_3, A}+\omega _{\G k_3,B}\ri)^{2}}%
\text{ .}
\end{eqnarray}%
The factor $2$ of $\mathcal{P}^{\G k_1 = \G k_2}_{A\rightarrow AAB}(\G{p}; \De t)$ combines
with the factor $-1$ in $\mathcal{P}^{\G k_1 \neq \G k_2}_{A\rightarrow AAB}(\G{p}; \De t)$ and give a positive definite result. Then, as anticipated, the vacuum term must be subtracted, leading to the final result
\begin{equation}
\mathcal{P}_{D}^{A}(\G{p};\Delta t) =\frac{m_{AB}^{4}}{\omega _{\G p,A}%
\omega _{\G p,B}}\left( \frac{\sin ^{2}%
\left[  \left( \omega _{\G p,A} -\omega _{\G p,B}\ri)%
\Delta t/2 \right] }{\left( \omega _{\G p,A} -\omega _{\G p,B}%
\ri)^{2}}+\frac{\sin ^{2}%
\left[  \left( \omega _{\G p,A} +\omega _{\G p,B}\ri)%
\Delta t/2 \right] }{\left( \omega _{\G p,A} +\omega _{\G p,B}%
\ri)^{2}}\right) \text{ .}
\end{equation}

In order to verify the correctness of the previous result one can calculate the survival probability, i.e. the probability of the process
\begin{equation}
\left\vert A,\textbf{p}\right\rangle \rightarrow \left\vert A,\textbf{p}%
\right\rangle \, .
\end{equation}%
The calculation proceeds as above and the details can be found in Ref. \cite{Blasone:2023brf}. The result
 (up to second order) reads%
\begin{equation}
\mathcal{P}^A_{S}(\textbf{p};\Delta t)  \ = \ 1- \frac{m_{AB}^{4}}{\omega _{\G p,A}%
\omega _{\G p,B}}\left( \frac{\sin ^{2}%
\left[  \left( \omega _{\G p,A} -\omega _{\G p,B}\ri)%
\Delta t/2\right] }{\left( \omega _{\G p,A} -\omega _{\G p,B}%
\ri)^{2}}+\frac{\sin ^{2}%
\left[  \left( \omega _{\G p,A} +\omega _{\G p,B}\ri) 
\Delta t/2\right] }{\left( \omega _{\G p,A} +\omega _{\G p,B}%
\ri)^{2}}\right) \, . 
\label{pSAnorm}
\end{equation}%
Therefore, the unitarity is still verified 
\begin{equation}
\mathcal{P}^A_{D}(\textbf{p};\Delta t) +\mathcal{P}^A_{S}(\textbf{p};\Delta t) \ =\ 1 \, . 
\end{equation}%

The structure of the above oscillation/survival formulas closely reminds the structure of Eq.\eqref{oscfor}, involving both a low-frequency and
the high-frequency term. Moving to the case of interest of fermion flavor oscillations, we will see that this is not just a trivial observation.

\subsection{Neutrino oscillations in the interaction picture} \label{neutsec}
Let us now move to the neutrino case, described by the Lagrangian \eqref{pneutlag}.
 
In the interaction picture neutrino fields can be expanded as free fields, evolving under the action of $\mathcal{L}_{\nu,0}$:
\begin{eqnarray}
\nu_{\si}(x) = \frac{1}{\sqrt{V}} \sum_{\G k,r}\,  \left[ u_{{\bf k},\si}^{r}(t) \, \alpha_{{\bf k},\si}^{r} + v_{-{\bf k},\si}^{r}(t) \, \bt_{-{\bf k},\si}^{r\dag}   \right]  e^{i{\bf k}\cdot {\bf x}}  \, ,
\label{fieldexint}
\end{eqnarray}
with $u^r_{{\bf k},\si}(t) \,= \, e^{- i \om_{\G k,\si} t}\, u^r_{{\bf k},\si}\;$,
$\;v^r_{{\bf k},\si}(t) \,= \, e^{ i \om_{\G k,\si} t}\, v^r_{{\bf k},\si}$,
 $\om_{\G k,\si}=\sqrt{|\G k|^2 + m_\si^2}$. 
 
 The perturbative vacuum is defined by
\be \label{vacint}
\al^r_{\G k, \si}|0 \rangle = 0 = \beta _{{\bf k},\si}^{r} |0 \rangle \  ,
\ee
while creation and annihilation operators satisfy the usual anticommutation relations
\be  \{\al ^r_{{\bf k},\rho}, \al ^{s\dag }_{{\bf q},\si}\} = \de_{\G k \G q}\de _{rs}\de _{\rho \si}  \quad \, , \quad \{\bt^r_{{\bf k},\rho},
\bt^{s\dag }_{{\bf q},\si}\} =
\de_{\G k \G q} \de _{rs}\de _{\rho \si} \, , 
\ee
and the spinors are normalized so that
\bea 
u^{r\dag}_{{\bf k},\rho} u^{s}_{{\bf k},\rho} =
v^{r\dag}_{{\bf k},\rho} v^{s}_{{\bf k},\rho} \ =  \ \de_{rs}
\quad, \quad u^{r\dag}_{{\bf k},\rho} v^{s}_{-{\bf k},\rho} = 0 \;.
\eea
Then, a one-particle state is defined as
\be
|\nu^r_{\G p,\si}\ran \equiv \al^{r\dag}_{\G p,\si}|0\ran \, , \qquad |\overline{\nu}^r_{\G p,\si}\ran \equiv \bt^{r\dag}_{\G p,\si}|0\ran \, .
\ee

The last ingredient is the interaction Hamiltonian: 

\bea
H_{int}(t)& = &  m_{e \mu}  \sum_{s,s'=1,2}\sum_{\G p}
 \Big[\bt^s_{\G p,\mu}\bt^{s\dag}_{\G p,e} \de_{s s'} W^*_\G p(t)+\al^{r\dag}_{\G p,\mu} \al^r_{\G p,e} \de_{s s'} W_\G p(t) \non \\[2mm]
& + & \bt^s_{-\G p,\mu}\al^{s'}_{e,\G p} \lf(Y^{s s'}_\G p(t)\ri)^*+\al^{s\dag}_{\G p,\mu}\bt^{s'\dag}_{-\G p,e} Y^{s s'}_\G p(t)\, + \, e \leftrightarrow \mu \Big] \, ,
\eea
where we defined the coefficients
\bea
W_\G p(t) & = & \overline{u}^s_{\G p,\mu} u^s_{\G p,e} e^{i\left( \omega_{\G k,\mu}-\omega_{\G k,e}\ri)t} \ = \ W_\G p \,  e^{i\left( \omega_{\G p,\mu}-\omega_{\G p,e}\ri)t}  \\[2mm]
Y^{s s'}_\G p(t) & = &  \, \overline{u}^{s}_{\G p,\mu} v^{s'}_{-\G p,e} e^{i\left( \omega_{\G k,\mu}+\omega_{\G k,e}\ri)t} \ = \ Y^{s s'}_\G p e^{i\left( \omega_{\G p,\mu}+\omega_{\G p,e}\ri)t}
\eea
It is useful to see the explicit expression of the time-independent parts:
\bea
W_\G p & = & \sqrt{\frac{\lf(\omega_{\G p,e}+m_{e}\ri)\lf(\omega_{\G p,\mu}+m_{\mu}\ri)}{4\omega_{\G p,e} \omega_{\G p,\mu}}}
\left(1-\frac{|\G p|^{2}}{(\omega_{\G p,e}+m_{e})(\omega_{\G p,\mu}+m_{\mu})}\right)  \, , \\[2mm]
Y^{22}_{\G p} & = & -Y^{11}_{\G p} \ = \ \frac{p_3}{\sqrt{4 \om_{\G p,e}\om_{\G p,\mu}}}
\lf(\sqrt{\frac{\om_{\G p,\mu}+m_\mu}{\om_{\G p,e}+m_e}}+\sqrt{\frac{\om_{\G p,e}+m_e}{\om_{\G p,\mu}+m_\mu}}\ri) \, , \\[2mm]
Y^{12}_{\G p} & = & \lf(Y^{21}_{\G p} \ri)^* \ = \ -\frac{p_1-i p_2}{\sqrt{4 \om_{\G p,e}\om_{\G p,\mu}}}
\lf(\sqrt{\frac{\om_{\G p,\mu}+m_\mu}{\om_{\G p,e}+m_e}}+\sqrt{\frac{\om_{\G p,e}+m_e}{\om_{\G p,\mu}+m_\mu}}\ri) \, .
\eea

\begin{figure}[h]
        \centering       \includegraphics[scale=0.6]{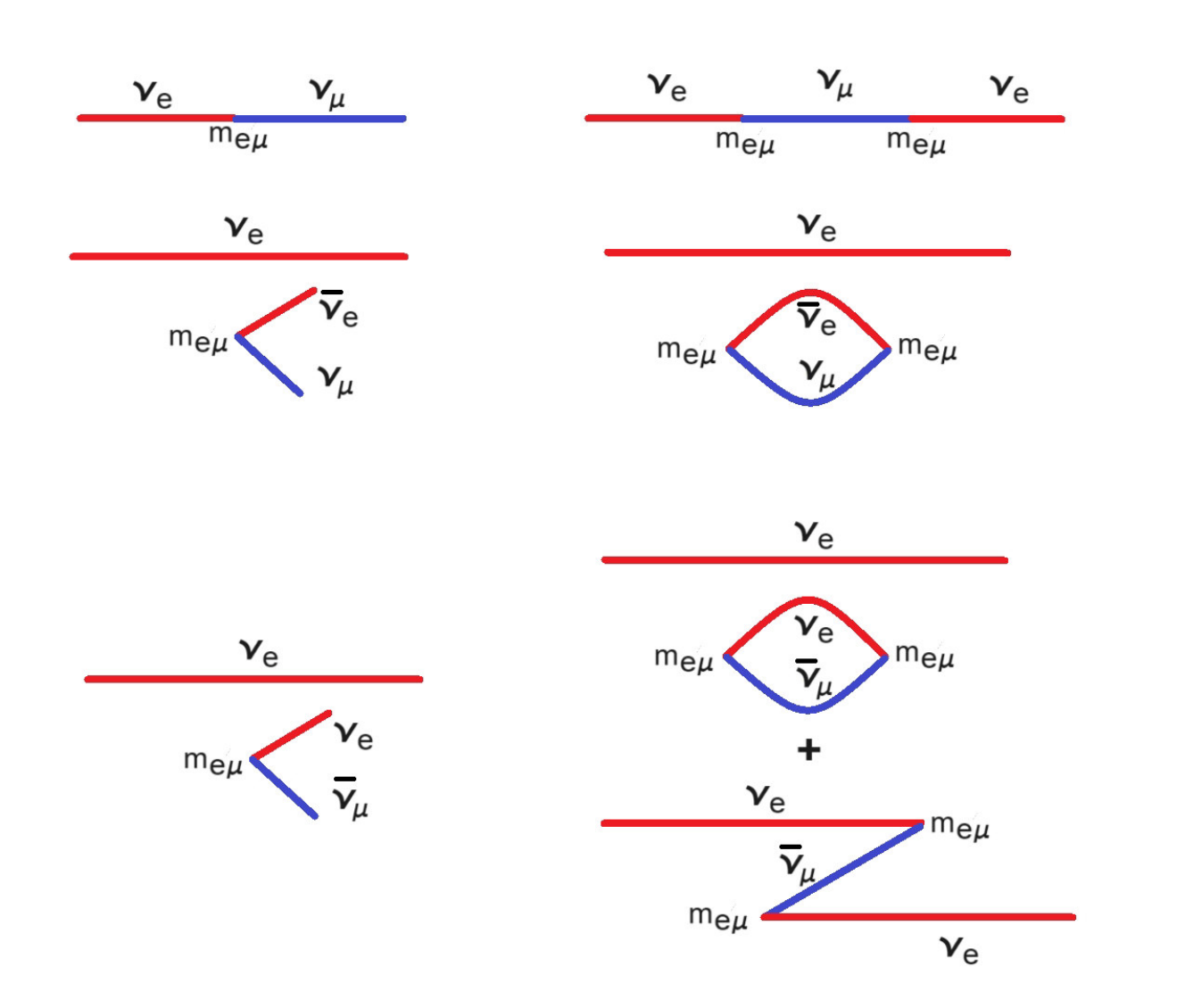}\\
        \caption{Schematic representation of the different contributions.
Left column: the first-order flavor-transition;
Right column: the corresponding second-order diagrams, whose imaginary part reduces to the contribution shown on the left.
In detail:
(i) The first diagram on the left represents the direct transition $\nu_e \rightarrow \nu_{\mu}$, as given in Eq.(\ref{wcon}). This corresponds to the standard oscillation mechanism, which depends on the energy difference between mass eigenstates.
(ii) The second diagram on the left corresponds to Eq.\eqref{second}, but its contribution vanishes because it must be subtracted—it is a disconnected vacuum diagram, as illustrated on the right-hand side.
(iii) The third diagram on the left corresponds to Eq.~(\ref{thirda}). This diagram contains both a subtractive (unphysical) component and a physical contribution that survives. The surviving part is shown on the right-hand side as two ``survival" diagrams, where the sum of energies appears.  }
        \label{diagrams}
   \end{figure}

Let us now analyze the flavor-transition and survival processes. The corresponding Feynman diagrams are depicted in figure \ref{diagrams}.

The simplest non-trivial flavor transition process is
\be
|\nu^r_{\G p,e}\ran \ \rightarrow \ |\nu^s_{\G k,\mu}\ran \, , 
\ee
which corresponds to the upper diagram on the left-column of figure \ref{diagrams}. Its amplitude reads
\bea \non
{}\hspace{-3mm} \mathcal{A}^{rs}_{e \to \mu}(\G p,\G k, ;t_i,t_f) \ &\approx & \ - i m_{e \mu}\de_{r s} \de_{\G k,\G p} W_{\G p}\, \ \int^{t_f}_{t_i} \!\! \dr t \, e^{i\lf(\om_{\G k,\mu}-\om_{\G p,e}\ri)t} \\[2mm] \non   
& = & m_{e \mu} \,  \de_{r s} \de_{\G k,\G p}  \, \,  \lf(e^{i\lf(\om_{\G p,\mu}-\om_{\G p,e}\ri)t_f}-e^{i \lf(\om_{\G p,\mu}-\om_{\G p,e}\ri)t_i}\ri)  \frac{W_{\G p}}{\om_{\G k,e}-\om_{\G k,\mu}} \\ 
&=&   \de_{r s} \de_{\G k,\G p}  \,  \tilde{\mathcal{A}}_{e \to \mu}(\G k; t_i,t_f) \, , 
\eea
where
\be
\tilde{\mathcal{A}}_{e \to \mu}(\G p; t_i,t_f) \ = \ \frac{m_{e \mu} \, W_{\G p}}{\om_{\G p,e}-\om_{\G p,\mu}} \, \lf(e^{i\lf(\om_{\G p,\mu}-\om_{\G p,e}\ri)t_f}-e^{i \lf(\om_{\G p,\mu}-\om_{\G p,e}\ri)t_i}\ri) \, . 
\ee

As in the boson case (see Eq.\eqref{probsum}), the oscillation probability is computed by taking the square and summing over the final density of states, now including the sum over the helicities:
\bea \non
&&\mathcal{P}_{e \to \mu}(\G p;\De t)  =  \sum_{\G k,s} |\mathcal{A}^{rs}_{e \to \mu}(\G p, \G k;t_i,t_f)|^2 \ = \ |\tilde{\mathcal{A}}_{e \to \mu}(\G p, t_i,t_f)|^2 \\[2mm] \label{wcon}
& &\qquad =   W_{\G p}^2 \, \frac{2 m^2_{e \mu} }{\lf(\om_{\G p,e}-\om_{\G p,\mu}\ri)^2}  \lf[1-\cos\lf[\lf(\om_{\G p,\mu}-\om_{\G p,e}\ri)\De t\ri] \ri]  \, ,   \quad  \De t \equiv t_f-t_i \, .
\eea

Another non-trivial process to be considered is the decay
\be
|\nu^r_{\G p,e}\ran \ \rightarrow \ |\nu^{s_1}_{\G k_1,e}\ran |\nu^{s_2}_{\G k_2,\mu}\ran |\overline{\nu}^{s_3}_{\G k_3,e}\ran \, ,
\ee
which corresponds to the middle diagram on the left-column of figure \ref{diagrams}.

The amplitude reads
\bea \non
&&\hspace{-6mm} \mathcal{A}^{r s_1 s_2 s_3}_{e \to e\overline{e} \mu}(\G p,\G k_1,\G k_2, \G k_3;t_i,t_f)  \approx
 -i \, m_{e \mu} \, \, Y_{\G k_2}^{s_3 s_2} \, \de_{\G k_1, \G p}  \de_{\G k_2, - \G k_3}\, \de_{r s_1}   \int^{t_f}_{t_i} \!\! \dr t \, e^{-i\lf(\om_{\G k_2,\mu}+\om_{\G k_2,e}\ri)t}   \\[2mm] \non
&& = \  -m_{e \mu}  \, \de_{r s_1} \,  \de_{\G k_1, \G p}  \de_{\G k_2, - \G k_3} \,  \lf(e^{-i\lf(\om_{\G k_2,\mu}+\om_{\G k_2,e}\ri)t_f}-e^{-i \lf(\om_{\G k_2,\mu}+\om_{\G k_2,e}\ri)t_i}\ri)  \frac{Y_{\G k_2}^{s_2 s_3}}{\om_{\G k_2,e}+\om_{\G k_3,\mu}} \\
&& =  \ \de_{\G k_1, \G p}  \de_{\G k_2, - \G k_3} \,  \de_{r s_1} \, \tilde{\mathcal{A}}^{s_2 s_3}_{e \to e\overline{\mu}\mu}(\G k_2;t_i,t_f) \, , 
\eea
where
\be
\tilde{\mathcal{A}}^{s_2 s_3}_{e \to e\overline{e} \mu }(\G k;t_i,t_f) \ = \ -\frac{m_{e \mu} \, Y^{s_2 s_3}_{\G k}}{\om_{\G k,e}+\om_{\G k,\mu}} \, \lf(e^{-i\lf(\om_{\G k,\mu}+\om_{\G k,e}\ri)t_f}-e^{-i \lf(\om_{\G k,\mu}+\om_{\G k,e}\ri)t_i}\ri) \, . 
\ee
Then, corresponding probability is
\bea \non
\mathcal{P}_{e \to e\overline{e} \mu}(\G p;\De t)  & = & \sum_{\G k_1,\G k_2,\G k_3} \sum_{s_1,s_2,s_3} |\mathcal{A}^{r s_1 s_2 s_3}_{e \to e\overline{e} \mu}(\G p,\G k_1,\G k_2, \G k_3;t_i,t_f) |^2  \\  &=&  
\sum_{\G k}\sum_{s_2,s_3}|\tilde{\mathcal{A}}^{s_2 s_3}_{e \to e\overline{e} \mu }(\G k;t_i,t_f)|^2\, .
\eea
In the large-$V$ limit we get
\be\label{second}
\mathcal{P}_{e \to e\overline{e} \mu}(\G p;\De t)  \ = \ V \sum_{s_2,s_3} \, \int \!\! \frac{\dr^3 \G k}{(2 \pi)^3} \, \frac{\lf(Y^{s_2 s_3}_\G k\ri)^2 }{\lf(\om_{\G k,e}+\om_{\G k,\mu}\ri)^2}  \sin^2\lf[\lf(\om_{\G k,\mu}+\om_{\G k,e}\ri)\De t/2\ri] \, . 
\ee
This is an entirely divergent contribution (vacuum diagram) and it must be subtracted.

The last first-order process is
\be
|\nu^r_{\G p,e}\ran \ \rightarrow \ |\nu^{s_1}_{\G k_1,e}\ran |\nu^{s_2}_{\G k_2,e}\ran |\overline{\nu}^{s_3}_{\G k_3,\mu}\ran \, , \qquad \G k_1 \neq \G k_2 \ \lor \ s_1 \neq s_2 \, ,
\ee
which corresponds to the lower diagram on the left-column of figure \ref{diagrams}.
The amplitude reads
\bea \non
\mathcal{A}^{r s_1 s_2 s_3}_{e \to e e \overline{\mu}}(\G p,\G k_1,\G k_2, \G k_3;t_i,t_f)
& = & \de_{\G k_1, \G p}  \de_{\G k_2, - \G k_3} \,  \de_{r s_1} \, \tilde{\mathcal{A}}^{s_2 s_3}_{e \to e e\overline{\mu}}(\G k_2;t_i,t_f)
\\
&&-\de_{\G k_2, \G p}  \de_{\G k_1, - \G k_3} \,  \de_{r s_2} \, \tilde{\mathcal{A}}^{s_1 s_3}_{e \to e e\overline{\mu}}(\G k_1;t_i,t_f) \, . 
\eea
where $\tilde{\mathcal{A}}^{s_2 s_3}_{e \to e e \overline{\mu} }(\G k;t_i,t_f)=\tilde{\mathcal{A}}^{s_2 s_3}_{e \to e\overline{e} \mu }(\G k;t_i,t_f)$. Note that it vanishes when $\G k_1=\G k_2$ and $s_1=s_2$.
We thus get probability
\bea
&& \mathcal{P}_{e \to e e \overline{\mu}}(\G p;\De t)  \ = \ \ha \sum_{\G k_1,\G k_2,\G k_3} \sum_{s_1,s_2,s_3} |\mathcal{A}^{r s_1 s_2 s_3}_{e \to e e \overline{\mu}}(\G p,\G k_1,\G k_2, \G k_3;t_i,t_f) |^2 \non \\[2mm]
&& =  \sum_{\G k,s_2,s_3} |\tilde{\mathcal{A}}^{ s_2 s_3}_{e \to e e\overline{\mu}}(\G k;t_i,t_f)|^2-\sum_{s_3} |\tilde{\mathcal{A}}^{r s_3}_{e \to e e\overline{\mu}}(\G p;t_i,t_f)|^2\, .
\eea
Now we have to face with a delicate point: we cannot simply subtract the first piece involving the sum over momenta: this procedure would give a negative probability as final result. The way to proceed is to
notice that, because of the Pauli principle, the vacuum should not carry the contribution with $\G k= \G p$, and then to isolate such  a contribution from the sum
\bea
&&\mathcal{P}_{e \to e e \overline{\mu}}(\G p;\De t)   =   \sum_{\G k \neq \G p,s_2,s_3} |\tilde{\mathcal{A}}^{ s_2 s_3}_{e \to e e\overline{\mu}}(\G k;t_i,t_f)|^2+ \sum_{s_2,s_3} |\tilde{\mathcal{A}}^{ s_2 s_3}_{e \to e e\overline{\mu}}(\G p;t_i,t_f)|^2
  \\ \non
  &&-\sum_{s_3} |\tilde{\mathcal{A}}^{r s_3}_{e \to e e\overline{\mu}}(\G p;t_i,t_f)|^2\ = \sum_{\G k \neq \G p,s_2,s_3} |\tilde{\mathcal{A}}^{ s_2 s_3}_{e \to e e\overline{\mu}}(\G k;t_i,t_f)|^2+\sum_{s_3} |\tilde{\mathcal{A}}^{r s_3}_{e \to e e\overline{\mu}}(\G p;t_i,t_f)|^2 \, .
\eea
Then, in the large-$V$ limit
\be
\mathcal{P}_{e \to e e \overline{\mu}}(\G p;\De t)  \ = \ V \sum_{s_2,s_3} \, \int \!\! \frac{\dr^3 \G k}{(2 \pi)^3} \, |\tilde{\mathcal{A}}^{ s_2 s_3}_{e \to e e \overline{\mu}}(\G k;t_i,t_f)|^2+\sum_{s_3} |\tilde{\mathcal{A}}^{r s_3}_{e \to e e\overline{\mu}}(\G p;t_i,t_f)|^2 \, . 
\ee
As in the previous case, the first piece diverges and must be subtracted, while the second piece gives a finite contribution we will keep. The finite contribution can be explicitly computed:
\be\label{thirda}
\mathcal{P}_{e \to e e \overline{\mu}}(\G p;\De t)  \ = \  \frac{4 m_{e \mu}^2 Y_\G p^2 }{\lf(\om_{\G p,e}+\om_{\G p,\mu}\ri)^2}  \sin^2\lf[ \lf(\om_{\G p,\mu}+\om_{\G p,e}\ri)\De t/2\ri]  \, ,
\ee
where we have defined
\be
Y_{\G p}^2 \ = \ \sum_{\G s} \lf(Y^{rs}_{\G p}\ri)^* Y^{rs}_{\G p} \, .
\ee
Its expression can be readily find:
\be
Y_{\G p} \ = \ \frac{|\G p|}{\sqrt{4 \om_{\G p,e}\om_{\G p,\mu}}}
\lf(\sqrt{\frac{\om_{\G p,\mu}+m_\mu}{\om_{\G p,e}+m_e}}+\sqrt{\frac{\om_{\G p,e}+m_e}{\om_{\G p,\mu}+m_\mu}}\ri) \, .
\ee

Finally, the total flavor-transition probability is obtained by summing the two finite contributions computed above:
\bea \non
\mathcal{P}^{e}_{D}(\G p; \De t)  &=&   \frac{4 m^2_{e \mu}W_{\G p}^2}{\lf(\om_{\G p,e}-\om_{\G p,\mu}\ri)^2}  \sin^2\lf[\lf(\om_{\G p,\mu}-\om_{\G p,e}\ri)\De t/2\ri]
\\ &&
+ \frac{4 m^2_{e \mu}Y_\G p^2 }{\lf(\om_{\G p,e}+\om_{\G p,\mu}\ri)^2}  \sin^2\lf[\lf(\om_{\G p,\mu}+\om_{\G p,e}\ri)\De t/2\ri]  \, . \label{wyprob}
\eea

Flavor transitions between neutrinos are forbidden in the asymptotic limit \( t_i \to -\infty \), \( t_f \to +\infty \), unless the neutrino masses involved are equal (e.g., \( m_e = m_\mu \)). In this regime, energy conservation is strictly imposed by four-momentum delta functions, which prohibit flavor-changing processes. 
Once more, this remarks the essential role of energy uncertainty in the phenomenon of neutrino oscillations, consistent with the interpretation offered by the flavor-energy uncertainty relation (see Section \ref{TEUR}), and closely parallels similar features observed in the quantum description of unstable particles.

Let us now observe that, at the first order in $m_{e \mu}$ (and thus is $\theta$), $m_e \approx m_1$ and $m_\mu \approx m_2$ (see Eqs.\eqref{met},\eqref{mmut}). Therefore, the Bogoliubov coefficients \eqref{ucoff},\eqref{vcoff} take the form
\bea 
 |U_\G p| & \approx &   \sqrt{\frac{\lf(\om_{\G p,e}+m_e\ri)\lf(\om_{\G p,\mu}+m_\mu\ri)}{4\om_{\G p,e}\om_{\G p,\mu}}}  \lf(1+\frac{|\G p|^2}{\lf(\om_{\G p,e}+m_e\ri)\lf(\om_{\G p,\mu}+m_\mu\ri)}\ri)\, ,\\[2mm]
|V_\G p| & \approx &  \sqrt{\frac{\lf(\om_{\G p,e}+m_e\ri)\lf(\om_{\G p,\mu}+m_\mu\ri)}{4\om_{\G p,e}\om_{\G p,\mu}}} \lf(\frac{|\G p|}{\om_{\G p,e}+m_e}-\frac{|\G p|}{\om_{\G p,\mu}+m_\mu}\ri)\, .
\eea
These are related to $W_\G p$ and $Y_\G p$ as
\be 
 |U_\G p| \, \approx \, W_{\G p}\frac{m_\mu-m_e}{\om_{\G p,e}-\om_{\G p,\mu}} \, ,\quad
|V_\G p| \,\approx \, Y_{\G p}\frac{m_\mu-m_e}{\om_{\G p,e}+\om_{\G p,\mu}} \, .
\ee
We can thus rewrite the probability \eqref{wyprob}, at the leading order in $m_{e \mu}$, in the form
\begin{small}
\be \label{mprob}
\mathcal{P}^{e}_{D}(\G p; \De t)  \ = \ \sin^2 2 \theta \lf[ |U_\G p|^2 \sin^2\lf[ \lf(\om_{\G p,\mu}-\om_{\G p,e}\ri)\De t/2 \ri]+ |V_\G p|^2  \sin^2\lf[ \lf(\om_{\G p,\mu}+\om_{\G p,e}\ri)\De t /2 \ri] \ri] \, . 
\ee
\end{small}
with $\theta=m_{e \mu}/(m_\mu-m_e) \approx \sin \theta$.

Therefore, in the approximation we used this expression coincides with the oscillation probability \eqref{oscfor}. This is a remarkable result because the present approach has not required the definition of flavor charges or the construction of a flavor Fock space, nevertheless the Bogoliubov coefficients emerged from the perturbation expansion.

Let us now compute the survival probability \(\mathcal{P}^e_{S}(\boldsymbol{k}, \Delta t)\). At the zeroth order, the contribution is simply \(\mathcal{P}^e_{S}(\boldsymbol{k}, \Delta t) = 1\). To obtain a non-trivial correction to the survival probability, we need to evaluate the second-order terms. In this case, it is convenient to write
\bea\non
&& U(t_i,t_f) \ = \ \ide -i \, m_{e \mu} \int^{t_f}_{t_i} \!\! \dr^4 x \, : \overline{\nu}_e(x) \nu_\mu(x)+\overline{\nu}_\mu(x) \nu_e(x): \\ [2mm] \non 
&& - \  \frac{m^2_{e \mu}}{2} \int^{t_f}_{t_i} \!\! \dr^4 x_1 \int^{t_f}_{t_i} \!\! \dr^4 x_2 \, \, \mathcal{T}\Big[   \lf(:\overline{\nu}_e(x_1) \nu_\mu(x_1)+\overline{\nu}_\mu(x_1) \nu_e(x_1):\ri) \times
\\
&& \qquad\qquad\qquad \times\lf( : \overline{\nu}_e(x_2) \nu_\mu(x_2)+\overline{\nu}_\mu(x_2) \nu_e(x_2):\ri)\Big] \, + \, \ldots \, . 
\eea

The second order piece can be expanded by means of the Wick theorem
\bea \non
&&U^{(2)}(t_i,t_f) =  -\frac{m_{e \mu}^2}{2} \int^{t_f}_{t_i} \!\! \dr^4 x_1 \int^{t_f}_{t_i} \!\! \dr^4 x_2 \, \,    \Big[:\overline{\nu}_e(x_1) \nu_\mu(x_1)\overline{\nu}_e(x_2) \nu_\mu(x_2):   \non \\ [2mm]
& &+:\overline{\nu}_e(x_1) \nu_\mu(x_1)\overline{\nu}_\mu(x_2) \nu_e(x_2):+ :\overline{\nu}_\mu(x_1) \nu_e(x_1)\overline{\nu}_e(x_2) \nu_\mu(x_2):
\non \\ [2mm]
& &
+:\overline{\nu}_\mu(x_1) \nu_e(x_1)\overline{\nu}_\mu(x_2) \nu_e(x_2): \non \\[2mm] \non
&&+  i \lf(S^e_{\al \bt}(x_1-x_2) \, :\overline{\nu}^\al_\mu(x_1) \nu^\bt_\mu(x_2):+ S^\mu_{\al \bt}(x_1-x_2) \, :\overline{\nu}^\al_e(x_1) \nu^\bt_e(x_2):\ri)  \\ [2mm]
&&- i \lf(S^e_{\bt \al}(x_2-x_1) \, :\nu^\al_\mu(x_1) \overline{\nu}^\bt_\mu(x_2) :+ S^\mu_{\bt \al}(x_2-x_1) \, : \nu^\al_e(x_1) \overline{\nu}^\bt_e(x_2):\ri) \non \\[2mm]
&&+ S^\mu_{\al \bt}(x_1-x_2)S^e_{\bt \al}(x_2-x_1) + S^e_{\al \bt}(x_1-x_2)S^\mu_{\bt \al}(x_2-x_1)  \Big] \, , \label{secu}
\eea
where
\be \label{perprop}
S^\si_{\al \bt}(x) \ = \ \int \!\! \frac{\dr^4 p}{(2 \pi)^4} \, e^{-i p x} \, \frac{\lf(\slashed{p}+m_\si\ri)_{\al \bt}}{p^2-m^2_\si+i \varepsilon} \, , \qquad \si=e,\mu \, ,
\ee
is the Dirac propagator. From this expression one can derive the diagrams on the right column of figure \ref{diagrams}. The employment of Wick theorem makes evident that the divergent vacuum contributions come from the terms with a full-contraction, which correspond to the diagrams with the vacuum bubble, and which must be subtracted.

The survival process is
\be
|\nu^r_{\G p,e}\ran \ \rightarrow \ |\nu^r_{\G p,e}\ran \, .
\ee
Saving only up to linear terms in $m_{e \mu}$, the amplitude can be written as
\be
\mathcal{A}^{r}_{e \to e}(\G p; t_i,t_f)=1+\ha \mathcal{A}^{(2) r}_{e \to e}(\G p; t_i,t_f) \, , 
\ee
where $A^{(2) r}_{e \to e}(\G k;t_i,t_f)$ is the second-order term, which is proportional to $m_{e \mu}^2$. Taking the square, and summing over the final momentum and helicity, and disregarding the terms containing  powers $m^n_{e \mu}$, with $n>2$, we get
\be
\mathcal{P}^{e}_{S}(\G p; \De t) \ \approx \ 1+ \Re e \lf(\mathcal{A}^{(2)}_{e \to e}(\G p;t_i,t_f)\ri) \, . 
\ee
Explicitly
\bea\non
\mathcal{P}^{e}_{S}(\G p; \De t) &=&    1-\sin^2 2 \theta\,  |U_\G p|^2 \sin^2\lf[\lf(\om_{\G p,\mu}-\om_{\G p,e}\ri)\De t/2 \ri]
\\ [2mm]
&&\quad -\sin^2 2 \theta \, |V_\G p|^2  \sin^2\lf[\lf(\om_{\G p,\mu}+\om_{\G p,e}\ri)\De t/2\ri]  \, . 
\eea
Then
\be
\mathcal{P}^{e}_{D}(\G p; \De t) + \mathcal{P}^{e}_{S}(\G p; \De t)\ = \ 1 \, ,
\ee
thus preserving the unitarity as it should be.

%

\section{Conclusions} \label{conclusion}

In this review, we have provided a detailed and pedagogical introduction of two complementary QFT approaches to neutrino mixing and oscillations: the non-perturbative flavor Fock space formalism and the perturbative interaction picture framework. 

Starting from a minimally extended Standard Model including (Dirac) neutrino masses and mixing, we have shown how neutrino flavor charges can be consistently introduced.
We have revisited the standard QM treatment of neutrino oscillations, presenting both the conventional derivation and an alternative first-quantized approach based on the Dirac equation. We then reviewed the construction of the flavor Fock space, highlighting its non-perturbative features, such as the emergence of unitarily inequivalent representations and the condensate structure of the flavor vacuum. These aspects lead to several physically relevant consequences, including the conservation of lepton number at tree level, the entangled nature of flavor states, and the appearance of non-trivial corrections to the standard oscillation formula in the non-relativistic regime. We also derived the time–energy uncertainty relation (TEUR) for neutrino oscillations within this formalism.

In parallel, we have explored a perturbative approach in which neutrino mixing is interpreted as an interaction among flavor fields. Using Dyson expansion techniques in the interaction (Dirac) picture, we computed flavor transition and survival probabilities. Remarkably, this approach reproduces the same oscillation formulas as the non-perturbative Fock space method, within the limits of the perturbative approximation. In both cases, it is crucial to work at finite times, rather than relying on the standard 
$S$-matrix framework, in order to properly describe flavor oscillations and maintain consistency with the time–energy uncertainty relation. We emphasized the structural similarities between this framework and the QFT treatment of unstable particles, particularly regarding the role of finite-time Feynman diagrams.  

Taken together, these two perspectives offer a unified and conceptually rich framework for understanding neutrino oscillations in QFT. The comparison between non-perturbative and perturbative methods not only clarifies foundational aspects of the theory, but also opens the door to new developments, both for a deeper understanding of the theoretical structure and for phenomenological applications. Open questions remain, particularly concerning the nature of the flavor vacuum and its dynamical origin, possible extensions to curved spacetime or thermal backgrounds, and the role of quantum entanglement and decoherence in realistic detection scenarios. We hope that this review will serve as a useful reference for further investigations into the QFT foundations of neutrino physics.

We have seen that flavor neutrino states can be interpreted as entangled states. This observation paved the way for an active research activity, investigating various aspects of quantum correlations associated with neutrino oscillations~\cite{Banerjee:2015mha,Alok:2014gya,Dixit:2018gjc,Wang:2020vdm,Ming:2020nyc,Blasone:2021cau,Li:2022mus,Bittencourt:2022tcl,Bittencourt:2023asd}.
Also, it has been shown that neutrino oscillation data can be used to probe violations of \emph{Leggett--Garg inequalities}~\cite{Formaggio2016}, a temporal analogue of Bell's inequalities. Since then, several works have explored how neutrino (and meson) oscillations may serve as a testing ground for foundational aspects of QM~\cite{Naikoo:2019gme,gango,Blasone:2022iwf,Blasone:2024jur}.
In particular, in Ref.~\cite{Blasone:2021mbc}, the flavor Fock-space formalism was employed to study the Leggett--Garg inequalities in the context of neutrino oscillations. It was found that the QFT-based approach is less compatible with the classical notion of realism than the standard QM treatment. The entanglement of (bosonic) flavor vacuum was investigated in Ref.\cite{Blasone:2021aiu}.

What about phenomenological implications? The corrections to the standard neutrino oscillation formula~\eqref{psirho}, which are introduced by the present formalism, could become relevant in the context of measurements of the so-called \emph{cosmic neutrino background} (C$\nu$B). This will be investigated by the PTOLEMY experiment~\cite{Betts:2013uya,Cocco:2017nax,Messina:2018xzi,PTOLEMY:2019hkd}, which aims to detect relic neutrinos from the C$\nu$B, neutrinos that decoupled approximately one second after the Big Bang, through the capture process on tritium:
\[
\nu_e + {}^3\text{H} \rightarrow e^- + {}^3\text{He}.
\]

In the standard framework of neutrino oscillations, the rate for this process is given by~\cite{PTOLEMY:2019hkd}
\begin{equation}
\Gamma_{\text{CNB}} = \sum_j |U_{e j}|^2 \, \bar{\sigma} \, v_\nu \, f_{e,j} \, n_0 \, ,
\end{equation}
where \( U \) is the PMNS mixing matrix, \( v_\nu \) is the neutrino velocity as measured on Earth, \( \bar{\sigma} \) is the averaged cross section, \( n_0 \) is the large-scale average neutrino number density, and \( f_{e,j} \) are the clustering factors for each mass eigenstate.
We expect that the analysis developed in this work leads to corrections to the above expression, through the inclusion of the Bogoliubov coefficients \( |U_{\boldsymbol{p}}| \) and \( |V_{\boldsymbol{p}}| \). This expectation is supported by the results of Ref.~\cite{Lee:2017cqf}, where the rate of \(\beta\)-decay,
\be
{}^3\text{H} \rightarrow e^- + {}^3\text{He} + \nu_e \, ,
\ee
was computed within the flavor Fock-space formalism. As discussed in the main text, the corrections in that case, and thus in the C$\nu$B case as well, become significant in the non-relativistic regime. A first attempt to compute the tritium capture rate within this framework was made in Ref.~\cite{Capolupo:2022hhr}.

The nature of flavor neutrino states has been investigated also from a theoretical point of view. Indeed, by calculating the decay rate of an accelerated proton, it was found that general covariance is preserved only when flavor neutrino states (and thus not massive neutrino states) are assumed to be produced in the weak interaction vertices \cite{Ahluwalia:2015kxa,Blasone:2018czm,Cozzella:2018qew,Blasone:2019agu,Blasone:2020vtm}.
On the other hand, flavor vacuum structure (both for bosons and for fermions) induces breakdown of thermality for Unruh radiation\cite{Blasone:2017nbf,Luciano:2021mto,Luciano:2021onl} and violations of the weak equivalence principle \cite{nonrel,Blasone:2023yqz}. In fact, if the flavor neutrino states have to be considered as the fundamental entities, as it emerges from the above discussion, one should consider how to fit such elementary particles in terms of irreducible representations of the Poincar\'e group\cite{Blasone:2003wf,Lobanov:2015esa,bigs2}.

To conclude, we would like to mention that both the perturbative and non-perturbative approaches reviewed in this paper can be employed to investigate the intriguing phenomenon of \emph{chiral oscillations}~\cite{Bittencourt:2024yxi,Blasone:2025hjw,Morozumi:2025gmw}. As we have seen, charged-current weak processes involve only left-handed leptons. However, chirality is not a conserved quantity for massive particles: this gives rise to the phenomenon of chiral oscillations~\cite{DeLeo:1996gt,fukugita2003physics,PhysRevD.71.076008,bernardini2011quantum,Bittencourt:2020xen,suekane2021quantum}, which is relevant in the non-relativistic regime and thus could affect neutrinos of  the C$\nu$B, see Refs.\cite{Bittencourt:2020xen,Ge:2020aen}.
Within the non-perturbative formalism, the study in QFT of chiral and flavor oscillations has been also recently unified~\cite{Blasone:2025dye}. The resulting chiral-flavor oscillation formula is in agreement with the one previously derived in Ref.~\cite{PhysRevD.71.076008}, in a framework based on the Dirac equation, as the one reviewed in Section~\ref{DiracEq}.


\appendix
\section{Unitarily inequivalent representations for fields with different masses} \label{ineqcar}
Let us consider two Dirac fields $\tilde{\psi}$ and $\psi$ with  masses  $\tilde{m} = 0$ and $m \neq 0$:
\be
i\gamma^\mu\partial_\mu \tilde{\psi}=0 \, , \qquad
(i\gamma^\mu\partial_\mu-m)\psi =0 \, .
\ee
At $t=0$, these fields can be expanded as
\bea \label{psi}
\psi(\textbf{x}) & = & \frac{1}{\sqrt{V}} \, \sum_{\G k} \sum_{r=\pm 1}[\al^r_{\textbf{k}} \, u^r_{\textbf{k}} \, e^{i\textbf{k} \cdot \textbf{x}} \, + \, \bt^{r\dagger}_{\textbf{k}} \, v^r_{\G k}\, e^{-i\textbf{k} \cdot \textbf{x}}]\, ,\label{spinfo} \\[2mm] \label{tildepsi}
\tilde{\psi}(\textbf{x}) & = & \frac{1}{\sqrt{V}} \, \sum_{\G k} \sum_{r=\pm 1}[\tilde{\al}^r_{\textbf{k}} \, \tilde{u}^r_{\textbf{k}} \, e^{i\textbf{k} \cdot \textbf{x}} \, + \, \tilde{\bt}^{r\dagger}_{\textbf{k}} \, \tilde{v}^r_{\G k}\, e^{-i\textbf{k} \cdot \textbf{x}}]\, .
\eea

We want to express the creation and annihilation operators of one field in terms of the other ones. To this aim we can impose the boundary condition $\psi(0)=\tilde{\psi}(0)$:
\cite{barton1963introduction,Miransky:1994vk}
\begin{equation}
\sum_{r=\pm} \, (\tilde{\al}^r_{\textbf k}\, \tilde{u}^r_{\G k} \, + \, \tilde{\bt}^\dagger_{-\textbf k} \, \tilde{v}^r_{-\textbf k}) \, = \, \sum_{r=\pm} \, (\al^r_{\textbf k}\, u^r_{\textbf k}+\bt^{r\dagger}_{-\textbf k} \, v^r_{-\textbf k}) \, .
\end{equation}
We can thus isolate $\al^r_{\textbf k}$ and $\bt^{r \dagger}_{-\textbf k}$ using the orthogonality relations of Dirac spinors:
\bea  \label{bogfer1}
\al^r_{\textbf k} & = & \cos \Theta_\G k \, \tilde{\al}^r_{\textbf k} \, + \, \epsilon^r \, \sin \Theta_\textbf k \, \tilde{\bt}^{r \dagger}_{-\textbf k} \, , \\[2mm] \label{bogfer2}
\bt^{r \dagger}_{-\textbf k} & = & \cos \Theta_\textbf k \, \tilde{\bt}^{r\dagger}_{-\textbf k} \,-\, \epsilon^r \, \sin \Theta_\G k \,  \tilde{\al}^r_{\textbf k}   \, ,
\eea
where $\Theta_\G k= \ha \, \cot^{-1}(|\G k|/m)$. This is a canonical transformation known as Bogoliubov transformation \cite{barton1963introduction,berezin1966method,umezawa1982thermo,umezawa1993advanced,Miransky:1994vk,blasone2011quantum}.

From Eqs. (\ref{bogfer1}) and (\ref{bogfer2}) one can see that the vacuum annihilated by the tilde operators $\tilde{\al}^{r}_{\textbf k} \, |\tilde{0}\rangle  = 0 = \tilde{\bt}^r_{\textbf k} \, |\tilde{0}\rangle$ ,  is not annihilated by the non-tilde ones, $\al^{r}_{\textbf k} \, |\tilde{0}\rangle \neq 0$,  $\bt^r_{\textbf k} \, |\tilde{0}\rangle \neq 0 $. To find the relation between $|0\rangle $  ($ \al^{r}_{\textbf k} \, |0\rangle  = 0 =  \bt^r_{\textbf k} \, |0\rangle $) and $|\tilde{0}\rangle$ and thus between the Fock spaces $\mathcal{H}$ and $\tilde{\mathcal{H}}$ to which they belong, we must find the generator $B$ for Eqs. \eqref{bogfer1} and \eqref{bogfer2}:
\begin{equation}
\al^r_{\G k}\ = \ B(m) \, \tilde{\al}^r_{\textbf k} \, B^{-1}(m) \, , \qquad \bt^{r}_{\textbf k}\ = \ B(m) \, \tilde{\bt}^r_{\textbf k} \, B^{-1}(m) \, . \label{trance}
\end{equation}
Employing the Baker–-Campbell–-Hausdorff formula, one can check that $B$ has the form
\begin{equation} \
B(m) \ = \ \exp\left[\sum_{r,\G k} \, \epsilon^r \, \Theta_\textbf k \, \left(\tilde{\al}^r_{\G k} \, \tilde{\bt}^r_{-\G k}-\tilde{\bt}^{r\dagger}_{-\G k} \, \tilde{\al}^{r\dagger}_{\G k} \right)\right].\label{generator}
\end{equation}
Thus, the vacuum state $|0\rangle $ can be written as\cite{PhysRev.122.345,PhysRev.124.246}
\begin{equation}
|0 \rangle \ = \ B(m) \,|\tilde{0}\rangle \ = \ \prod_{r,\textbf k}\lf[\cos \Theta_\textbf k\, - \, \epsilon^r \, \sin \Theta_\textbf k \, \tilde{\al}^{r\dag}_{\G k} \, \tilde{\bt}^\dagger_{-\textbf k}\ri] \, |\tilde{0}\rangle \, . \label{coherfer}
\end{equation}

The vacuum-vacuum amplitude is
\begin{equation}
\langle \tilde{0}|0\rangle \ \equiv \ \langle \tilde{0}|B(m)|\tilde{0}\rangle  \ = \  \exp \left(2\sum_\textbf k \log \cos \Theta_\textbf k\right) \, ,
	\end{equation}
which, in the large-$V$ limit reads 
\begin{equation}
\langle \tilde{0}|0\rangle \ = \  \exp \left(2 \, V \,  \int \!\! \frac{\mathrm{d}^3 \G k}{(2\pi)^3} \,  \log \cos \Theta_\textbf k\right) \, .
\end{equation}
Taking into account that $\log \Theta_\textbf k \sim {-m^2}/({8 \, |\G k|^2})$ when $|\G k|\rightarrow \infty$, and introducing an ultraviolet cut-off $\Lambda$, we get
\begin{equation}
\langle \tilde{0}|0 \rangle \ = \  \exp \left(-\frac{\Lambda m^2 V}{8\pi^2} \right)\, .
\end{equation}
which goes to zero in the limit
$V \to \infty$ or $\La \to \infty$: $\langle \tilde{0}|0\rangle \to 0$. This means that in such a limit $|\tilde{0}\ran$ does not belong to the domain of $B(m)$ or, in other words, $|0\ran$ does not belong to $\tilde{\mathcal{H}}$. This means that the two representations of CAR are \emph{unitarily inequivalent} \cite{barton1963introduction,berezin1966method,umezawa1982thermo,umezawa1993advanced,Miransky:1994vk,blasone2011quantum} and the Bogoliubov transformation \eqref{bogfer1}, \eqref{bogfer2} is an \emph{improper canonical transformation}.  Expressions such as Eqs. (\ref{Bogol}) and (\ref{coherfer}), which relate one vacuum to the other, are only formal in the above limits \cite{berezin1966method}.







\bibliographystyle{ws-ijmpa}
\bibliography{libraryNeutrino}

\end{document}